# Insights into Ionic Diffusion in C–S–H Gel Pore from Molecular Dynamics Simulations: Spatial Distributions, Energy Barriers, and Structural Descriptor

Weiqiang Chen[a,b,c] and Kai Gong[a,b,c,*]

[a] Department of Civil and Environmental Engineering, Rice University, Houston, Texas 77005, United States

[b] Rice Advanced Materials Institute, Rice University, Houston, Texas 77005, United States

[c] Ken Kennedy Institute, Rice University, Houston, Texas 77005, United States

* Corresponding author. E-mail: kg51@rice.edu

## Abstract

Understanding transport behavior in nanoconfined environments is critical to many natural and engineering systems, including cementitious materials, yet its molecular-level mechanisms remain poorly understood. Here, molecular dynamics (MD) simulations were used to investigate $Na^+$, $Cl^-$, and water diffusion inside a 4 nm calcium-silicate-hydrate (C–S–H) pore channel over temperatures ranging from 300 K to 360 K. Spatially resolved analysis revealed strong suppression of diffusivity near the solid-liquid interface and gradual recovery toward the pore center. Arrhenius analysis further quantified the spatial variation of activation energy barriers and intrinsic mobilities across the pore channel, showing distinct confinement effects. The spatially resolved structural analysis uncovers a mechanistic transition from structure-controlled to hydrodynamics-controlled transport regimes with increasing distance from the pore surface. A structural descriptor, total coordination strength (TCS), was introduced, providing a predictive link between local liquid structure and molecular mobility within ~1 nm of the interface. Beyond ~1 nm, suppressed diffusivities were well captured by an empirical model inspired by the Darcy–Brinkman framework. To the best of our knowledge, this is the first MD study to comprehensively resolve the spatial heterogeneity of transport, thermal kinetics, and structure within cementitious nanopores. These findings deepen the fundamental understanding of nanoscale transport phenomena and suggest that tailoring the nanochannel structure and interfacial chemistry of cementitious gels, e.g., surface coordination environments, pore size distributions, and adsorption sites, may offer a promising strategy to suppress ionic ingress and enhance the durability of cement-based materials.

# 1 Introduction

**Background and motivation.** Nanoscale ionic transport is of significant interest in numerous natural processes and engineering applications, including mineral weathering,[1, 2] soil nutrient cycling,[3] signal transduction at the cell membrane,[4] water desalination and purification,[5, 6] battery energy storage,[7-9] geological nuclear waste disposal,[10, 11] enhanced oil recovery in tight reservoirs,[12, 13] and cementitious materials.[14, 15] In cement-based materials, ionic transport through micro- and nanoscale pores of binder gels directly impact their permeability to aggressive chemical species often present in building environments, thereby influencing their long-term durability.[16] For example, the ingress of chloride-bearing chemicals, prevalent in marine environments, through the porous network of cement-based cover in reinforced concrete can significantly accelerate the corrosion of the embedded steel rebars, ultimately compromising their structural integrity.[17, 18] Moreover, the transport of water and ions within cementitious gel pores is also intricately linked to other fundamental physicochemical processes in cement-based materials, such as leaching, drying, shrinkage, cracking, and creep.[19] Ionic transport also plays a crucial role in engineering applications where concrete serves as a containment barrier, such as in sanitation infrastructure and nuclear waste disposal systems.[20, 21]

Water and ionic transport in cement-based materials has been extensively studied using a range of experimental techniques, including quasi-elastic neutron scattering (QENS),[22] neutron imaging,[23] nuclear magnetic resonance (NMR),[24] and ionic migration measurement.[25, 26] While these methods have provided valuable insights into transport phenomena, direct molecular-level characterization of ion migration within cementitious nanopores remain elusive. This is primarily due to the inherent complexity of cementitious pore network, which feature a broad range of pore sizes, irregular geometries, diverse pore solution chemistries, and varying levels of connectivity, tortuosity, and relative humidity.[27] These challenges are further compounded by the emergence of alternative low-$CO_2$ binders with increasingly complex chemistries. In this context, molecular dynamics (MD) simulations offer a powerful and complementary approach to address these challenges. By allowing precise control over pore surfaces and solution chemistries, MD simulations provide a unique capability to probe ionic transport and solid-liquid interactions at atomic resolution under various environmental conditions. This molecular-level insight is essential for advancing the design of more durable and sustainable cementitious systems.

**Recent advances in nanoconfinement and temperature effects.** Recent MD studies have significantly advanced our understanding of transport phenomena within different cementitious nanopores, including calcium-silicate-hydrate (C-S-H) gels with a Ca/Si ratio of ~1.7, prevalent in the widely used ordinary Portland cement (OPC) systems,[21, 28, 29] and sodium-aluminosilicate-hydrate (N-A-S-H),[30-34] and calcium-aluminosilicate-hydrate (C-(N)-A-S-H) gels[35, 36] found in alternative alkali-activated cements (an important class of low-$CO_2$ binders [37]). These studies have revealed the critical role of nanoconfinement and temperature in governing adsorption and transport behaviors. Specifically, MD simulations reveal that water and ionic diffusivities within C-S-H gel pores (Ca/Si = ~1.7)[28, 29] and their C-S-H analogues (e.g., tobermorite-based with lower Ca/Si ratios)[15] are significantly reduced compared to bulk conditions due to the nanoconfinement effect, with reduction becoming more pronounced as channel size decreases.[14, 20, 29] In both C-S-H gels (Ca/Si = ~1.7)[38] and tobermorite/jennite-based gels,[39-42] the mobility of ions and water molecules is markedly suppressed near the gel surfaces due to strong solid-fluid interfacial interactions. Nanoconfinement is also seen to cause anisotropic diffusivity in N-A-S-H gel, where the movements of alkali ions ($Cs^+$, $K^+$, and $Na^+$) and water molecules are slower perpendicular to the gel surface than parallel to it.[32] Temperature also plays a crucial role in ionic transport, with increasing temperature generally enhancing ionic diffusivity, as observed for NaCl in tobermorite/jennite-based C-S-H nanopores.[14, 43] However, the elevated temperature has been shown to reduce the capillary transport of sodium sulfate by weakening their hydration shells, promoting ion pairing and adsorption on the tobermorite-based C-S-H pore surface.[44] Moreover, the activation energy for water diffusion in C-S-H gel (Ca/Si = ~1.67) nanopore exhibits a non-monotonic dependence on basal spacing: it is lower than bulk values in ultrafine pores (basal spacing < ~1.8 nm), reaching a maximum near ~2.8 nm, and gradually returns to bulk-like values as the basal spacing continue increasing.[28]

**Research gaps.** Despite these recent advances in understanding nanoscale ionic transport processes in cementitious binder gels, key knowledge gaps remain regarding their spatial heterogeneity and the underlying molecular-level physicochemical mechanisms. Most prior MD studies have reported pore-averaged transport properties, providing limited insight into spatial variations near solid–liquid interfaces. Only a few have examined the spatial evolution of water diffusivity in cementitious nanopores,[38-40] and even fewer have explored the spatial variations in thermal kinetics or liquid structure. This lack of spatially resolved insight hinders the elucidation of the critical interfacial phenomena governing transport behavior. Such insights are also essential for developing mechanistic models and multiscale simulations that rely on accurate nanoscale inputs,[45, 46] as well as for explaining anomalous transport behaviors observed in other nanoconfined systems and at solid-liquid interfaces, as recently reported.[47, 48] In addition, while many studies have quantified the structure attributes of nanoconfined solutions, a unified structural descriptor that link local liquid structure to transport behavior remains elusive.

To address these critical gaps, this study employs force-field MD simulations to resolve the spatial heterogeneity of water and ion transport in aqueous NaCl solutions confined within 4 nm-wide C-S-H gel nanopore (Ca/Si = 1.67), representative of those found in OPC-based systems, across temperatures from 300 K to 360 K. The simulations yield high-resolution spatial profiles of ion and water diffusivity, thermal kinetics and local liquid structure across the nanopore, allowing the atomic origin controlling water and ion diffusivity to be elucidated. The paper is organized as follows: **Section 2** introduces the MD simulation models and methodologies. **Section 3** evaluates water and ionic diffusivity within C-S-H nanopores, including both pore-scale averaged values and their spatial distributions. The temperature-dependent spatial evolution of water and ionic diffusivity is then used to quantify activation energy barriers and intrinsic mobilities. To explain the observed spatial variations in diffusion behaviors, a comprehensive analysis of the liquid structure across the pore is performed, including atomic density distribution, radial distribution functions, and coordination number distribution. A new liquid structure descriptor, total coordination strength (TCS), is introduced to capture the relationship between local coordination environments and water and ion diffusivity near the gel surface, while transport away from the pore surface is described by continuum hydrodynamics. **Section 4** discusses the broader implications and limitations of the studies, followed by a summary of key findings in **Section 5.** Overall, this work deepens our understanding of nanoscale ionic transport in C-S-H gels, and provides mechanistic insight that may inform the design of more sustainable and durable cementitious materials.

## 2 Computational Methods

### 2.1 C-S-H pore model

To study ionic transport in the nanoscale pores of OPC-based cementitious systems, we constructed a slit-shaped nanochannel with a width of ~4 nm (**Figure 1**b) based on a C-S-H gel in **Figure 1**a with a Ca/Si ratio of 1.67, representing the poorly crystalline binder phase in OPC systems.[49] The C-S-H structure in **Figure 1**a was obtained from a widely used molecular model developed by Mohamed et al.,[50] generated from defective 14 Å tobermorite using a building block description and optimized with classical MD simulations and density functional theory (DFT) calculations.[50] The chosen channel size (~4 nm) falls within the experimentally measured gel pore size range of cementitious materials (0.5–10 nm).[27, 51] To create the nanochannel, the interlayer distance was expanded while preserving the intralayer atomic configuration, a common strategy in MD studies of cementitious gel pores.[21, 52]

Once the nanochannel was created, an aqueous NaCl solution at 2 M concentration and a $H_2O/Na^+$ ratio of about 27.75:1 was packed into the nanochannel using the PACKMOL software.[53] The number of water molecules and ion pairs was estimated based on the volume of the channel space and the experimental density of the aqueous NaCl solution. After system equilibration at 300 K, the NaCl solution density in the middle of the nanochannel (well known to approximate bulk state) was found to be $\sim 1.07 \mathrm{g/cm^3}$, which closely matched the interpolated experimental value for 2 M NaCl solution ($\sim 1.07 \mathrm{g/cm^3}$).[54] The high ionic concentration was chosen to improve statistical sampling and ensure reliability in the subsequent MD analyses. The dimensions and views of the final model for the C-S-H gel pore containing 2 M NaCl solution are shown in **Figure 1**b, with each model consisting of ~13,000 atoms.

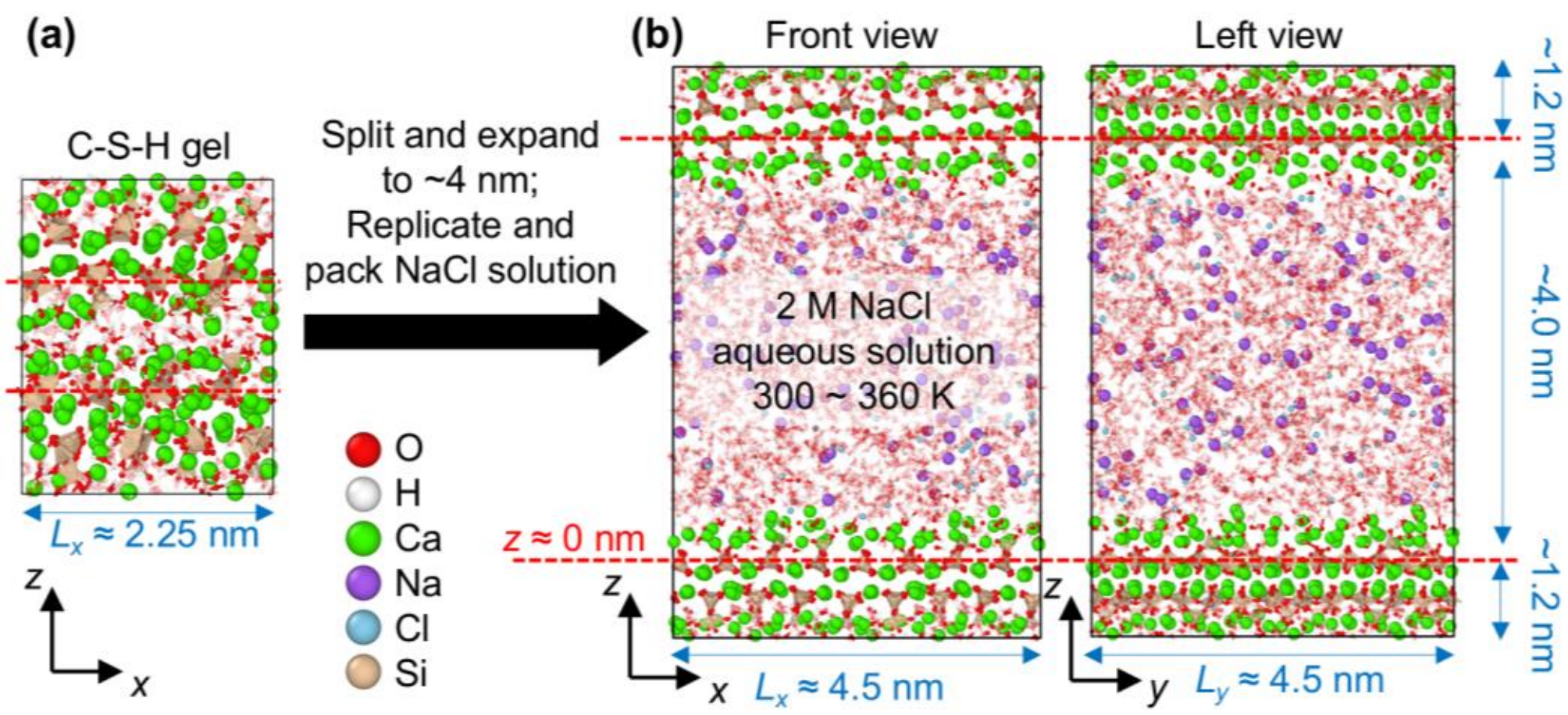

**Figure 1.** Structure model of (a) the C-S-H gel with a Ca/Si ratio of 1.67, constructed using atomic coordinates reported in ref. [50], and (b) the ~4 nm-wide C-S-H nanochannel filled with 2 M NaCl aqueous solution, where $z$ = 0 nm corresponds to the average position of surface Si atoms. Water molecules are shown with partial transparency to enhance visualization.

### 2.2 Simulation details

Based on the nanopore model constructed in **Section 2.1**, force field MD simulations were conducted to examine ionic transport in C-S-H binder gel using the Large-scale Atomic/Molecular Massively Parallel Simulator (LAMMPS),[55] with visualizations performed using the Open Visualization Tool (OVITO).[56] Periodic boundary conditions were applied along the $x$-, $y$-, and $z$-directions of the pore model, and the equations of motion were integrated using the velocity Verlet algorithm with a timestep of 1 fs. The MD simulation process began with an energy minimization step on the liquid-filled nanochannel using the Polak-Ribiere conjugate gradient (CG) algorithm. This was followed by a 4 ns relaxation in the isothermal and isobaric NPT ensemble, equilibrating at 1 atm pressure and target temperatures of 300, 320, 340, or 360 K. The 4 ns relaxation time was found to be adequate for system equilibration according to the potential energy evolution profiles (**Figure S1**, **Supporting Information**). Subsequently, a 4 ns production run was conducted in the canonical NVT ensemble at the same temperature as the equilibration stage to collect thermodynamic data, with atomic trajectories sampled every 4 ps for analysis. For both the NPT and NVT ensembles, the Nosé-Hoover thermostat and barostat[57, 58] were used, with relaxation times of 0.1 ps and 1 ps, respectively. Notably, our simulated pore system differs from most previous MD studies on C-S-H-type nanopores, which focus mostly on tobermorite C-S-H pore with Ca/Si ratio close to 1,[14, 39, 42, 44, 52, 59-63] and primarily modeled "capillary pores" simulating the ingress of external solutions into initially dry cementitious pores. In contrast, this study focuses on ionic transport within fully saturated C-S-H gel pores (Ca/Si = ~1.67), representative of OPC system, providing a complementary perspective to previous studies.

In total, four nanochannel simulations were performed, encompassing four temperature conditions (300, 320, 340, and 360 K). This variation of temperature enables the investigation of thermal effects on ionic transport, as well as the calculation of thermal kinetics like diffusion energy barriers and intrinsic mobility. It should be noted that, in real-world systems, elevated temperatures may induce desorption phenomena and variations in interlayer water content of C–S–H gels. These effects are not considered in the present study; instead, the simulations are restricted to constant water content under undrained conditions. For comparison, control simulations of bulk 2 M NaCl solutions were performed under the same MD simulation steps within a 4 nm × 4 nm × 4 nm simulation cell at each temperature. The comparison of the nanoconfined solutions with their bulk counterparts provided insights into the effects of nanoconfinement on ionic transport and enabled validation against existing literature.

All MD simulations were carried out using the classical ClayFF force field,[64] with a detailed description of the force field and potential parameters given in **Table S1–S3** and **Section S2** of the **Supporting Information**. The potential parameters and partial charge assignments for $Na^+$ and $Cl^-$ ions were sourced from ref. [65], while parameters for other atom types were obtained from ref. [64]. The Lorentz-Berthelot mixing rule[66, 67] was applied to determine the non-bonded Lennard-Jones parameters between dissimilar atom types, as adopted by the ClayFF force field. Long-range electrostatic interactions were solved using the particle-particle-particle-mesh (PPPM) algorithm[68] with an accuracy of $10^{-5}$. A cut-off radius of 1.2 nm was used for both short-range van der Waals interaction (including tail corrections), and Coulombic interactions. The ClayFF force field has been widely adopted for simulating ionic transport in cementitious nanopores,[14, 15, 29, 31, 32, 34, 39, 40, 42, 44, 52, 59-63, 69, 70] due to its demonstrated accuracy in modeling hydrated cementitious systems, while offering significantly lower computational cost compared to alternative force fields (e.g.,

ReaxFF) and DFT methods.[21] To ensure statistical reliability and robustness for the analysis, each simulation has been independently repeated three times with randomized initial configurations.

# 3 Results & Discussion

## 3.1 Ionic diffusion in bulk and nanoconfined NaCl solution

### *3.1.1 Average self-diffusion coefficient*

Mean square displacement (MSD), a measure of how far particles travel away from their initial positions over time, is a widely used metric to study the mobility of water and ions in various systems.[14, 28, 29, 59, 71-73] MSD can be directly calculated from MD trajectories using **Equation** (1):

$$\mathrm{MSD}(\tau + t_0) = \langle(\boldsymbol{r}_i(\tau + t_0) - \boldsymbol{r}_i(t_0))^2\rangle, \tag{1}$$

where, $\boldsymbol{r}_i(t_0)$ and $\boldsymbol{r}_i(\tau + t_0)$ are the coordinates of particle $i$ at times $t_0$ and $\tau + t_0$, respectively, and $\langle\cdot\rangle$ represents the ensemble average over the period $\tau$. **Figure 2**a shows typical MSD profiles of $Na^+$ ions in the 2 M bulk NaCl solution over a 2 ns timeframe, extracted from the 4 ns NVT production stage, with each profile corresponding to a different initiation time ($t_0$ = 0, 1, and 2 ns, respectively). As expected, these MSD values increase almost linearly with time due to the random, uncorrelated motion of ions in the solution. Based on these MSD profiles, the self-diffusion coefficients ($D$) of the ions were calculated by fitting the MSD data to the well-known Einstein's diffusion equation,[74]

$$D = \lim_{\tau\to\infty} \frac{1}{2d}\frac{\mathrm{MSD}(\tau)}{\tau}, \tag{2}$$

where, $d = 1, 2, 3$, depending on the dimensionality of the studied diffusivity. $d = 3$ is used in this work to quantify the three-dimensional mobility of solution species. By adopting three different initiation times, three diffusion coefficients were calculated from the corresponding MSD profiles, following a similar method used in a previous study.[75] These were averaged to give a diffusion coefficient for each production run. For each case, three independent production runs with distinct initial configurations were performed. The reported diffusion coefficient represents the average across the three runs, with the standard deviation among them shown as the error bar.

The resulting diffusion coefficients of $Na^+$, $Cl^-$, and water molecules under bulk conditions at 300 K are summarized in **Figure 2**b, alongside literature values for the same concentration of NaCl solutions for comparison. The results are reasonably consistent with the previously reported values from experiments[76-79] and MD simulations,[71, 80, 81] confirming the reliability of our calculation. For example, the calculated average diffusion coefficient of $Na^+$ in the 2 M NaCl solution at room temperature is about $1.07 \times 10^{-9}\ \mathrm{m^2/s}$, which is in good agreement well with the experimentally measured (isotope tracer) self-diffusion coefficient of $Na^+$ in 2 M NaCl solution (1.175 $\times 10^{-9}\ \mathrm{m^2/s}$).[77] This value also represents an improvement over previous MD simulations, such as $0.817 \times 10^{-9}\ \mathrm{m^2/s}$ reported in ref. [80]. The MD simulation results show that $Cl^-$ ions diffuse faster than $Na^+$ ions in the aqueous NaCl solution, while both ions diffuse more slowly than water molecules, consistent with previous MD simulations and experimental data.[77, 80] The slower ionic diffusion can be attributed to the formation of hydration shells around the ions, where water molecules drag and slow the ions down relative to free water molecules. Furthermore, $Na^+$ ions act as "structure makers", forming tightly bound and ordered hydration shells due to their higher hydration energy, whereas $Cl^-$ ions are "structure breakers" with looser, and more dynamic hydration shells due to lower hydration energy.[82] As a result, $Cl^-$ ions experience less solvent drag and diffuse faster than $Na^+$.

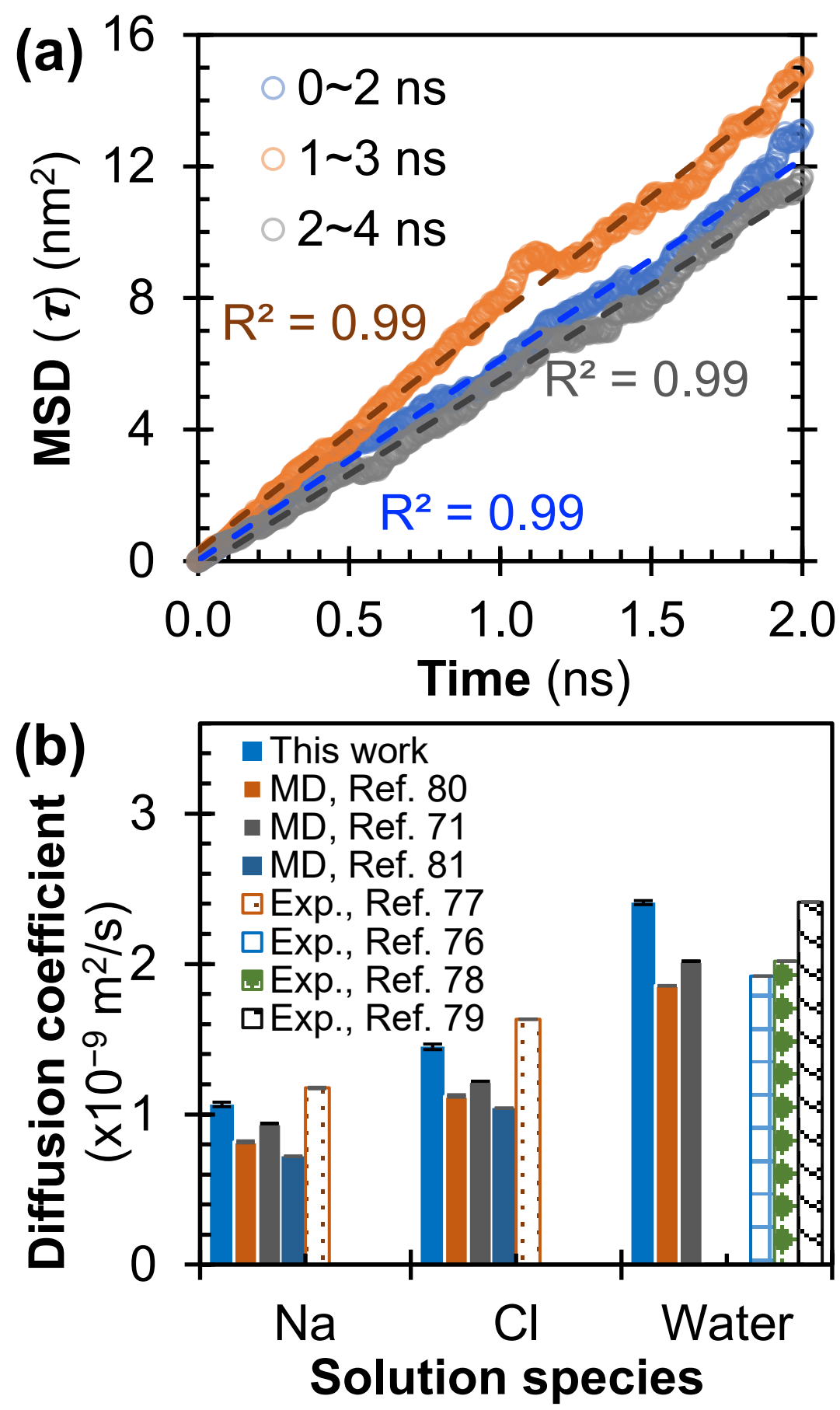

**Figure 2.** (a) Typical mean square displacement (MSD) curves for $Na^+$ in a 2 M NaCl bulk solution at 300 K, calculated from MD trajectories using three different time windows during 4 ns MD production run. $R^2$ values in (a) represents the goodness-of-fit for linear regression of MSD using **Equation** (2). (b) Comparison of the resulting average diffusion coefficients for $Na^+$, $Cl^-$, and water molecules in the 2 M NaCl bulk solution at 300 K with literature values from MD simulations[71, 80, 81] and experimental measurements.[76-79] The error bars represent the standard deviations from three independent simulations with different initial configurations.

Using the same method described above, we calculated the average self-diffusion coefficients (*xyz*) of water molecules, $Na^+$, and $Cl^-$ ions confined within the C-S-H nanochannel and the results are shown in **Figure 3**, alongside their corresponding bulk solution values for comparison. The results show that nanoconfinement within the 4 nm-wide C-S-H pore leads to a significant reduction (nearly 50%) in the average three-dimensional (*xyz*) diffusivity of all species. Such nanoconfinement-induced suppression of ionic and water diffusion has been widely reported.[15, 28, 29, 72, 73] The relative diffusion trend $Na^+ < Cl^- <$ Water persists in the C-S-H pore, consistent with previous MD simulations on nanoconfined NaCl solutions,[14, 83-85] suggesting that the underlying mechanisms governing ionic mobility, such as hydration structure and ion-solvent interactions, remain influential under nanoconfinement.

Due to slit-pore confinement in the *z*-direction, the corresponding MSD saturates at long times, rendering diffusion non-Fickian.[28] At sufficiently long times, the contribution from the confined direction vanishes, and the three-dimensional (*xyz*) MSD-based diffusion coefficient effectively reduces to the in-plane (*xy*) dynamics. To quantify this effect, we also calculated the two-dimensional in-plane (*xy*), and the one-dimensional normal-to-surface (*z*) diffusion coefficients for comparison (**Figure 3**). The results show that nanoconfinement markedly decreases the diffusivities of both water and ions along the *z* direction compared with the in-plane directions (*xy*) within the 4 ns time window studied. Nevertheless, the relative diffusion trend ($Na^+ < Cl^- <$ Water) remains consistent across all dimensionalities (*xyz*, *xy* and *z*). For this reason, all subsequent analyses in this article are based on the three-dimensional MSD (*xyz*), which describes the effective and isotropic mobilities of the nanoconfined solution species.

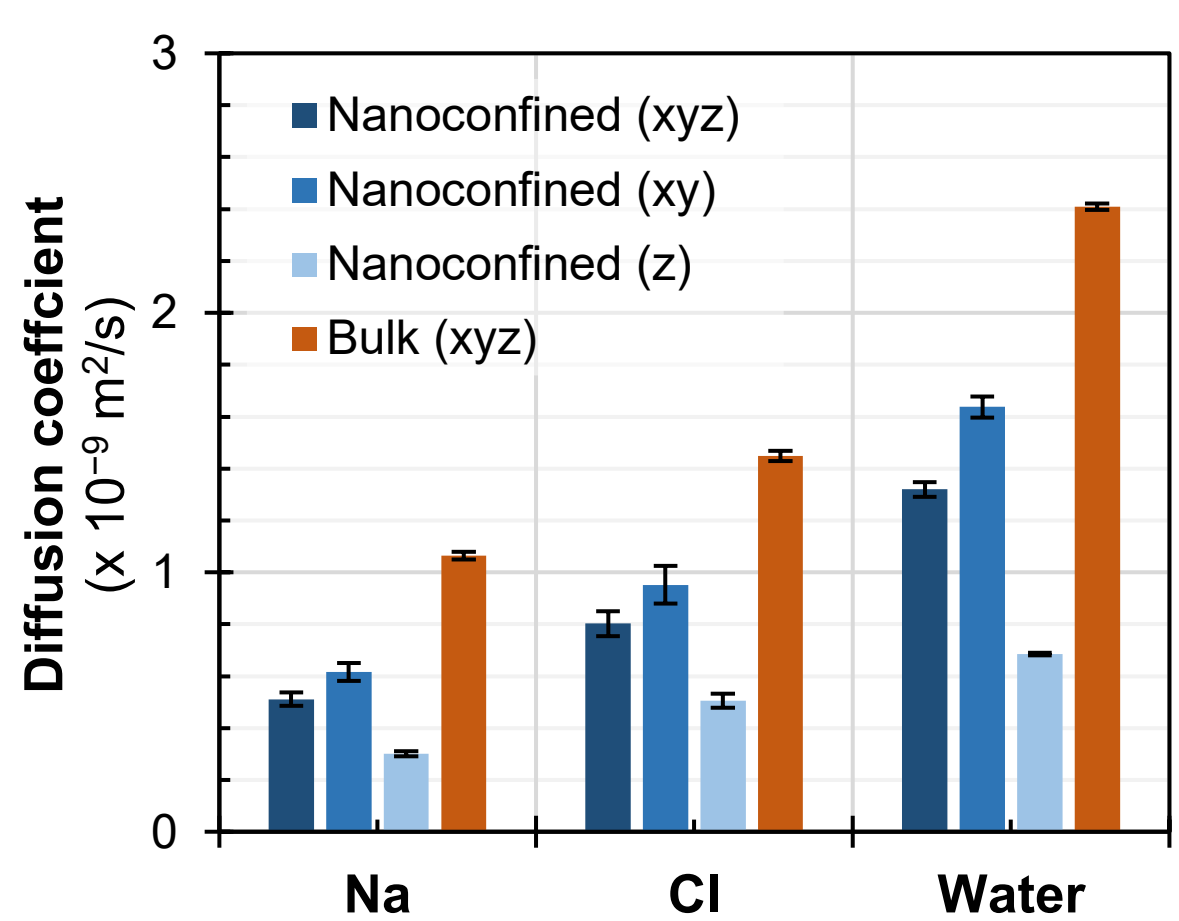

**Figure 3.** Impact of nanoconfinement within the 4 nm C-S-H pore channel on the average self-diffusion coefficients of $Na^+$, $Cl^-$, and water molecules in the 2 M NaCl solution at 300 K.

Direction-resolved diffusivities of the nanoconfined solution species, i.e., the two-dimensional in-plane (*xy*), and the one-dimensional normal-to-surface (*z*) components, are shown alongside the three-dimensional (*xyz*) values. The error bars represent standard deviations from three independent simulations.

### *3.1.2 Spatial evolution of ionic diffusion across the C-S-H pore channel*

To further elucidate the mechanisms underlying the reduced diffusivity observed in the C-S-H nanochannel, we quantified the spatial distribution of diffusion coefficients for water, $Na^+$, and $Cl^-$ ions along the *z*-direction, perpendicular to the channel surface (see detailed calculation methods in **Section S3** of **Supporting Information**). **Figure 4**a shows the resulting diffusion profiles (averaged over the *x*-, *y*- and *z*-directions) at 300 K, spanning from the average position of surface Si atoms at the C-S-H surface ($z = 0$ nm) to the center of the nanochannel ($z = \sim 2.5$ nm). For comparison, the corresponding diffusion coefficients for bulk solution are indicated by horizontal dashed lines. The profiles clearly show that diffusivity drops to near-zero values in the vicinity of the C-S-H surface ($z = 0$–$0.5$ nm) and gradually increases with distance from the wall, approaching bulk-like behaviors in the central region of the pore ($z = \sim 2.5$ nm). These trends are consistently observed across all examined temperatures (320, 340, and 360 K), as seen by the corresponding diffusivity profiles in **Figure S2** of **Supporting Information**. This pronounced retardation near the surface confirms that interfacial interactions strongly suppress molecular mobility in this region. Such suppression of diffusion near solid interfaces has been previously reported in MD studies of water on silica surface,[86] within silica,[73, 87] C-S-H gel (Ca/Si = ~1.7) nanopores,[38] and tobermorite/jennite-based gel nanopores,[39, 40] as well as ion transport in silicon nitride nanopores.[88] These results suggest that the widely observed reduction in overall diffusivity under nanoconfinement (as seen in **Figure 3**) primarily arises from strongly hindered transport near the solid-liquid interface. Notably, the relative diffusion trend—$Na^+ < Cl^- < H_2O$—is preserved across the entire channel, consistent with the average values presented in **Figure 2** and **Figure 3**.

We further computed two-dimensional diffusion profiles (averaged over the *x*- and *y*-directions, parallel to the pore surface) and compared them to the three-dimensional (xyz) diffusion profiles from **Figure 4**a and **Figure S2**. The comparison, presented in **Figure S3** of the **Supporting Information**, reveals only minor differences between the two within the timeframe probed, confirming the robustness of the observed diffusion trends. It should be noted that the diffusion coefficients and their spatial distributions were evaluated under NVT conditions in this work. To ensure that the NVT thermostat does not introduce fictitious forces affecting the calculated dynamics, we performed an additional NVE test at 300 K following equilibration. A comparison of the resulting diffusion coefficient profiles is given in **Figure S4** of **Supporting Information**, which reveal negligible differences. These results confirm that the use of NVT ensemble here for evaluating the MSD is appropriate, in agreement with previous MD studies[89].

Importantly, our spatially resolved analysis of ion and water dynamics can provide mechanistic insights into anomalous transport phenomena previously inferred from pore-scale averages and bulk quantifications. For instance, earlier studies have reported in-plane subdiffusive behavior of water molecules in small C–S–H pores based on pore-scale averaged MSD profiles.[28, 90] Our spatially resolved MSD results (**Figure S5** of the **Supporting Information**) rationalize this phenomenon by showing that strong adsorption near the gel surface suppresses mobility and induces subdiffusive dynamics, whereas diffusion gradually approaches the Fickian regime toward the pore center. The superposition of these spatially distinct behaviors leads to the pore-scale subdiffusion observed in prior studies. More discussion on the complete MSD profiles and scaling analyses is presented in **Section S6** of the **Supporting Information**.

To assess which species are more strongly immobilized by the C-S-H pore surface, we calculated the spatial evolution of diffusivity ratios ($H_2O/Na^+$, $H_2O/Cl^-$, and $Na^+/Cl^-$) across the nanopore using the data from **Figure 4**a. The results are presented in **Figure 4**b, alongside the corresponding bulk solution ratios (indicated by horizontal dashed lines) for comparison. To ensure numerical stability and avoid artificially inflated ratios in regions where diffusivity approaches zero, only data points with diffusion coefficient exceeding 10% of the respective bulk values are plotted in **Figure 4**b. The significantly elevated $H_2O/Na^+$ and $H_2O/Cl^-$ ratios near the pore surface suggest that ionic mobility is more strongly suppressed than that of water due to interfacial interactions. Between the two ions, $Na^+$ exhibits a more pronounced reduction in diffusivity compared to $Cl^-$ near the surface, suggesting stronger adsorption of $Na^+$ onto the C-S-H surface. These trends are consistently observed across all examined temperatures (320, 340, and 360 K), as seen by their corresponding diffusivity ratio profiles (**Figure S6**) in the **Supporting Information**.

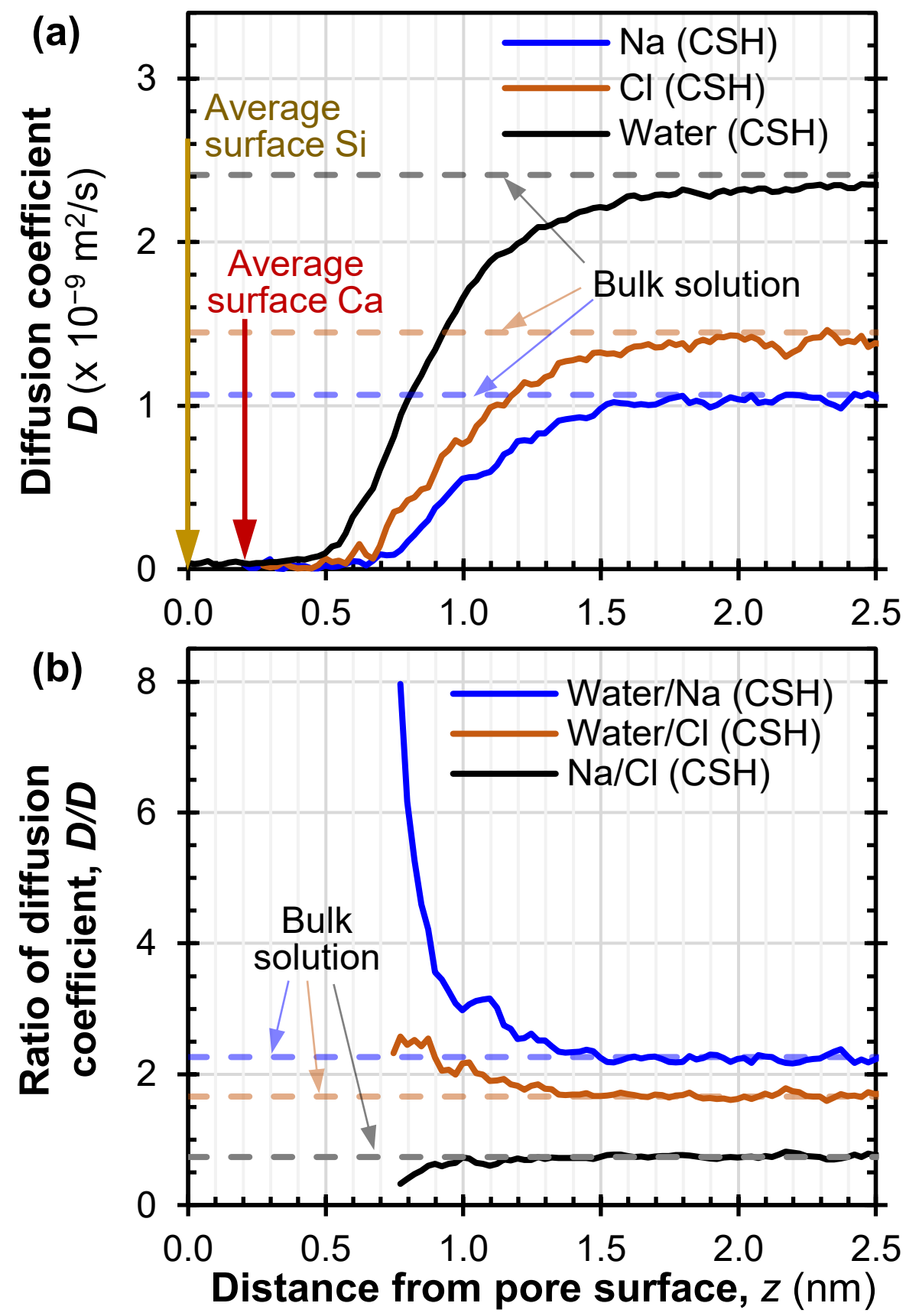

**Figure 4.** (a) Spatial evolution of the self-diffusion coefficient ($D$, in $10^{-9}$ $m^2/s$) for $Na^+$, $Cl^-$, and water molecules across the C-S-H nanopore filled with 2 M NaCl solution at 300 K, from the pore surface ($z$ = 0 nm, defined by the average position of surface Si atoms) to the center of the channel ($z$ = 2.5 nm). The region $z$ = 0–0.5 nm corresponds to the surface roughness of the C-S-H pore, where the average position of surface Ca atoms is located at 0.2 nm, as shown in **Figure 5**b. (b) Spatial evolution of the diffusion coefficient ratios of water/$Na^+$, water/$Cl^-$, and $Na^+$/$Cl^-$ across the nanopore. Horizontal dashed lines in the figures represent the corresponding values for the bulk 2 M NaCl solution at 300 K.

To better understand the diffusivity profiles in **Figure 4**, we further analyzed the spatial distribution of atomic species across the C-S-H pore channel by calculating the number density profiles of both the confined NaCl solution and the gel pore surface, as shown in **Figure 5** for the 300 K scenario. Specifically, **Figure 5**a displays the number density profiles of water oxygen (Ow), water hydrogen (Hw), sodium ($Na^+$), and chloride ($Cl^-$) ions along the *z*-direction (perpendicular to pore surface), while **Figure 5**b presents the corresponding profiles of the C-S-H gel surface, including surface silicon (St), interlayer calcium (Ca), bridging oxygen (Ob), hydroxyl oxygen (Oh), and hydroxyl hydrogen (Hh). The $z = 0$ position marks the average location of surface Si atoms, and the horizontal dashed lines indicate the corresponding bulk densities for comparison.

The water (Ow and Hw) density profile in **Figure 5**a reveals three well-defined adsorption layers near the C–S–H surface, located approximately at 0.2–0.5 nm, 0.5–0.8 nm, and 0.8–1.1 nm, with an interlayer spacing of ~0.3 nm—consistent with previous reports on interfacial water structuring.[91, 92] The $Na^+$ and $Cl^-$ profiles exhibit a similarly layered structure near the surface, with significantly elevated densities compared to those in the pore center and bulk solution, indicating strong adsorption of both ions on the gel surface. Notably, the adsorption of water precedes that of $Na^+$ and $Cl^-$ ions, with the first $Na^+$ layer peaking between the first and second water layers. A comparison of **Figure 5**a and 5b suggests that $Na^+$ ions are stabilized through coordination with surface hydroxyl groups (Oh and Hh, peaking at ~0.3–0.4 nm) and water oxygen atoms in the first two water layers. This shows that $Cl^-$ ions appear to adsorb secondarily, coordinated with pre-adsorbed $Na^+$, interlayer $Ca^{2+}$, and surface hydroxyl groups (Hh), as further confirmed by the coordination number profiles in **Section 3.3.2**.

A small population of water molecules is also observed within the 0–0.2 nm range, attributable to the surface roughness, as this range corresponds to the average positions of surface Si and Ca atoms (**Figure 5**b). This surface roughness, along with the strong ion and water adsorption, likely contributed to the pronounced suppression of molecular mobility near the solid–liquid interface seen in **Figure 4**a. Comparing the diffusivity profiles (**Figure 4**a) with the density profiles (**Figure 5**a) shows that the first layer of water ($z <$ ~0.5 nm) and the adsorbed ion layer ($z <$ ~0.7 nm) exhibit near-zero diffusivity within the timescale probed by our MD simulations. Beyond ~1.1 nm, both water and ion density profiles converge toward their respective bulk values, indicating the formation of a homogeneous, bulk-like solution structure in the central region of the nanopore.

However, full recovery of bulk-like diffusivity is not achieved until $z > \sim 1.5$ nm (**Figure 4**a), suggesting a spatial lag between structural and dynamic equilibration.

These spatial trends are generally consistent across both surfaces of the pore channel and at elevated temperatures (320, 340, and 360K), as shown in the full density profiles in **Figure S7** in the **Supporting Information**. Minor differences are observed between the two surfaces, attributable to their inherent structural asymmetry. Furthermore, a comparison of the density profiles across temperatures suggests that elevated temperatures enhance $Na^+$ adsorption onto the gel surface, which in turn promotes secondary $Cl^-$ adsorption. This increased interfacial ion concentration may have intensified the local electrostatic field normal to the surface, thereby strengthening the dipole alignment of interfacial water molecules and increasing the attraction of water oxygen atoms toward the C–S–H surface, as illustrated in **Figure S8**. Consistently, the dipole orientation of the interlayer water is also seen to recover its bulk value only beyond $z > \sim 1.5$ nm from the surface.

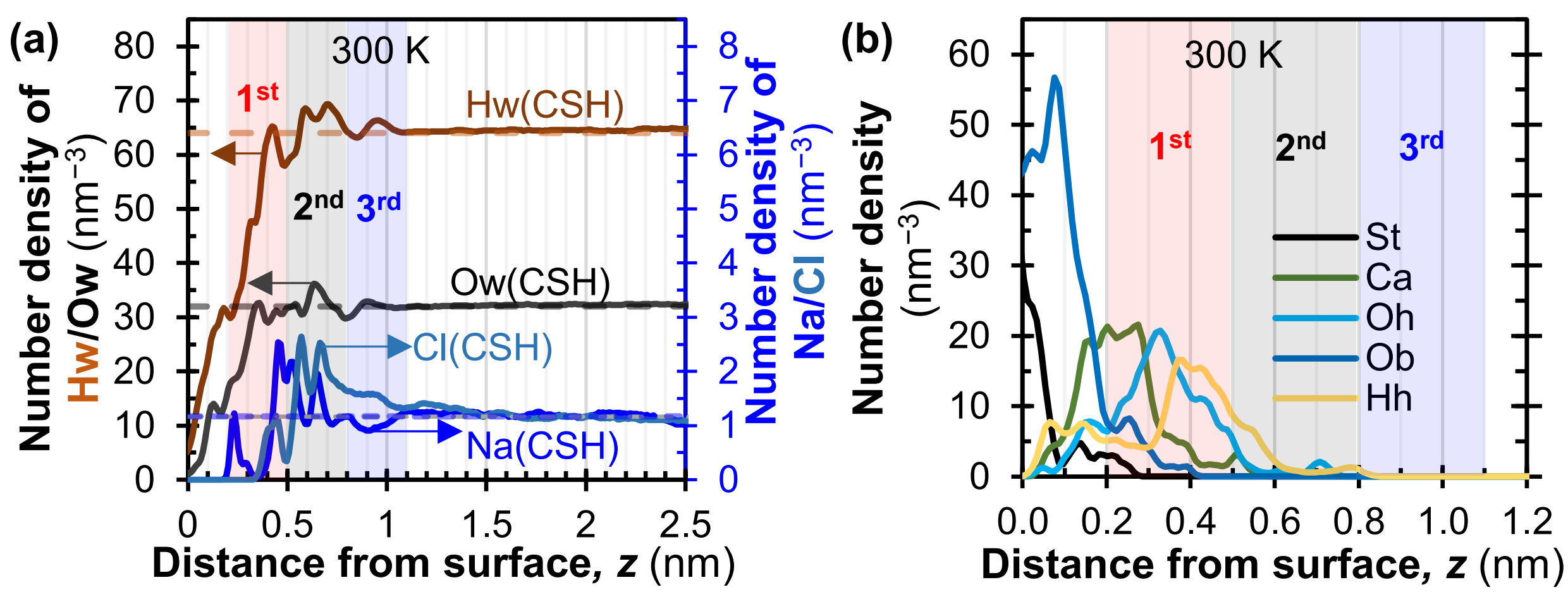

**Figure 5.** Number density profiles (in $nm^{-3}$) across the C–S–H nanopore at 300 K for (a) interlayer solution species, including water hydrogen (Hw), water oxygen (Ow), sodium ion ($Na^+$), and chloride ion ($Cl^-$), and (b) gel surface atoms, including tetrahedral silicon (St), interlayer calcium (Ca), bridging oxygen (Ob), hydroxyl oxygen (Oh), and hydroxyl hydrogen (Hh). $z = 0$ nm corresponds to the average position of surface Si atoms. Horizontal dashed lines in (a) represents the respective bulk solution densities for comparison.

## 3.2 Energy barrier and intrinsic mobility of diffusion

### *3.2.1 Average energy barrier of self-diffusion*

To better understand the temperature dependence of diffusion in bulk NaCl solution and that confined in the C-S-H nanopore, we calculated the average self-diffusion coefficients of $Na^+$, $Cl^-$, and $H_2O$ across a range of temperatures under both bulk and nanoconfined conditions. The corresponding Arrhenius plots, depicting the natural logarithm of the diffusion coefficient (ln *D*) versus the inverse temperature (1000/*T*), are shown in **Figure 6**. These data were fitted using the Arrhenius-type expression as usually done for liquid diffusivity:[11, 43, 93-96]

$$\ln(D) = \frac{-E_a}{R}\frac{1000}{T} + \ln(D_0), \tag{3}$$

where, *D* is the diffusion coefficient ($m^2/s$), $E_a$ is the activation energy barrier (kJ/mol), *R* is the gas constant ($8.3145\ \mathrm{J \cdot mol^{-1} \cdot K^{-1}}$), *T* is the absolute temperature (K), and $D_0$ is a pre-exponential factor describing the intrinsic mobility of the diffusing species.

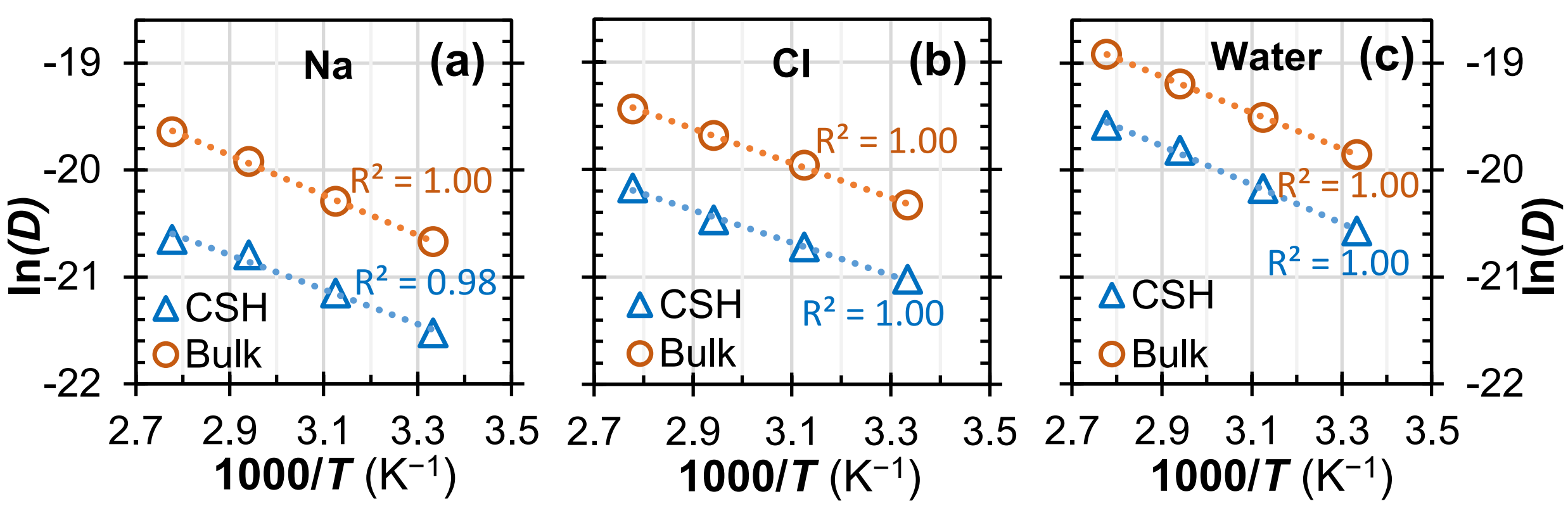

**Figure 6.** Correlation between ln*(D)* and 1000/*T* for (a) Na, (b) Cl, and (c) water molecules in the 2M NaCl solution under both nanoconfined (C-S-H pore) and bulk conditions. *D* represents the average self-diffusion coefficient (in $m^2/s$), calculated using **Equation (2)** based on the MSD of all Na, Cl, or water in the pore or simulation cell (for bulk solution). *T* denotes the temperature of the MD simulations. The goodness-of-fit ($R^2$ values) for the linear regressions is also provided in the figures.

The Arrhenius plots in **Figure 6** exhibit strong linear correlations for all species under both bulk and nanoconfined conditions, confirming that their temperature dependence follows Arrhenius behavior over the probed range. As expected, molecular diffusivity increases with temperature, reflecting thermally activated transport. Across all temperatures, diffusion coefficients under nanoconfined conditions are consistently lower than those in bulk, consistent with spatially resolved diffusivity data shown in **Figure S2**. The established diffusivity hierarchy—$H_2O$ > $Cl^-$ > $Na^+$—is preserved under both conditions and across all temperatures.

The fitted activation energy barriers ($E_a$) and intrinsic mobility prefactors ($D_0$) for each species are summarized in **Table 1**. Together, these parameters govern the temperature dependence of diffusion. The activation energy $E_a$ reflects the energy required to overcome physicochemical interactions such as hydration shells, electrostatic traps, or adsorption onto gel surfaces.[43, 97] In contrast, $D_0$ represents the theoretical diffusion coefficient in the limit of infinite temperature, or when no energy barriers are present.[43, 97] A lower $D_0$ suggests higher frictional resistance, geometric confinement, or steric hindrance, whereas a higher $E_a$ implies stronger thermally activated interactions that impede mobility. As shown in **Table 1**, the intrinsic mobility, $D_0$, decreases significantly for all three species ($Na^+$, $Cl^-$, and $H_2O$) when confined within the C–S–H gel. For instance, the $D_0$ values for $Na^+$ and $Cl^-$ drops nearly threefold, from ~348–378 × $10^{-9}$ m²/s in bulk to ~99–117 × $10^{-9}$ m²/s under confinement. This marked reduction reflects the strong suppression of intrinsic mobility due to confinement-induced friction, steric hindrance, and restricted transport pathways.

Interestingly, the activation energy barrier ($E_a$) is seen to behave differently for ions and water under confinement. For both $Na^+$ and $Cl^-$, $E_a$ values slightly decrease relative to their bulk values, from ~14.4–13.8 kJ/mol in bulk to ~13.0–12.4 kJ/mol under confinement. In contrast, the $E_a$ of water increases from ~13.7 kJ/mol in bulk to ~14.3 kJ/mol in confined pores. These trends are consistent with previous MD simulations of smectite nanochannels,[11] where 4 nm confinement led to a slight reduction in $E_a$ for $Na^+$ diffusion (from 17 to 16.7 kJ/mol) and a noticeable increase in $E_a$ for water diffusion (from 15.5 to 19 kJ/mol). Our calculated $E_a$ for bulk water (~13.7 kJ/mol) is within the literature-reported values 11.3–19.7,[11, 98-102] but generally lower than experimental measurements around 14.7 (over temperature range of 30–80 °C),[98] 17.8,[76, 79] and 17.6–19.7 kJ/mol,[103] which can be attributed to the known underestimation of activation energy by non-polarizable water models such as the simple point charge (SPC) and the simple point charge extended (SPC/E).[11, 104] The computed activation energies for $Na^+$ and $Cl^-$ diffusion fall within the broad range of experimentally reported values for cement pastes, which span from 11.9 to 83.7 kJ/mol depending on cement type, water-to-cement ratio, experimental technique, saturation state, and specimen age.[105-107] To date, few MD studies have reported activation energies for diffusion within C-S-H-type gel pores. One prior MD study using a ClayFF-like "core-only" potential (CSH-FF) and the SPC water model reported $E_a$ values of 11.3–8.2 kJ/mol for water diffusion in C-S-H gel with Ca/Si ratios of ~1.2–2.1.[108] Another MD study[43] employing the COMPASS forcefield reported $E_a$ values of 4.54–4.76 kJ/mol for $Na^+$ and 3.06–3.30 kJ/mol for $Cl^-$ diffusion within ~8 nm-wide C-S-H nanochannels based on tobermorite and jennite structures. In other nanopore systems, a previous MD study[85] reported $E_a$ values of 5.51–10.52 kJ/mol for $Na^+$ and 5.46–12.42 kJ/mol for $Cl^-$ diffusion within 7–16 Å-wide neutral, negatively, or positively charged graphene nanochannels over a temperature range of 283–333 K and a pressure of $300 \pm 150$ atm. Under these thermodynamic conditions, their reported $E_a$ values for $Na^+$ and $Cl^-$ in 2 M NaCl bulk solution are 17.95 kJ/mol and 14.76 kJ/mol, respectively. The observed trend of $E_a(Na^+) > E_a(Cl^-)$ is consistent with our results in **Table 1**, while differences in magnitude may be attributed to their usage of rigid SPC/E water model and larger environmental pressures.

**Table 1.** A summary of activation energy barriers ($E_a$, in kJ/mol), and temperature-independent prefactors ($D_0$, in $\times 10^{-9}$ m$^2$/s) for bulk and nanoconfined solution species, obtained by fitting the Arrhenius expression (**Equation** (3)) to the average diffusion coefficients across different temperatures (**Figure 6**).

| Solution species | Bulk | | Nanoconfined | |
|---|---|---|---|---|
| | $E_a$ (kJ/mol) | $D_0$ ($\times 10^{-9}$ m$^2$/s) | $E_a$ (kJ/mol) | $D_0$ ($\times 10^{-9}$ m$^2$/s) |
| Na | $14.42 \pm 0.30$ | $348.47 \pm 43.29$ | $13.03 \pm 0.79$ | $98.62 \pm 27.44$ |
| Cl | $13.81 \pm 0.40$ | $377.77 \pm 52.93$ | $12.39 \pm 0.40$ | $117.17 \pm 15.01$ |
| Water | $13.70 \pm 0.23$ | $586.52 \pm 47.10$ | $14.31 \pm 0.31$ | $414.80 \pm 45.48$ |

### *3.2.2 Spatial evolution of activation energy barrier and intrinsic mobility across the C-S-H pore*

Based on the spatial profiles of diffusion coefficients at different temperatures (**Figure S2**) and their fitting to the Arrhenius relation (**Equation (3)**), we estimated the spatial evolution of the activation energy barrier, $E_a(z)$, and intrinsic mobility, $D_0(z)$, for each species across the C-S-H nanochannel, as shown in **Figure 7**. The profiles presented include only the regions where the Arrhenius relation holds with an acceptable goodness-of-fit ($R^2 > 0.8$) on both sides of the pore channel. This excludes interfacial regions near the solid surface where strongly adsorbed ions and water molecules exhibit non-Arrhenius behavior within the timescale of our simulations (see **Figure S9** and detailed discussion in **Section S10** of the **Supporting Information**). Also, in these excluded regions, diffusivities approach zero ($D \approx 0$), leading to extreme values in the logarithmic transformation ($\ln D \rightarrow -\infty$) and large uncertainties in the linear fits. Specifically, the fitting results shown in **Figure 7** include only data beyond $z >$ ~0.65 nm for $Na^+$, $z >$ ~0.85 nm for $Cl^-$, and $z >$ ~0.25 nm for water. While differences exist between the profiles on the left and right interfaces due to subtle asymmetries in the local gel structure, the overall trends are consistent across both sides of the pore.

**Figure 7** shows that in the center of the nanochannel, all three species ($Na^+$, $Cl^-$, and $H_2O$) exhibit both lower activation energy barriers and lower intrinsic mobilities compared to their bulk solution. The reduced $E_a$ suggests that thermal activation for diffusion is less hindered in the channel center relative to the bulk, likely due to a weaker water-ion coordination environment, as will be illustrated in **Section 3.3.2**. However, the intrinsic mobility $D_0$ is also lower than in bulk, reflecting steric hindrance and increased friction caused by confinement and solid-liquid interactions. As a result, although diffusion in the channel center appears less thermally activated, it remains slower overall, particularly at elevated temperatures, than in the bulk, as seen in **Figure S2**.

In contrast, near the solid-liquid interface, specifically within the second and third water layers, both $E_a$ and $D_0$ increase significantly for $Na^+$ and $Cl^-$. The elevated $E_a$ indicates that ion diffusivity in these regions is strongly constrained at lower temperatures, possibly due to strong adsorption on the C–S–H surface. In these regions, ions may become temporarily trapped at specific binding sites, coordinated with surface atoms, and must overcome high energy barriers to hop between adjacent sites. Once sufficient thermal energy is available to overcome these barriers, the ions can desorb from these sites and hop/exchange to nearby similar sites, resulting in enhanced mobility. Such surface-mediated site-to-site hopping or gliding mechanisms have been reported in other material surfaces, such as the transport of adsorbate particles (atoms, molecules, and clusters) on the metal and semiconductor surfaces.[109, 110]

Water molecules exhibit a similar trend in this interfacial region—showing elevated $E_a$ and $D_0$ in the second and third hydration layers. However, as they approach the first interfacial water layer, their transport behavior appears to shift sharply to a regime characterized by both low $D_0$, and low $E_a$. In this highly ordered zone, water molecules are strongly hydrogen-bonded to the C–S–H surface and become quasi-immobilized, with minimal translational freedom. The low $E_a$ reflects the absence of long-range hopping events, which typically involve higher energy barriers than localized vibrations or molecular reorientations, while the significantly reduced $D_0$ arises from pronounced geometric confinement and strong surface interactions, likely exacerbated by nanoscale surface roughness in the region below $z < 0.5$ nm. The observed spatial variation of $E_a(z)$ for water appears to be consistent with a recent MD study on water diffusion in C–S–H pores of different sizes,[28] which reported that the water diffusion energy barrier is lower than bulk in ultrafine pores (basal spacing < ~1.8 nm), increases to values above the bulk as the basal spacing approaches ~2 nm, and gradually returns to bulk-like values as the basal spacing exceeds ~4 nm. These spatially resolved insights into both the activation energy barrier and intrinsic mobility of ions and water under nanoconfinement may be important to understanding the anomalous ionic transport behaviors observed in nanoconfined systems and at solid-liquid interfaces, as recently reported.[47, 48]

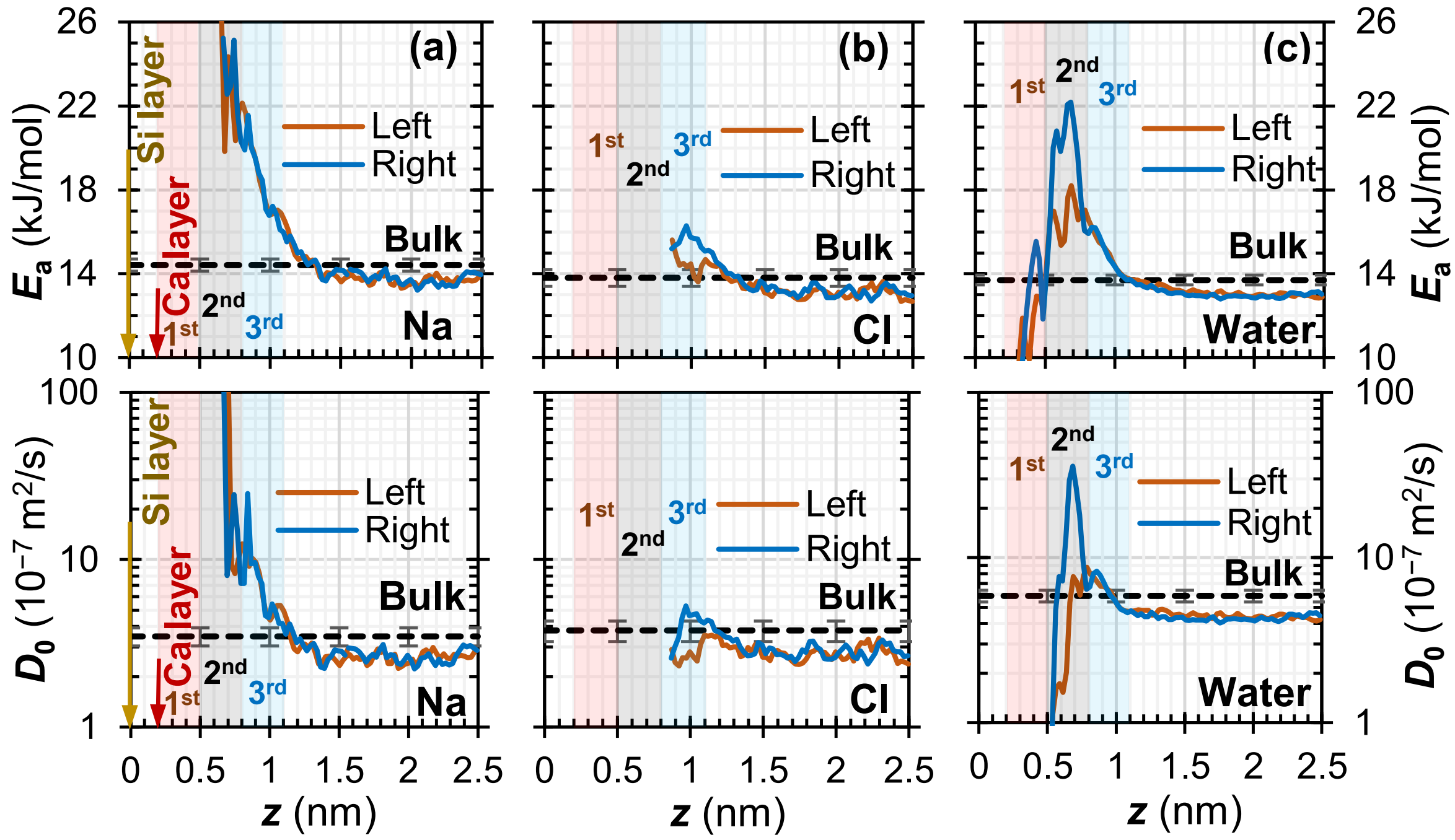

**Figure 7.** Spatial evolution of the activation energy barrier ($E_a$, in kJ/mol, upper panel), and intrinsic mobility ($D_0$, in $\times 10^{-7}$ $m^2/s$, lower panel) for the diffusion of (a) $Na^+$, (b) $Cl^-$, and (c) water molecules in the nanoconfined NaCl solution, shown as a function of distance from the "Left" and "Right" surfaces of the C-S-H pore. *z* = 0 nm and *z* = ~2.5 nm correspond to the average position of the surface Si layer and the pore center, respectively. Horizontal dashed lines indicate the corresponding values for bulk NaCl solution for comparison. All profiles represent averages from three independent simulations to improve statistical stability.

## 3.3 Structural analysis

### *3.3.1 Radial distribution function (RDF) and Individual bond strength*

To gain deeper insights into the diffusion behaviors observed in previous sections, we performed a detailed structural analysis of both bulk and nanoconfined NaCl solutions across different temperatures. **Figure 8** presents the RDF profiles for selected atom-atom pairs at 300 K, revealing the local atomic arrangements and coordination environments of key species. The partial RDFs involving $Na^+$ (Na-X, as shown in **Figure S10**a) indicate that its nearest neighbors with attractive interactions are dominated by O and Cl atoms. **Figure 8**a further compares the Na-X partial RDFs for different types of O atoms, i.e., those from water (Ow), surface hydroxyl groups (Oh), and surface bridging O (Ob), as well as $Cl^-$. The position of the first peak in each partial RDF corresponds to the most probable Na–X nearest-neighbor distance (i.e., bond length), as summarized in **Table 2**. The trend follows the sequence: Na–Oh (2.27 Å) < Na–Ob (2.33 Å) < Na–Ow (2.37 Å) <Na–Cl (2.85 Å), indicating that $Na^+$ interacts most strongly with surface hydroxyl O (Oh), followed by surface bridging O (Ob), water O (Ow) and finally $Cl^-$. Nearest-neighbor distances for other atom-atom pairs, i.e., Cl-X, Ow-X, and Hw-X, were also determined and presented in **Table 2**. The obtained atomistic distances are consistent with previous experimental/computational studies on the nearest interatomic distances in the first coordination shell of $Na^+$, $Cl^-$, and water molecules in NaCl solutions, including Na–Ow, Cl–Hw, Na–Cl, and Ow–Hw pairs in aqueous NaCl solutions.[71, 81, 82, 111, 112]

To quantify the relative strength of these individual interactions, we estimated the bond force and bond energy of each Na-X pair based on their respective bond lengths (**Table 2**) and interatomic force field parameters (see **Section S2** in the **Supporting Information**). The calculated individual bond strengths are summarized in **Table 3**. Notably, the Na–Oh interaction exhibits the highest bond strength (4.55 nN), followed by Na–Ob (3.69 nN) and Na–Ow (2.38 nN). Combined with the significantly higher peak intensity of the Na–Oh RDF relative to Na–Ob (**Figure 8**a), these results suggest that $Na^+$ near the surface is predominantly immobilized through adsorption onto surface hydroxyl O (Oh) sites. Meanwhile, the strong Na–Ow and Na–Cl RDF peaks (**Figure 8**a) reflect the formation of a well-defined hydration shell around $Na^+$ and the presence of contact ion pairing with $Cl^-$.

The partial RDFs involving $Cl^-$ and its nearest neighbors, shown in **Figure 8**b, reveal strong correlations with $Na^+$, $Ca^{2+}$, and water H (Hw) atoms. The computed interaction strengths (**Table 3**) demonstrate that Cl-Ca exhibits a significantly higher bond strength (3.90 nN) compared to Cl–Na (1.96 nN) and Cl–Hw (1.92 nN). This suggests that $Cl^-$ near the C-S-H surface is primarily immobilized through adsorption onto surface $Ca^{2+}$ sites. The partial RDFs involving water O (Ow) and H (Hw) atoms with their nearest neighbors are presented in **Figure 8**c and d, respectively, with their corresponding bond lengths and strengths also summarized in **Table 2** and **Table 3**. A comparison of RDF peak intensities and bond strengths suggests that water molecules near the surface are mainly immobilized through interactions with surface $Ca^{2+}$ sites (via Ow-Ca coordination) and surface hydroxyl O sites (via Hw-Oh hydrogen bonding). In contrast, other types of hydrogen bonds, such as Hw-Ob, Ow-Hw, and Ow-Hh, exhibit significantly lower peak intensities and/or bond strengths, indicating weaker contributions to water immobilization near pore surface.

**Figure 8**e–f presents the partial RDFs for corresponding atom-atom pairs in the bulk NaCl solution, which show similar intensities compared to those in the confined solution. A comparison of bond lengths (**Table 2**) reveals nearly identical values between the bulk and confined systems, suggesting that within the studied 4 nm nanopores, nanoconfinement does not significantly perturb the nearest-neighbor coordination distances among atoms in the NaCl solutions. The corresponding partial RDFs at all temperatures (300, 320, 340, and 360 K) are provided in **Figures S11–14** in the **Supporting Information,** with the associated nearest neighbor bond distances given in **Table 2**. A comparison of these bond distances across different temperatures shows that they remain largely unchanged within the studied 4 nm nanopores, with the largest variation being less than ~2%. Therefore, for a given bond type, the same bond distance and, hence, the same corresponding individual bond strengths were adopted, as given **Table 3**, for subsequent analysis.

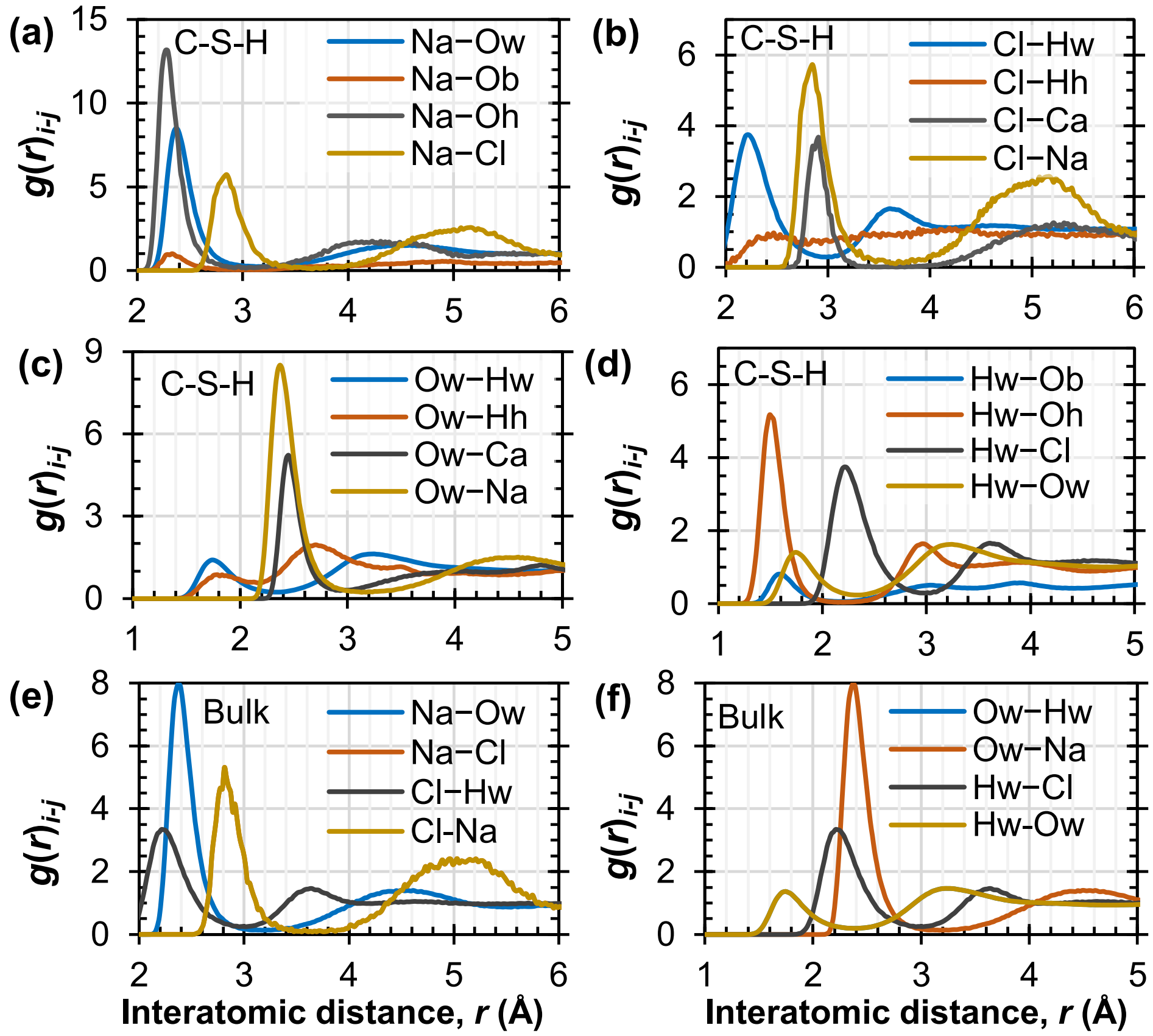

**Figure 8.** (a–d) Partial RDFs in the nanoconfined solution for (a) $Na^+$, (b) $Cl^-$, (c) Ow, and (d) Hw with their respective nearest neighbors. (e–f) Partial RDFs involve (g) $Na^+$ and $Cl^-$, and (h) Ow and Hw in the bulk solution and their nearest neighbors. Atom labels are as follows: Ow = oxygen atoms in water molecules; Ob = bridging oxygen atoms on the C-S-H surface; Oh = oxygen atoms in surface hydroxyl groups; Hw = hydrogen atoms in water molecules; Hh = hydrogen atoms in surface hydroxyl groups.

### *3.3.2 Coordination number (CN)*

While the nearest-neighbor interatomic bond distances remain largely insensitive to temperature variations between 300 and 360 K, we further analyzed the impact of temperature and confinement on the number of nearest neighbors (i.e., coordination number (CN)) around $Na^+$, $Cl^-$, Ow, and Hw atoms. The CN was calculated by integrating the corresponding partial RDFs up to the first minimum beyond the first peak, as commonly performed in the literature.[71, 73, 81, 94, 113-115] The cutoff distances corresponding to these first minima are summarized in **Table S4**, which shows only minor variations (< 4 %) across different temperatures and under nanoconfinement for the same atom-atom pairs. Hence, an average cutoff distance was adopted for each pair to facilitate consistent comparison, in line with the approach used in previous studies.[94, 113] To validate this approach, we also calculated CNs using temperature-specific cutoff distances and found negligible differences compared to results based on the average values (see **Table S5** and **Figure S15** in the **Supporting Information**).

The CN results are summarized in **Table 4**, along with literature data included in the caption for comparison,[71, 81, 82, 115-117] which reveals noticeable impacts of both temperature and nanoconfinement on specific atom-atom interactions. In bulk solution, increasing the temperature slightly reduces the number of Ow atom surrounding $Na^+$ (from 5.55 to 5.31) and Hw atom surrounding $Cl^-$ (from 6.59 to 6.28), while the CN for Na-Cl pairs increases (from 0.20 to 0.31). Similar temperature-induced trends are observed under nanoconfinement, with a decrease in Ow-Na and Ow-Cl coordination and an increase in Na–Cl ion pairing. These trends suggest that elevated temperatures lead to slight dehydration of $Na^+$ and $Cl^-$ and enhance Na-Cl ion pairing in both bulk and confined systems within the studied 4 nm nanopores. For water molecules, the CN for Ow–Hw pairs decreases with increasing temperature in both bulk and nanoconfined solutions, dropping from 1.63 (bulk) and 1.45 (confined) at 300 K to 1.55 (bulk) and 1.41 (confined) at 360 K. This reduction indicates a temperature-induced decrease in hydrogen bonding, with the effect being slightly attenuated under confinement. Regarding nanoconfinement effects, although the CN of Na–Ow slightly decreases (from 5.55 in bulk to 5.27 at 300 K), additional interactions with surface oxygen atoms lead to an increase in the total oxygen CN around $Na^+$ to 5.74, due to the emergence of Na–Oh and Na–Ob coordination. Interestingly, the CNs for Na–Oh and Na–Ob increase with temperature, indicating stronger surface adsorption at elevated temperature. In contrast, other surface-related coordination numbers, such as Cl–Ca, Cl–Hh, Ow–Ca, and Ow–Hh, remain largely unchanged across the temperature range studied.

**Table 2.** Summary of the nearest interatomic distances (bond length, $r_0$) in the first coordination shell of $Na^+$, $Cl^-$, and water molecules in bulk and nanoconfined NaCl solutions at different temperatures, derived from the peak positions of the partial radial distribution functions (RDF) shown in **Figure 8** and **Figures S11–14**. The bond lengths obtained for bulk solutions are consistent with literature values for Na–Ow (2.33–2.49 Å[82], ~2.4 Å[71]), Cl–Hw (~2.2 Å[71], ~2.26 Å[111], 2.14[112]), Na–Cl (2.9 Å[81]), and Ow–Hw (~1.8 Å[71], 1.72 Å[112]) pairs in aqueous NaCl solutions. Atom labels are defined as follows: Ow = oxygen atoms in water molecules; Ob = bridging oxygen atoms on the C-S-H surface; Oh = oxygen atoms in surface hydroxyl groups; Hw = hydrogen atoms in water molecules; Hh = hydrogen atoms in surface hydroxyl groups.

| NaCl solutions | Temperature (K) | Bond length: $r_{0,\mathrm{Na-j}}$ | | | | Bond length: $r_{0,\mathrm{Cl-j}}$ | | | | Bond length: $r_{0,\mathrm{Ow-j}}$ | | | | Bond length: $r_{0,H\mathrm{w-j}}$ | | | |
|---|---|---|---|---|---|---|---|---|---|---|---|---|---|---|---|---|---|
| | | Ow | Ob | Oh | Cl | Hw | Hh | Ca | Na | Hw | Hh | Ca | Na | Ob | Oh | Cl | Ow |
| Bulk | 300 | 2.37 | 0 | 0 | 2.81 | 2.21 | 0 | 0 | 2.81 | 1.75 | 0 | 0 | 2.37 | 0 | 0 | 2.21 | 1.75 |
| | 320 | 2.37 | 0 | 0 | 2.83 | 2.23 | 0 | 0 | 2.83 | 1.75 | 0 | 0 | 2.37 | 0 | 0 | 2.23 | 1.75 |
| | 340 | 2.37 | 0 | 0 | 2.83 | 2.23 | 0 | 0 | 2.83 | 1.75 | 0 | 0 | 2.37 | 0 | 0 | 2.23 | 1.75 |
| | 360 | 2.35 | 0 | 0 | 2.81 | 2.23 | 0 | 0 | 2.81 | 1.75 | 0 | 0 | 2.35 | 0 | 0 | 2.23 | 1.75 |
| Confined in C-S-H pore | 300 | 2.37 | 2.33 | 2.27 | 2.85 | 2.21 | 2.47 | 2.91 | 2.85 | 1.75 | 1.79 | 2.45 | 2.37 | 1.59 | 1.49 | 2.21 | 1.75 |
| | 320 | 2.37 | 2.31 | 2.27 | 2.83 | 2.23 | 2.45 | 2.89 | 2.83 | 1.73 | 1.79 | 2.45 | 2.37 | 1.57 | 1.51 | 2.23 | 1.73 |
| | 340 | 2.37 | 2.33 | 2.27 | 2.83 | 2.23 | 2.41 | 2.87 | 2.83 | 1.75 | 1.85 | 2.45 | 2.37 | 1.57 | 1.51 | 2.23 | 1.75 |
| | 360 | 2.37 | 2.33 | 2.27 | 2.81 | 2.21 | 2.49 | 2.85 | 2.81 | 1.77 | 1.87 | 2.45 | 2.37 | 1.59 | 1.51 | 2.21 | 1.77 |
| | Average | 2.37 | 2.33 | 2.27 | 2.83 | 2.22 | 2.46 | 2.88 | 2.83 | 1.75 | 1.83 | 2.45 | 2.37 | 1.58 | 1.50 | 2.22 | 1.75 |

**Table 3.** Interatomic bond strengths (expressed as bond forces and bond energies) for different atom-atom pairs within the first coordination shell of $Na^+$ ($F_{\mathrm{Na-j}}$), $Cl^-$ ($F_{\mathrm{Cl-j}}$), and the Ow ($F_{\mathrm{Ow-j}}$) and Hw ($F_{\mathrm{Hw-j}}$) atoms of water molecules, calculated using the average bond lengths for each atom-atom pair from **Table 2** and the interatomic potential parameters given in **Table S1**. Additional details on the methodology used to calculate bond forces and energies are provided in **Section S2** of the **Supporting Information**.

| | Bond Strength: $F_{\mathrm{Na-j}}$ | | | | Bond Strength: $F_{\mathrm{Cl-j}}$ | | | | Bond Strength: $F_{\mathrm{Ow-j}}$ | | | | Bond Strength: $F_{\mathrm{Hw-j}}$ | | | |
|---|---|---|---|---|---|---|---|---|---|---|---|---|---|---|---|---|
| | Ow | Ob | Oh | Cl | Hw | Hh | Ca | Na | Hw | Hh | Ca | Na | Ob | Oh | Cl | Ow |
| Bond length (Å) | 2.37 | 2.33 | 2.27 | 2.83 | 2.22 | 2.46 | 2.88 | 2.83 | 1.75 | 1.83 | 2.45 | 2.37 | 1.58 | 1.50 | 2.22 | 1.75 |
| Bond force ($10^{-9}$ N) | 2.38 | 3.69 | 4.55 | 1.96 | 1.92 | 1.62 | 3.90 | 1.96 | 2.53 | 2.39 | 4.52 | 2.38 | 4.41 | 5.99 | 1.92 | 2.53 |
| Bond energy ($10^{-19}$ J) | −7.84 | −11.34 | −14.19 | −7.98 | −4.26 | −3.97 | −15.68 | −7.98 | −4.43 | −4.38 | −15.14 | −7.84 | −6.97 | −8.98 | −4.26 | −4.43 |

**Table 4.** Average coordination numbers (CNs) for different atom-atom pairs within the first coordination shell of $Na^+$ ($\mathrm{CN_{Na-j}}$), $Cl^-$ ($\mathrm{CN_{Cl-j}}$), and the Ow ($\mathrm{CN_{Ow-j}}$) and Hw ($\mathrm{CN_{Hw-j}}$) of water molecules in bulk and nanoconfined NaCl solution at various temperatures, calculated using the averaged cutoff distances given in **Table S4** in the **Supporting information**. The CNs obtained for bulk solutions at 300 K are consistent with literature values for Na–Ow (5.4[71], 5.2–5.6[82], ~5.2[116]), Cl–Hw (~6.24[116], ~6.4[115]), Na–Cl (0.159[81]), and Ow–Hw (1.81 for pure water, calculated by integrating the corresponding RDF in ref.[117] and using the estimated number density for water hydrogen atom at 1 g/cm$^3$, 0.0668/Å$^3$) pairs in aqueous solutions.

| NaCl solutions | Temperature (K) | $\mathrm{CN_{Na-j}}$ | | | | $\mathrm{CN_{Cl-j}}$ | | | | $\mathrm{CN_{Ow-j}}$ | | | | $\mathrm{CN_{Hw-j}}$ | | | |
|---|---|---|---|---|---|---|---|---|---|---|---|---|---|---|---|---|---|
| | | Ow | Ob | Oh | Cl | Hw | Hh | Ca | Na | Hw | Hh | Ca | Na | Ob | Oh | Cl | Ow |
| Bulk | 300 | 5.55 | 0 | 0 | 0.20 | 6.59 | 0 | 0 | 0.20 | 1.63 | 0 | 0 | 0.20 | 0 | 0 | 0.12 | 0.81 |
| | 320 | 5.48 | 0 | 0 | 0.25 | 6.50 | 0 | 0 | 0.25 | 1.61 | 0 | 0 | 0.20 | 0 | 0 | 0.12 | 0.81 |
| | 340 | 5.40 | 0 | 0 | 0.28 | 6.40 | 0 | 0 | 0.28 | 1.58 | 0 | 0 | 0.20 | 0 | 0 | 0.12 | 0.79 |
| | 360 | 5.31 | 0 | 0 | 0.31 | 6.28 | 0 | 0 | 0.31 | 1.55 | 0 | 0 | 0.19 | 0 | 0 | 0.11 | 0.77 |
| Confined in C-S-H pore | 300 | 5.27 | 0.13 | 0.34 | 0.21 | 6.37 | 0.06 | 0.12 | 0.21 | 1.45 | 0.03 | 0.14 | 0.19 | 0.05 | 0.06 | 0.12 | 0.72 |
| | 320 | 5.00 | 0.25 | 0.43 | 0.27 | 6.17 | 0.08 | 0.15 | 0.27 | 1.44 | 0.03 | 0.14 | 0.18 | 0.05 | 0.06 | 0.11 | 0.72 |
| | 340 | 4.80 | 0.39 | 0.46 | 0.27 | 6.16 | 0.09 | 0.13 | 0.27 | 1.43 | 0.02 | 0.13 | 0.17 | 0.05 | 0.05 | 0.11 | 0.71 |
| | 360 | 4.62 | 0.45 | 0.55 | 0.29 | 5.97 | 0.08 | 0.12 | 0.29 | 1.41 | 0.02 | 0.12 | 0.17 | 0.05 | 0.05 | 0.11 | 0.71 |

To better understand the CN trends observed in **Table 4**, we further calculated the spatial evolution of CN across the C-S-H channel (denoted as CN($z$)) for different atom-atom pairs based on the corresponding spatially resolved partial RDFs. The resulting CN($z$) profiles at 300 K are shown in **Figure 9**. These profiles reveal significant variations in coordination behavior near the surface, while convergence towards bulk values is observed beyond approximately three water layers ($z >$ ~1.1 nm). Similar trends are observed at other temperatures, as shown in **Figures S16–S19** in the **Supporting Information**. The CN($z$) profiles for $Na^+$ (**Figure 9**a) reveal that the dehydration of $Na^+$ occurs primarily near the surface, evidenced by a reduction in the CN for Na–Ow from ~5.5 in the bulk region to ~4 near the surface. Despite this dehydration, the total oxygen CN around $Na^+$ near the surface remains considerably higher (~6.5) than in the bulk, due to additional coordination with surface Ob and Oh. This increase in the overall oxygen CN, combined with the higher bond strengths of Na-Oh and Na-Ob interactions (**Table 3**), provides a mechanistic explanation for the significant reduction in $Na^+$ diffusivity (**Figure 4**a) and the increase in its activation energy barrier (**Figure 7**a) near the solid-liquid interface. The immobilization of $Na^+$ near the surface is thus primarily attributed to strong adsorption onto Ob and Oh sites of the C-S-H gel surface.

The CN profiles for $Cl^-$ (**Figure 9**b) similarly show that the dehydration primarily occurs near the surface, evidenced by a significantly lower Cl–Hw CN (~4) near the surface ($z <$ ~0.5 nm) compared to the bulk (~6.6), followed by convergence to bulk values further from the surface. The relatively high CN of Ca around $Cl^-$, combined with its high bond strength (**Table 3**), confirms that $Cl^-$ is immobilized primarily via strong adsorption with surface $Ca^{2+}$ sites, with additional contribution from enhanced CNs for Cl–Na and Cl–Hh interactions.

For the Ow and Hw atoms in water molecules, the CN($z$) profiles show that both Ow–Hw and Hw–Ow CNs are significantly lower near the surface compared to the bulk, progressively increasing with distance from the surface and approaching bulk-like values beyond approximately three water layers ($z >$ ~1.1 nm). However, the CN profiles for Ow–Ca and Hw–Oh indicate that the markedly reduced diffusivity of water molecules in the first hydration layer ($z <$ ~0.5 nm), as shown in **Figure 4**a, is primarily due to strong interactions with surface Ca and Oh atoms. These interactions substantially weaken in the second water layer, whereas contributions from Ow–Na, Hw–Cl, and water–water hydrogen bonding become more prominent. In the third water layer, the main deviation from bulk behavior appears to arise from a persistent increase in Hw–Cl coordination compared with the bulk.

Similar trends are also observed at other temperatures, as shown in the corresponding CN(z) distribution profiles in **Figures S16–S19** in the **Supporting Information**. In addition, the coordination numbers for Ow-Na and Hw-Cl at the channel center decrease relative to their bulk values as temperature increases, indicating a progressively weakened coordination environment in this region at increasing temperature. This effect may have contributed to the reduced activation energy barrier in the middle of the nanochannel relative to its bulk as observed in **Figure 7** and discussed in **Section 3.2.2**.

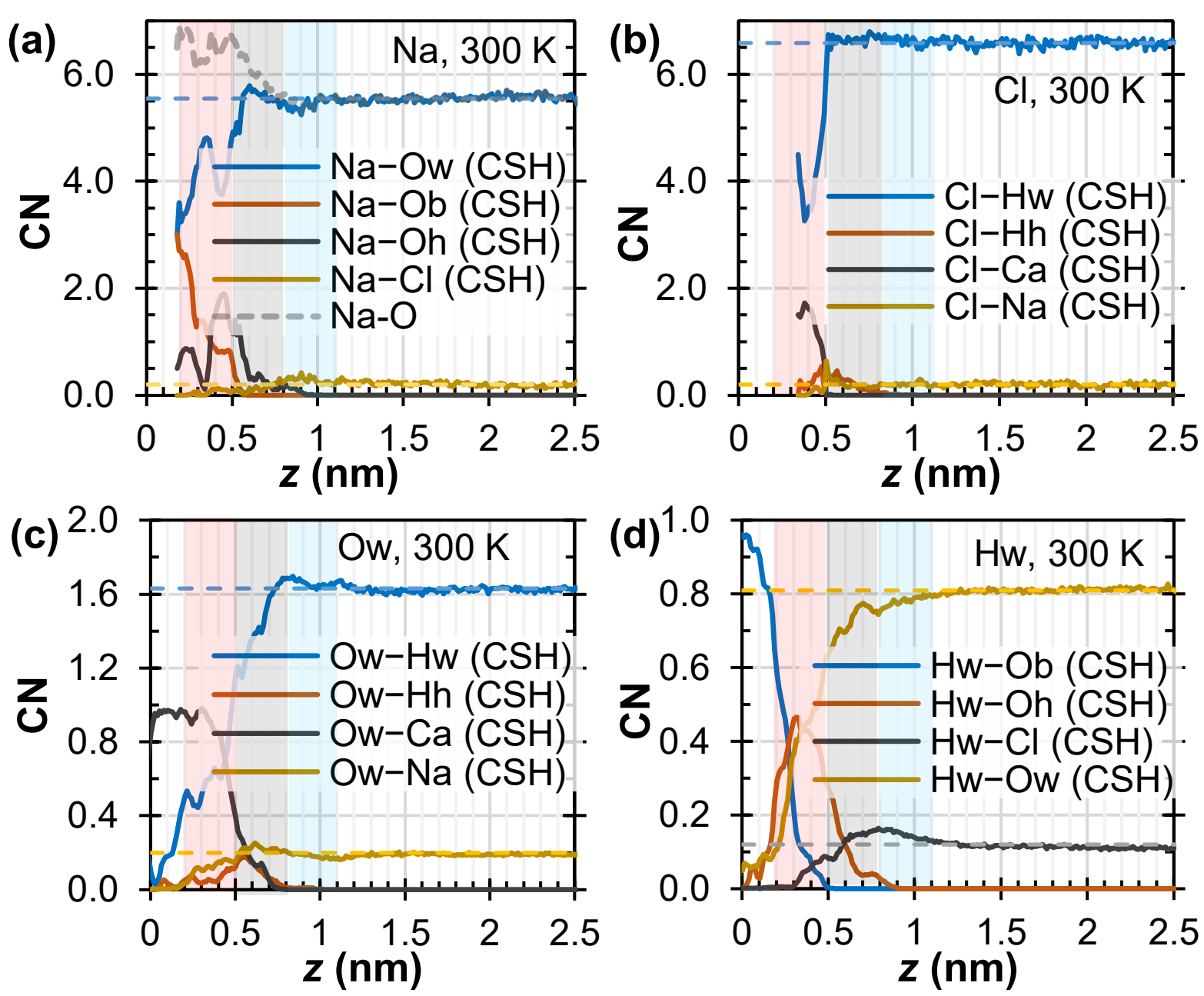

**Figure 9.** Spatial evolution of coordination numbers (CN) across the C-S-H nanopore, measured from the average position of surface Si atoms ($z = 0$ nm) to its pore center ($z = \sim 2.5$ nm), for (a) $Na^+$, (b) $Cl^-$, (c) oxygen atoms in water molecules (Ow), and (d) hydrogen atoms in water molecules (Hw), with respect to their nearest neighbors. The horizontal dashed lines represent the corresponding CNs values in the bulk NaCl solution.

### 3.4 Structural descriptor

Building on the detailed structural analysis presented in the previous section, we calculated several structural descriptors to evaluate their ability to capture diffusivity trends, inspired by our recent studies on the development of structural predictors for glass reactivity.[94, 113, 118] Here, we introduced a structural descriptor termed **total coordination strength** (TCS), which provides an estimate of the average local bond strength that an ion or molecule must overcome to diffuse away from its nearest neighbors. Specifically, the total coordination strengths experienced by $Na^+$ ($TCS_{Na}$), $Cl^-$ ($TCS_{Cl}$), and water molecules ($TCS_{Water}$) are defined in **Equations (4)–(6)**, respectively:

$$TCS_{Na}(z) = \sum CN_{Na-j}(z) \cdot F_{Na-j}, \quad (4)$$

$$TCS_{Cl}(z) = \sum CN_{Cl-j}(z) \cdot F_{Cl-j}, \quad (5)$$

$$TCS_{Water}(z) = \sum CN_{Ow-j}(z) \cdot F_{Ow-j} + 2 \times \sum CN_{Hw-j}(z) \cdot F_{Hw-j}, \quad (6)$$

where $CN_{Na-j}$, $CN_{Cl-j}$, $CN_{Ow-j}$, and $CN_{Hw-j}$ are the coordination numbers between $Na^+$, $Cl^-$, Ow, and Hw atoms and their neighboring species j, respectively, and $F_{Na-j}$, $F_{Cl-j}$, $F_{Ow-j}$, and $F_{Hw-j}$ denote the corresponding individual bond strengths, quantified by either bond force or bond energy. Absolute values of negative bond energies are used in the calculation. In the main text, results based on bond energy are presented, while those based on bond force are presented in **Supporting Information** for comparison. Both approaches yield consistent trends, confirming that the main observations are robust regardless of the specific metric used to quantify TCS.

The average CNs and their spatial distributions for each atom-atom pair are presented in **Table 4** and **Figure 9**, as well as **Figures S16–S19**, while their corresponding bond strengths are given in **Table 3**. Based on these data and **Equations (4)–(6)**, we calculated the spatial evolution of TCS for $Na^+$, $Cl^-$, and water molecules in both bulk and nanoconfined NaCl solutions at various temperatures. **Figure 10**a compares the TCS values of $Na^+$, $Cl^-$, and water molecules in bulk solutions with their corresponding diffusion coefficients across different temperatures. The results show that the bulk diffusivities of both ions and water are inversely correlated with their TCS values. This is consistent with expectations, as a lower TCS corresponds to reduced resistance to breaking local bond environments, thereby facilitating easier diffusion and leading to higher diffusivity. Additionally, increasing temperature is seen to decrease TCS values for both ions and water, suggesting that elevated temperatures promote diffusion not only by providing higher thermal energy but also by weakening the local structural constraints.

**Figure 10**b shows the spatial evolution of TCS($z$) for $Na^+$, $Cl^-$, and water molecules across the C-S-H nanochannel at 300 K, compared to their respective bulk values. Near the surface, the TCS values for both ions and water are significantly elevated, indicating stronger local structural constraints. This increase in TCS near the surface explains the markedly lower diffusivity (**Figure 4**a) and higher activation energy barriers (**Figure 7**) observed in this region. The breakdown of TCS contribution from individual atom-atom pairs (**Figure S20**) reveals that this increase of TCS near the surface is mainly due to (i) Na-Oh and Na-Ob interactions for $Na^+$, (ii) Cl-Ca interactions for $Cl^-$, and (iii) Ow-Ca, Hw-Ob and Hw-Oh interactions for water molecules. It is seen from **Figure 10**b that TCS decreases progressively with distance from the surface, approaching bulk values beyond $z = {\sim}1.0$ nm. Similar spatial trends are observed at other temperatures (320 K, 340 K, and 360 K), as shown in **Figures S21–S23** in the **Supporting Information**.

To directly evaluate whether TCS can predict the spatial evolution of diffusivity within the nanoconfined C–S–H pore, we plotted the normalized TCS (i.e., $TCS/TCS_{bulk}$) against the normalized diffusion coefficient (i.e., $D/D_{bulk}$) in **Figure 10**c. Here, normalization was performed by dividing the local TCS and diffusion coefficients by their corresponding bulk values at the same temperature, allowing the impact of confinement to be isolated. Remarkably, the data for $Na^+$, $Cl^-$, and water molecules across all temperatures collapse onto a common master trend, demonstrating that increasing normalized TCS leads to an exponential decay in normalized diffusion coefficients. However, a closer examination of the region with $TCS/TCS_{bulk} \approx 1$ reveals substantial variations of $D/D_{bulk}$, indicating that the diffusivity of nanoconfined ions and water can remain significantly lower than bulk values even when their TCS approaches that of the bulk (i.e., in the region $z >$ ~1 nm). This suggests that, beyond ~1 nm from the surface, the reduction in diffusivity under nanoconfinement is no longer governed by local coordination strength.

**Figure 10**d–f further compares normalized TCS ($TCS/TCS_{bulk}$) against the normalized diffusion coefficient ($D/D_{bulk}$) for $Na^+$, $Cl^-$, and water molecule, respectively, specifically within the near-surface region (~0.5 nm $< z <$ 1 nm). The results clearly demonstrate that, in this region, the diffusion coefficients for both ions and water are inversely and exponentially correlated with TCS.

These observations suggest that in this surface-influenced region, the reduction in diffusivity due to nanoconfinement is primarily governed by enhanced local coordination constraints, as captured by TCS. The corresponding results using TCS estimated from bond force are presented in **Figure S24**, showing the same observations.

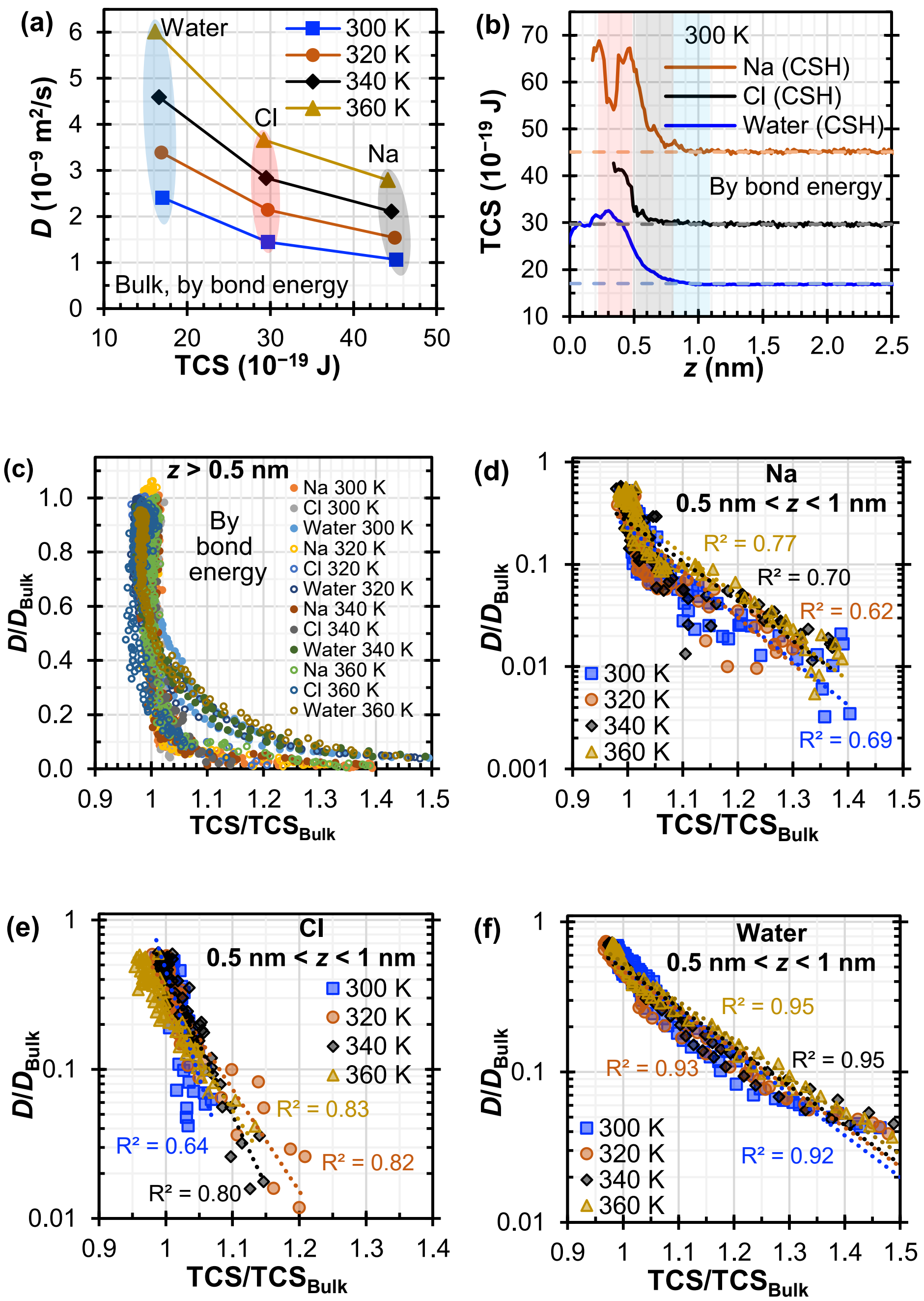

**Figure 10.** (a) Correlation between the total coordination strength (TCS) and diffusion coefficients for $Na^+$, $Cl^-$, and water molecules under bulk condition. (b) Spatial profiles of TCS($z$) for each solution species across the C–S–H nanochannel at 300 K. (c) Comparison of normalized diffusivity ($D/D_{bulk}$) and normalized TCS ($TCS/TCS_{bulk}$) for nanoconfined $Na^+$, $Cl^-$, and water molecules across all temperatures studied. (d–f) Exponential correlations between normalized diffusivity and normalized TCS within the near-surface region (0.5 nm < $z$ < 1.0 nm) for (d) $Na^+$, (e) $Cl^-$, and (f) water molecules. $R^2$ values in (d–f) are from exponential regression fits. All TCS values shown are based on bond energy, while those based on bond force are shown in **Figure S24** in **Supporting Information**.

To further investigate why, in the region ~1 nm < $z$ < ~1.5 nm, both ion and water diffusivities in the C–S–H nanochannel remain significantly lower than their bulk values (**Figure 4a** and **Figure S2**), despite the TCS approaching bulk levels, we performed empirical fitting of the diffusivity profile beyond $z$ = 1nm using a model inspired by the analytical solution of the one-dimensional steady-state Darcy–Brinkman equation under a partial slip boundary condition. This functional form was chosen to enforce the correct bulk asymptote while capturing the exponential decay behavior near the interface. The analogy is motivated by parallels between pressure-driven fluid flow and self-diffusion arising from thermal fluctuations in confined systems. For context, previous studies have reported a correlation between self-diffusivity and shear viscosity through the Stokes–Einstein relation for nanoconfined water dynamics in C–S–H nanopores.[119] A detailed derivation of the Darcy–Brinkman solution with a partial slip boundary condition,[119, 120] is provided in **Section S16** of the **Supporting Information**. Here, we adopt a mimicked exponential decay model with the same functional form as **Eq. (7)** to fit the MD-derived spatial distributions of water and ionic diffusivities.

$$D = D_{\mathrm{Bulk}}(T)\left(1 - a \cdot e^{-\frac{z-b}{c}}\right), \tag{7}$$

where, $D_{\mathrm{Bulk}}(T)$ is the bulk diffusivity at temperature $T$, $a$ is a fitting parameter describing the wall-induced suppression amplitude near the solid surface, $b$ accounts for the positional shift of the profile, and $c$ characterizes the decay length.

The fitting results and corresponding $R^2$ values for 300 K and 360 K within the range of $z > 1$ nm are presented in **Figure 11**a and b, respectively, while results for other temperatures are provided in **Figure S25** in the **Supporting Information**. The exponential decay model (**Equation** (7)) accurately captures the spatial evolution of water and ion diffusivities for $z > \sim 1.0$ nm, suggesting that, in this region, diffusivity suppression may be explained by continuum-like viscous resistance, in a manner consistent with the Darcy–Brinkman framework. Although water molecules and ions experience hydrodynamic perturbations—manifesting as reduced diffusivity—their local structural environment (e.g., TCS) closely resembles that of the bulk. This implies that the remaining suppression of diffusivity arises from long-range viscous coupling (as evidenced by the delayed recovery of water dipole orientation to bulk value beyond $z > 1.5$ nm, as shown in **Figure S8**) or confinement-induced hydrodynamic effects, rather than from local structural constraints. In contrast, for $z < 1.0$ nm, the exponential model in **Equation** (7) deviates from the MD results due to the dominance of interfacial structural effects, specifically the enhanced total coordination strength (TCS) that strongly governs transport near the solid surface. To test robustness, the fitting was also extended to $z > 0.75$ nm and $z > 0.5$ nm (**Figure S26–27)**. These extended fittings show generally good agreements down to 0.5 nm but still exhibit systematic deviation within $z = 1$ nm, especially for $Cl^-$ and water at 360 K. This analysis suggests that the region between ~0.5 and 1.0 nm represents a transitional regime, where structural ordering begins to relax while hydrodynamic resistance is not yet fully dominant, bridging the interfacial ($z < \sim 0.5$ nm) and bulk-like ($z > \sim 1$ nm) regimes.

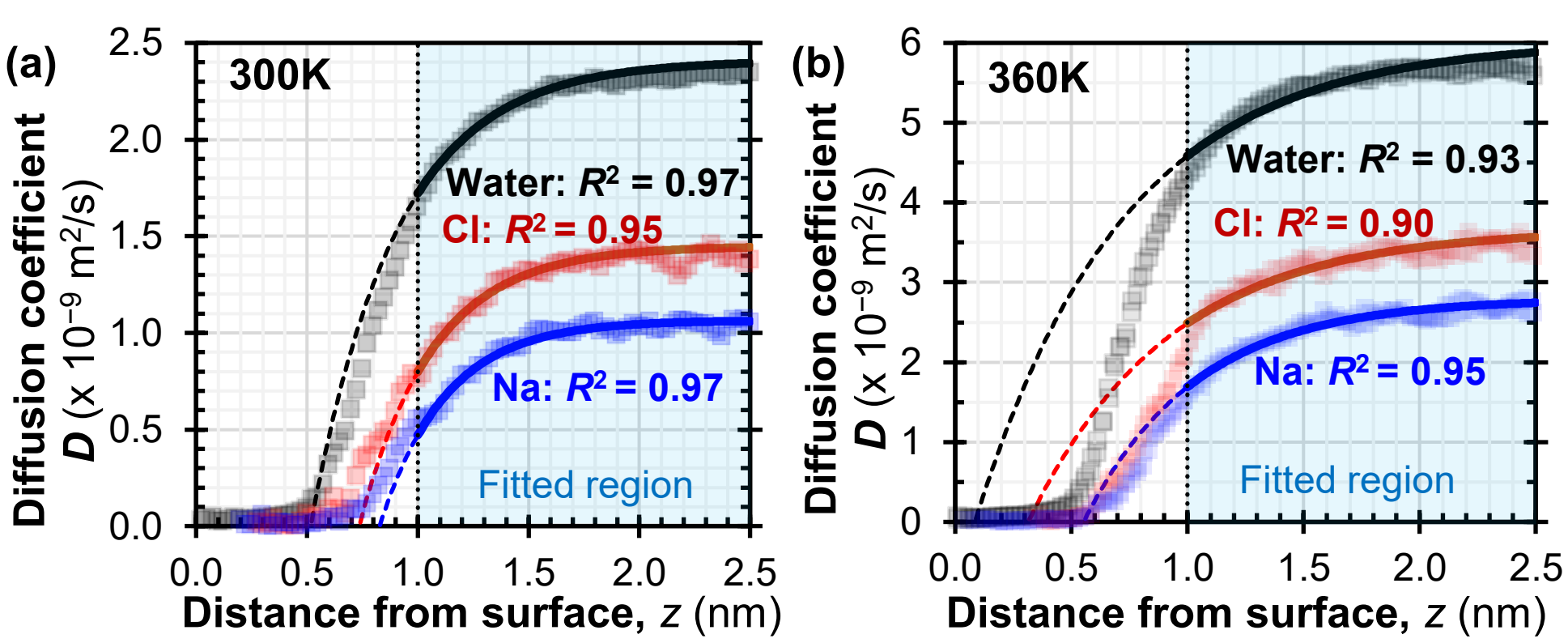

**Figure 11.** Fitting of the spatial distribution of diffusion coefficient $D$ ($10^{-9}$ $m^2/s$) for water, $Cl^-$ and $Na^+$ (black, red and blue squares, respectively) in the C–S–H nanochannel for the region $z >$ 1.0 nm at (a) 300 K and (b) 360 K using the mimicked exponential decay model (**Equation** (7), as given by the solid and dashed lines). $R^2$ values indicating goodness-of-fit for the fitted region are reported in each panel.

## 4 Broader Impact and Limitations

**Broader Impact.** A fundamental understanding of ionic transport in cementitious materials is essential for predicting long-term durability, modeling degradation mechanisms, and designing sustainable infrastructure with extended service life. This study presents a detailed mechanistic investigation of ion and water transport in C–S–H gel, the primary binding phase in OPC, the most widely used binder in concrete construction. While our simulations focus on a 4-nm C–S–H pore channel with Ca/Si = 1.67, the spatially resolved analytical framework and the structural descriptor introduced (i.e., total coordination strength (TCS)) are transferable to a broader range of binder chemistries, including blended cements and alkali-activated materials (AAMs), as well as varied pore sizes, solution chemistries, and environmental conditions (e.g., temperature and pressure). By linking spatially resolved diffusion profiles with local coordination environments and continuum-scale hydrodynamics, this study provides critical insights into the nanoconfined ionic transport behavior. The results may serve as key inputs for multiscale transport modeling in disordered porous media. Moreover, the observed interfacial immobilization mechanisms suggest that tuning nanoscale features, such as surface groups, pore sizes, and adsorption sites, may offer effective strategies to hinder the ingress of deleterious species, and enhance long-term durability of cementitious materials under aggressive environments.

Beyond cementitious systems, the methodologies demonstrated here—including spatially resolved transport analysis, and coordination-based structural descriptors—can be broadly transferable to other nanoconfined materials systems where ion mobility plays a critical role. These include nanoporous materials (e.g., clays,[11] zeolites,[33] and MOFs[121, 122]), and complex environments encountered in geochemical, biological, and energy-related applications—such as mineral weathering,[1, 2] nutrient cycling,[3] signal transduction at the cell membrane,[4] water desalination and purification,[5, 6] battery energy storage,[7-9] and subsurface waste containment.[10, 11] The transferability of this framework may therefore support advances across a wide spectrum of scientific and engineering domains where nanoscale transport processes are pivotal.

**Limitations.** Despite the valuable insights gained, several limitations associated with this study, and force-field-based MD simulations in general, warrant discussion. First, real-world cementitious gels are structurally and chemically heterogeneous, exhibiting variable pore surface chemistries, multiscale pore structures, and dynamically evolving solution compositions.[123] While our model isolates key mechanisms within a representative C-S-H pore system, caution should be exercised when generalizing the detailed findings to the full complexity of hydrated cement pastes. This work considers only one electrolyte (NaCl) and a single pore size (4 nm). Future studies should systematically explore the influence of gel composition, pore size distribution, and solution chemistry (e.g., $K^+$, $SO_4^{2-}$, $Ca^{2+}$, $Al^{3+}$, and varying ionic strengths) on transport properties, as well as the generalizability of structural descriptors such as TCS. In fact, smaller gel pores are known to exist in cementitious materials.[27, 51, 124] As the pore size decreases toward the ultraconfined regime—where the available space becomes comparable to the hydration size of interlayer water and ions, or even smaller than their molecular or ionic size (on the order of a few Å)—the transport behavior may change qualitatively. In such regimes, diffusion becomes highly anisotropic or non-Fickian, with strongly suppressed mobility, as reported in previous studies.[28, 90] This highlight the need for systematic extensions beyond the present model.

Second, the simulations were conducted using the classical, non-reactive CLAYFF force field, which does not account for bond breaking or formation at the solid-liquid interface. While this approach is well-suited for modeling the physical aspects of ion transport, it cannot capture interfacial chemical reactions that may occur under reactive or aggressive environmental conditions. To capture such effects, future work could employ reactive force fields (e.g., ReaxFF) or *ab initio* methods (e.g., density functional theory), which offer higher chemical fidelity at the expense of significantly higher computational costs and limited scalability. Balancing chemical realism with computational efficiency will be essential for extending this framework to more complex and reactive cementitious systems.

## 5 Conclusions

In this study, molecular dynamics (MD) simulations were employed to investigate the diffusion mechanisms of $Na^+$, $Cl^-$, and water molecules confined within a ~4-nm calcium-silicate hydrate (C–S–H) channel, representative of pore structure in ordinary Portland cement (OPC) binder, across temperatures from 300 K to 360 K. Spatially resolved analysis revealed strong suppression of diffusion near the solid–liquid interface, with gradual recovery toward the pore center. Arrhenius analysis showed that nanoconfinement reduces the intrinsic mobility ($D_0$) of all species, while affecting activation energy barriers ($E_a$) differently: lowering $E_a$ for $Na^+$ and $Cl^-$, but slightly increasing it for water. This indicates altered transport kinetic and energetic under confinement. Further spatially resolved analysis revealed elevated $E_a$ and $D_0$ in the second and third water layers, suggesting that although interfacial adsorption imposes strong energetic barriers at lower temperatures, enhanced surface-mediated transport mechanisms such as lateral hopping or gliding become possible once thermally activated. In contrast, reduced $E_a$ and $D_0$ near the channel center relative to the bulk indicate that transport in this region is limited primarily by steric hindrance and viscous friction rather than energetic trapping. For the innermost interfacial layer within surface roughness, water molecules displayed near-zero mobility, suggesting quasi-immobilization due to strong local bonding and geometric confinement.

To uncover the structural origins, we analyzed partial radial distribution functions (RDFs) and spatially resolved coordination number (CN) profiles across the pore channel. Near the C–S–H surface, enhanced local coordination was observed, primarily arising from strong adsorption of $Na^+$, $Cl^-$, and water onto surface oxygen and/or calcium sites. These enhanced interactions explain the observed suppression of diffusivity and increase in activation energy near the interface. Building upon these insights, we introduced a physically motivated structural descriptor, i.e., total coordination strength (TCS), to quantify local resistance to diffusion. A strong inverse correlation between TCS and diffusivity was observed across all species and temperatures. In the near-surface region ($z < 1.0$ nm), reductions in diffusivity were primarily governed by elevated TCS, reflecting the dominant influence of interfacial interactions. Beyond ~1.0 nm, where TCS values recovered to bulk levels, diffusivity remained suppressed. This behavior was accurately captured by fitting the diffusivity profiles beyond ~1.0 nm using an empirical exponential decay model inspired by the Darcy–Brinkman framework, indicating that transport in this region is potentially governed by continuum-scale viscous resistance rather than local structural constraints.

Together, these results reveal a mechanistic transition from structure-controlled to hydrodynamics-controlled transport regimes under nanoconfinement in C-S-H gel. TCS emerges as a physically meaningful descriptor for predicting the spatial and thermal evolution of ionic and water transport. This work not only provides fundamental insights into molecular transport under nanoconfinement but also suggest the potential of tailoring interfacial chemistry and nanoscale structure—such as surface coordination environments, pore size distributions, and adsorption sites—to modulate diffusion kinetics and dynamics in cementitious gels and other nanoporous materials.

## 6 Supporting Information

Additional results can be found in Supporting Information, including: temporal evolution of the system's energy during the relaxation stage, potential parameters of the ClayFF force field, spatial distribution of diffusivity, comparison between 3D and 2D diffusivity, comparison between NVT and NVE ensemble, spatially resolved in-plane mean square displacement (MSD) profiles, spatial distribution of diffusivity ratio, number density distribution of different solution species, dipole orientation distribution of interlayer water molecules, spatial distribution of activation energy barrier and intrinsic mobility, the RDF at different temperatures, the cutoff distance for coordination number (CN) calculation, the spatial distribution of the CN, the spatial distribution of the total coordination strength, the correlation between diffusivity and the total coordination strength determined by the bond force, analytical solution of the one-dimensional steady-state Darcy–Brinkman equation, and the empirical fitting of diffusivity distribution.

## 7 Use of AI-Assisted Technologies

The authors utilized ChatGPT (GPT-4o) to assist with language refinement during the manuscript preparation. All content has been carefully reviewed and revised as needed by the authors to ensure accuracy and clarity, and the authors take full responsibility for the final submitted work.

## 8 Declaration of Competing Interest

The authors declare that they have no known competing financial interests or personal relationships that could have appeared to influence the work reported in this paper.

## 9 Acknowledge

This work was supported in part by the Big-Data Private-Cloud Research Cyberinfrastructure MRI-award funded by NSF under grant CNS-1338099 and by Rice University's Center for Research Computing (CRC).

## 11 TOC Graphic

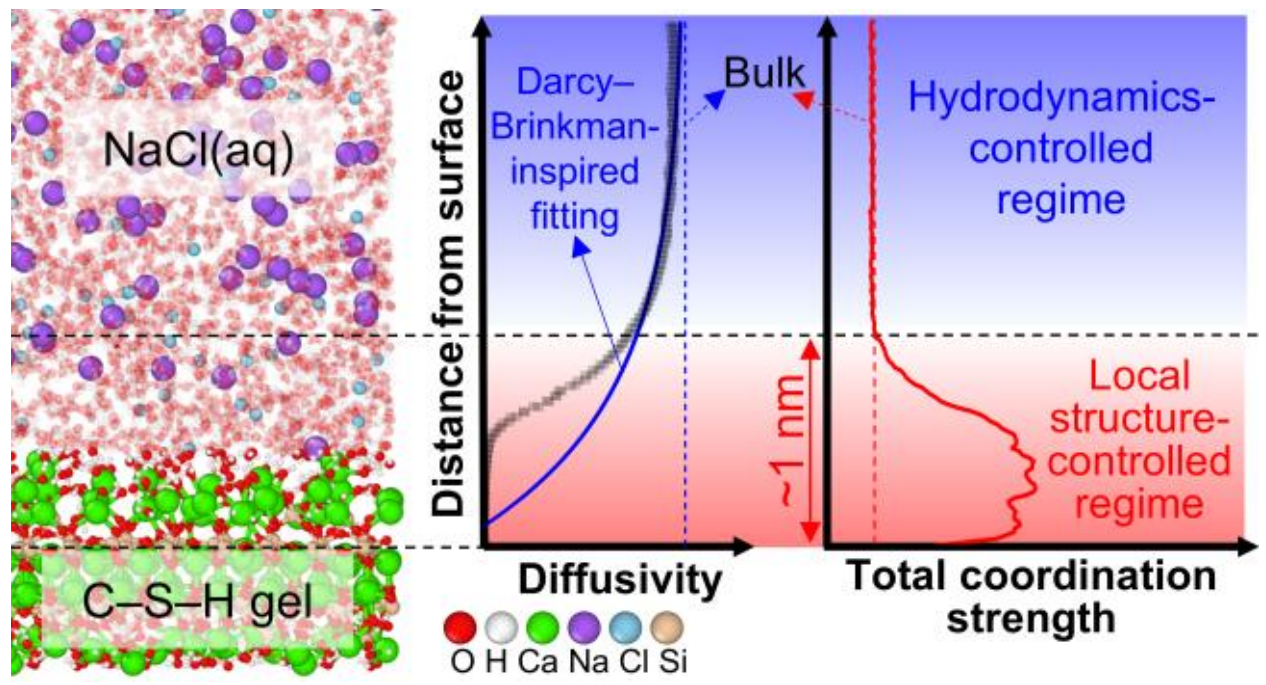

# Supplemental Materials for

# Insights into Ionic Diffusion in C–S–H Gel Pore from Molecular Dynamics Simulations: Spatial Distributions, Energy Barriers, and Structural Descriptor

Weiqiang Chen[a,b,c] and Kai Gong[a,b,c,*]

[a] Department of Civil and Environmental Engineering, Rice University, Houston, Texas 77005, United States

[b] Rice Advanced Materials Institute, Rice University, Houston, Texas 77005, United States

[c] Ken Kennedy Institute, Rice University, Houston, Texas 77005, United States

* Corresponding author. E-mail: kg51@rice.edu

**Section S1 The temporal evolution of the system's potential energy during the relaxation stage**

**Figure S1** shows the temporal evolution of the bulk solution and C–S–H nanopore system's potential energies during the relaxation stage at different temperatures, indicating minimal fluctuations. This confirms that the 4 ns relaxation period is sufficient for the system to reach equilibrium, and is suitable for initiating the subsequent production runs.

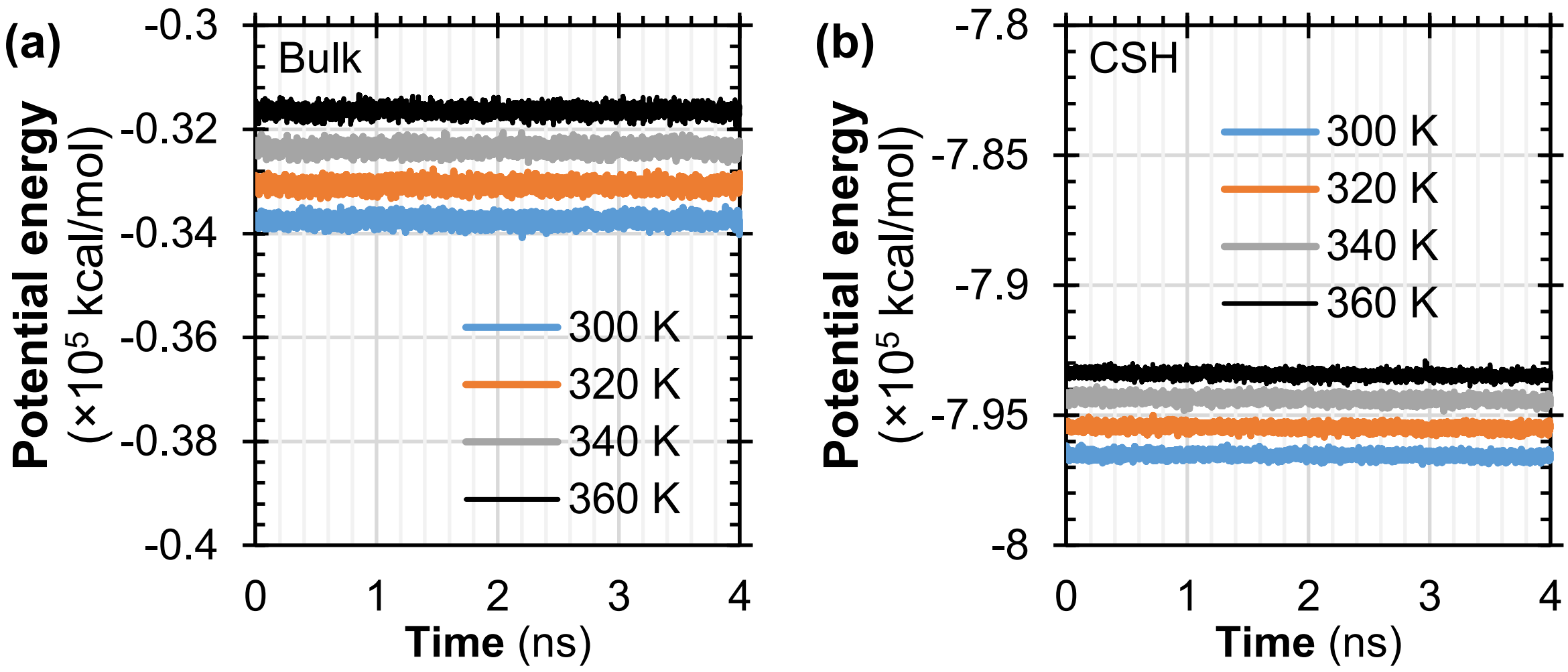

**Figure S1.** The temporal evolution of potential energy for (a) bulk solution system and (b) C–S–H nanopore system during the 4 ns relaxation stage at different temperature.

## Section S2 The potential parameters of the ClayFF force field

The total energy of the simulated system described by the ClayFF force field[1] can be obtained by the equation below:

$$E_{\text{Total}} = E_{\text{vdW}} + E_{\text{Coul}} + E_{\text{Bond stretch}} + E_{\text{Angle bend}}, \quad \text{(S1)}$$

where, $E_{\text{vdW}}$, $E_{\text{Coul}}$, $E_{\text{Bond stretch}}$, and $E_{\text{Angle bend}}$ are terms describing the short-range van der Waals, Coulombic (electrostatic), bond stretch, and bond angle bend interactions, respectively.

The first term can be calculated by the conventional Lennard-Jones 12-6 potential,

$$E_{\text{vdW}} = \sum 4D_{0,ij}\left[\left(\frac{R_{0,ij}}{r_{ij}}\right)^{12} - \left(\frac{R_{0,ij}}{r_{ij}}\right)^{6}\right], \quad \text{(S2)}$$

where, $r_{ij}$ is the interatomic distance between species $i$ and species $j$, and $D_{0,ij}$ and $R_{0,ij}$ are the empirical potential paramters. These two paramters for atoms with the same atom types can be direclty obtained from the **Table S1** below. For interaction between dissimilar species,

$$D_{0,ij} = \sqrt{D_{0,ii}D_{0,jj}},$$
$$R_{0,ij} = \frac{1}{2}(R_{0,ii} + R_{0,jj}), \quad \text{(S3)}$$

For the second term,

$$E_{\text{Coul}} = \frac{e^2}{4\pi\varepsilon_0}\sum\frac{q_i q_j}{r_{ij}}, \quad \text{(S4)}$$

where, $e$ is the electronic charge, $\varepsilon_0$ is the dielectric permittivity of vacuum ($8.85419 \times 10^{-12}\text{F/m}$), $q_i$ and $q_j$ are the partial charges of species $i$ and $j$ derived from quantum mechanics calculations, respectively.

For the third term,

$$E_{\text{Bond stretch}} = \sum B_{0,ij}(r_{ij} - r_{0,ij})^2, \quad \text{(S5)}$$

where, $B_{0,ij}$ is the force constant, $r_{0,ij}$ is the bond length at equilibrium state.

For the fourth term,

$$E_{\text{Angle bend}} = \sum A_{0,ijk}(\theta_{ijk} - \theta_{0,ijk})^2, \quad \text{(S6)}$$

where, $A_{0,ijk}$ is the force constant, $\theta_{ijk}$ and $\theta_{0,ijk}$ are the bond angles under current and equilibrium state, respectively. The corresponding parameters used in this work are listed below from **Table S1** to **Table S3**.

The bond/coordination strengths between different species, e.g., Na-Ow, represented by either the interatomic bond energy or bond force, can be calculated using **Equation** (S1) (energy method) and its first derivative (force method). The equilibrium bond lengths (i.e., the nearest interatomic distances) derived from radial distribution functions (RDFs) are taken as $r_{ij}$ in the calculation. The total coordination strength (TCS) is calculated as the sum of bond/coordination strengths from all neighboring species within the local coordination environment, as introduced in **Section 3.4** of the main text.

**Table S1.** Nonbond parameters.

| **Species $i$** | $q_i$* | $D_{0,ii}$ **(kcal/mol)** | $R_{0,ii}$ **(Å)** |
|---|---|---|---|
| Intralayer calcium (Cai) | +1.05 | 0.0000050219 | 5.5624343529 |
| Aqueous interlayer calcium ion (Ca) | +2 | 0.0999999980 | 2.8719902024 |
| Tetrahedral silicon (St) | +2.1 | 0.0000018402 | 3.3019566252 |
| Bridging oxygen (Ob) | −1.05 | 0.1554164124 | 3.1655200879 |
| Water oxygen (Ow) | −0.82 | 0.1554164124 | 3.1655200879 |
| Water hydrogen (Hw) | +0.41 | 0.0000000000 | 0.0000000000 |
| Hydroxyl oxygen (Oh) | −0.95 | 0.1554164124 | 3.1655200879 |
| Hydroxyl hydrogen (Hh) | +0.425 | 0.0000000000 | 0.0000000000 |
| Aqueous sodium ion (Na)[2] | +1 | 0.1300999871 | 2.3500126639 |
| Aqueous chloride ion (Cl)[2] | −1 | 0.1000998933 | 4.3999709805 |

*Minor adjustments to the partial charges were made following Ref. [3] in order to maintain overall charge neutrality of the simulation system.

**Table S2.** Bond stretch parameters.

| **Species $i$** | **Species $j$** | $B_{0,ij}$ **(kcal/mol/Å$^2$)** | $r_{0,ij}$ **(Å)** |
|---|---|---|---|
| Ow | Hw | 553.9350 | 1.0000 |
| Oh | Hh | 553.9350 | 1.0000 |

**Table S3.** Bond angle bend parameters.

| **Species $i$** | **Species $j$** | **Species $k$** | $A_{0,ijk}$ **(kcal/mol)** | $\theta_{0,ijk}$ **(Å)*** |
|---|---|---|---|---|
| Hw | Ow | Hw | 45.7530 | 109.4700 |

*LAMMPS covert degrees to radians internally.

## Section S3 The spatial distribution of water and ion diffusivity at different temperatures

To evaluate the spatial heterogeneity of water and ion diffusivity across the entire C–S–H gel nanopore, a binning method is employed along the $z$-direction. The nanopore is divided into multiple bins of fixed width $b$. For each bin centered at position $z$, its spatial range is defined as spanning from $z - b/2$ to $z + b/2$.

During the simulation, water molecules and ions may transiently enter and exit each bin. For any particle $i$ (either a water molecule or ion) that enters the bin centered at position $z$ at time $t_0$ and remains continuously within the bin for the entire interval from $t_0$ to $t_0 + \tau$, their average diffusivity at position $z$ is calculated as,

$$D(z) = \langle \frac{(\boldsymbol{r}_i(\tau+t_0) - \boldsymbol{r}_i(t_0))^2}{2d\tau} \rangle_{i,t_0,\tau}, \tag{S7}$$

where, $\langle \cdot \rangle_{i,t_0,\tau}$ represents an average running over particles, time origins, and residence times, $\boldsymbol{r}_i(t_0)$ and $\boldsymbol{r}_i(\tau + t_0)$ are the positions of particle $i$ at times $t_0$ and $\tau + t_0$, respectively. The parameter $d = 1, 2, 3$, depending on the dimensionality of the diffusivity being measured. In this study, $d = 3$ is used to represent the three-dimensional mobility of solution species. To balance computational efficiency with spatial resolution, a bin size of 1 Å was selected.

The binning procedure was performed by post-processing atomic trajectories exported from LAMMPS using an in-house MATLAB (R2023a, The MathWorks, Inc., Natick, MA, USA) script.[4] The unwrapped coordinates of each interlayer solution species were exported every 4 ps over a total duration of 4 ns, yielding 1001 atomic configurations. The MATLAB script follows **Eq. (S7)**, looping over all particles, time origins, and residence times in each bin to record qualified atomic motion events. To reduce the computational cost of this demanding task, the *parfor* (parallel for loop) command in MATLAB's Parallel Computing Toolbox was employed, enabling parallel execution of loop iterations across multiple CPU cores. It should be noted that the binning procedure along the $z$ direction primarily captures ionic and molecular motion parallel to the pore surface ($x$ and $y$ directions), while it does not account for appreciable motion perpendicular to the gel surface.

Following this method, the spatial distributions of water and ion diffusivity at various temperatures are quantified and presented in **Figure S2**, with the 300 K results shown in **Figure 4**a of the main article. At higher temperatures, overall trends remain consistent with those at 300 K: diffusivities drop to near-zero values near the C–S–H surface ($z$ = 0–0.5 nm) and gradually increases with distance from the wall, approaching bulk-like behavior near the pore center ($z$ = ~2.5 nm). Notably, **Figure S2** also reveals that, at elevated temperatures, the diffusivities of both ions and water increasingly deviate from their respective bulk values and become noticeably lower than bulk values even in the central region of the nanochannel.

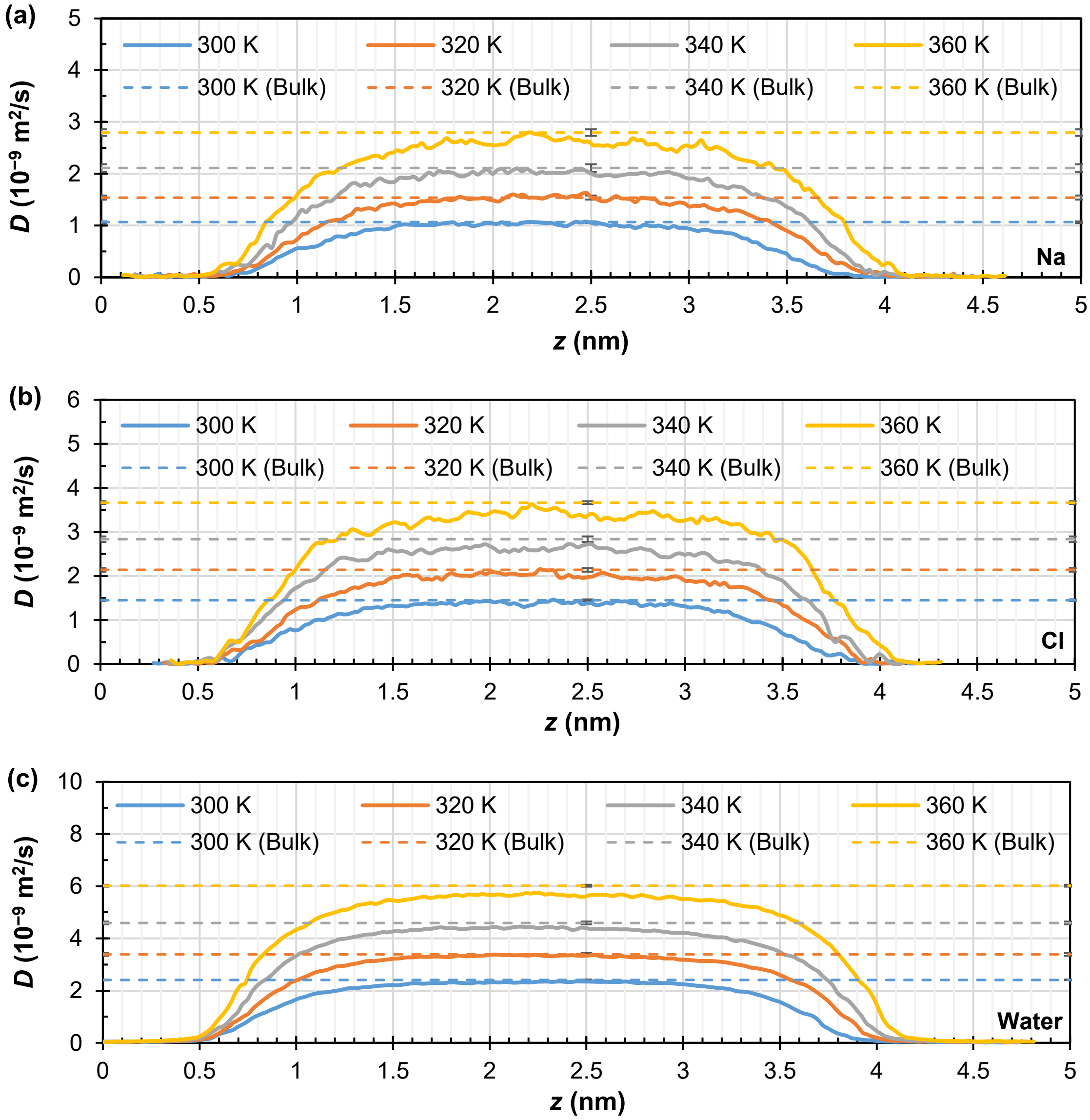

**Figure S2**. The spatial distribution of (a) $Na^+$, (b) $Cl^-$, and (c) water diffusivity at different temperatures, where a bin size of 1 Å is used along the *z*-direction to produce the spatial distribution profiles. Dashed lines indicate the corresponding average bulk diffusivity values, with error bars representing one standard deviation from three independent production runs.

## Section S4 Comparison of three-dimensional (3D) and two-dimensional (2D) diffusion coefficients

Three-dimensional (3D) diffusion refers to molecular motion in all spatial directions, i.e., *x*, *y*, and *z* directions, while two-dimensional (2D) diffusion captures motion parallel to the C–S–H gel surface, i.e., *x* and *y* directions only. **Figure S3** shows that the difference between 2D and 3D diffusivities is trivial across all temperatures. However, near the solid surface, a slightly larger discrepancy is observed due to the reduced mobility of ions and water molecules in the *z*-direction, where motion perpendicular to the gel surface is more constrained by solid-fluid interfacial interactions.

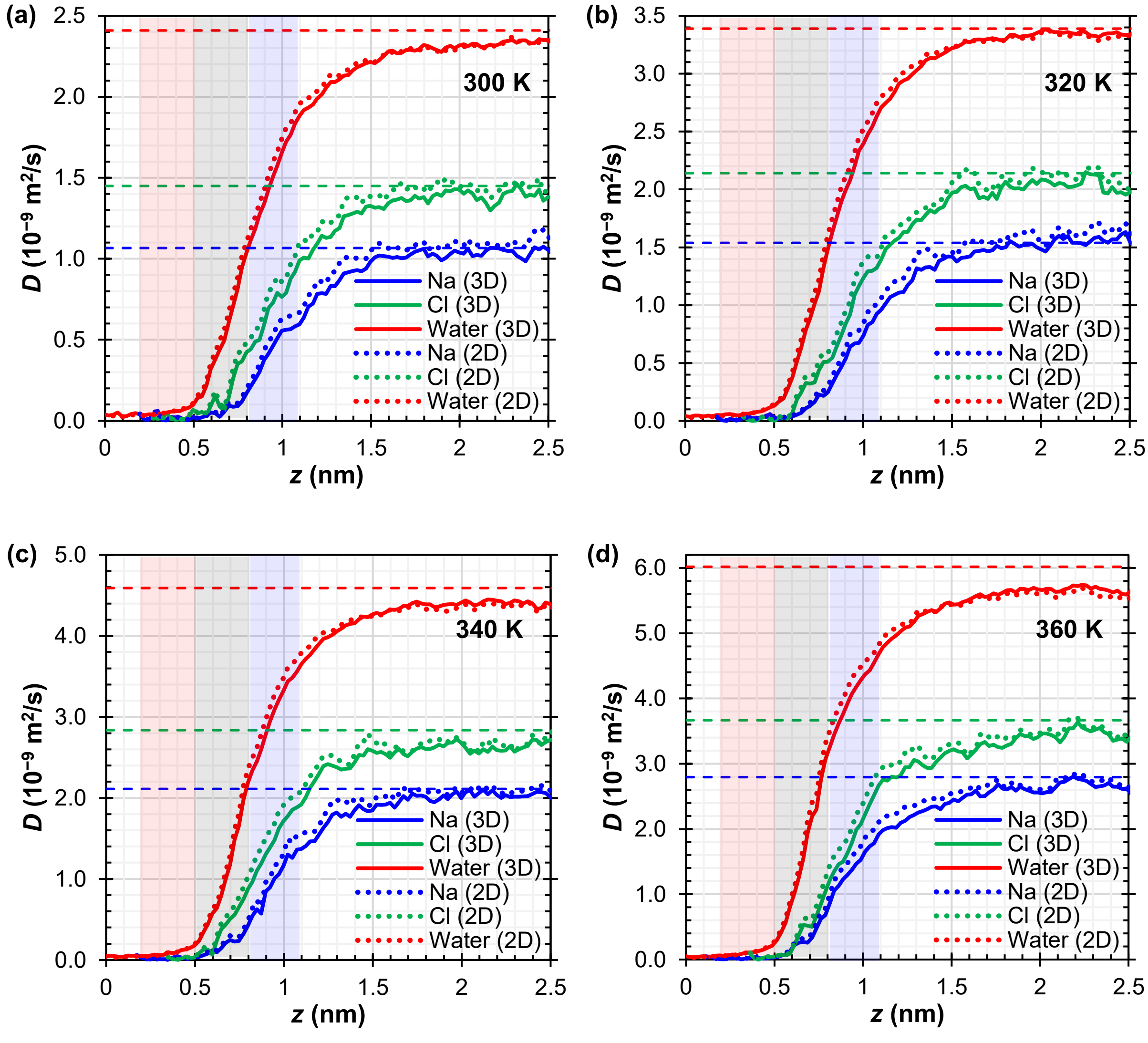

**Figure S3**. Spatial distribution of three-dimensional (3D) and two-dimensional (2D) diffusion coefficients for different solution species at various temperatures: (a) 300 K, (b) 320 K, (c) 340 K, and (d) 360 K. The horizontal dashed lines represent the corresponding 3D bulk diffusion coefficients for 2 M NaCl solution at each temperature.

## Section S5 Effect of NVT and NVE ensembles on dynamical calculations

In this work, diffusion coefficients were evaluated based on trajectories obtained under NVT conditions. From a rigorous standpoint, employing the NVE ensemble after proper equilibration provides the most appropriate framework for calculating the mean square displacement (MSD) and the corresponding diffusion coefficient.[5] Nevertheless, previous MD studies of ionic diffusion have shown that results obtained under NVT ensemble are generally very close to those from NVE simulations.[6] To explicitly verify that the thermostat does not introduce any fictitious forces affecting the calculated dynamics, we conducted an additional NVE simulation at 300 K. A comparison of the spatially resolved diffusion coefficient profiles is presented in **Figure S4**. The results demonstrate that the influence of the thermostat is negligible within the adopted simulation protocol.

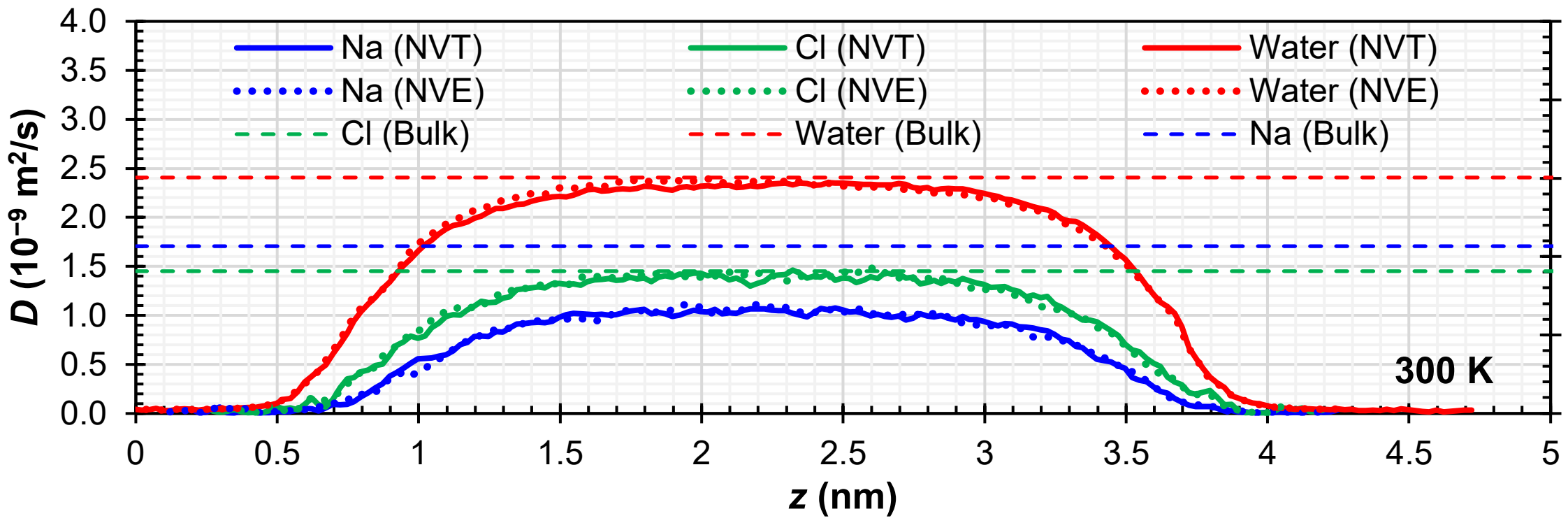

**Figure S4**. A comparison of the spatial distribution of diffusion coefficients for different solution species within the C–S–H channel under NVT (solid lines) and NVE (dotted lines) conditions. The horizontal dashed lines represent the corresponding bulk diffusion coefficients for 2 M NaCl solution at 300 K.

**Section S6 Spatially resolved in-plane mean square displacement (MSD) profiles of nanoconfined solution species**

In previous studies,[3, 7] where self-diffusion was investigated as a function of pore size and the pore-scale averaged diffusion coefficient was quantified, subdiffusive behavior along the pore-plane directions was observed in small meso- and micropores of C–S–H. To better elucidate this phenomenon, we calculated the spatially resolved mean square displacements (MSDs) of $Na^+$, $Cl^-$, and water molecules along the pore-plane directions (*xy*), as shown in **Figure S5**. These MSDs were obtained by post-processing the atomic trajectories with a sampling interval of 0.4 ps for ions and 1 ps for water molecules using MATLAB scripts.

The characteristic MSD scalings corresponding to the Fickian regime ($\mathrm{MSD} \propto t$), the sub-diffusive regime ($\mathrm{MSD} \propto t^a$ with $0 < a < 1$, e.g., $a = 0.5$), and the super-diffusive regime ($\mathrm{MSD} \propto t^a$ with $a > 1$, e.g., ballistic motion with $a = 2$) are also shown in the figure for reference. The results reveal that the slope ($a$ value) is close to unity at the pore center, indicating near-Fickian behavior. With increasing proximity to the pore surface, the slope progressively decreases below unity and then below 0.5, reflecting a growing deviation from Fickian scaling and the predominance by sub-diffusive behavior. Immediately adjacent to the gel surface ($z \approx 0$ nm), the slope approaches zero, corresponding extremely hindered, almost immobilized dynamics. This provides a microscopic explanation for the previously reported subdiffusion observed in pore-averaged diffusion coefficients within very small C–S–H nanopores.[3, 7]

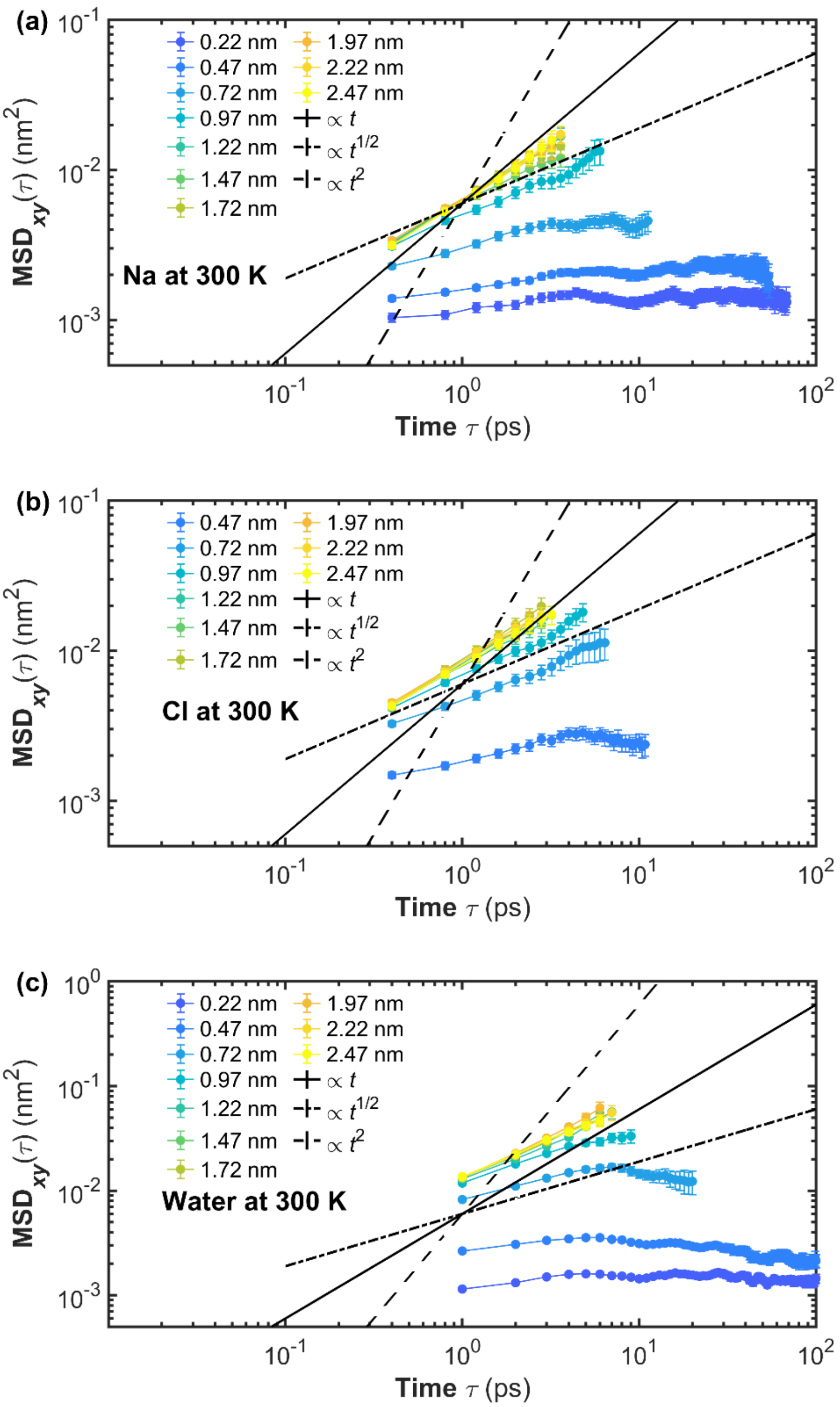

**Figure S5.** Spatially resolved mean square displacement (MSD) profiles along the in-plane directions (*xy*), plotted on a log–log scale. The black solid line, black dash–dot line, and black dashed line denote the characteristic MSD scalings corresponding to the Fickian regime (MSD $\propto t$), a representative sub-diffusive regime (MSD $\propto t^{\frac{1}{2}}$), and a representative super-diffusive regime (MSD $\propto t^2$), respectively.

**Section S7 The spatial evolution of the diffusion coefficient ratio for water/$Na^+$, water/$Cl^-$, and $Na^+$/$Cl^-$**

Based on the spatial diffusivity profiles in **Figure S2**, we calculated the spatial distribution of the diffusion coefficient ratio for $H_2O$/$Na^+$, $H_2O$/$Cl^-$, and $Na^+$/$Cl^-$ at various temperatures. The 300 K results are presented in **Figure 4**b of the main article, while **Figure S6** presents the results for all temperature. The results consistently show that the $H_2O$/$Na^+$ and $H_2O$/$Cl^-$ ratios are significantly elevated near the pore surface, indicating that ionic mobility is more strongly suppressed than that of water due to interfacial interactions. Among the two ions, $Na^+$ exhibits a greater reduction in diffusivity near the surface compared to $Cl^-$, suggesting stronger adsorption of $Na^+$ onto the C–S–H surface.

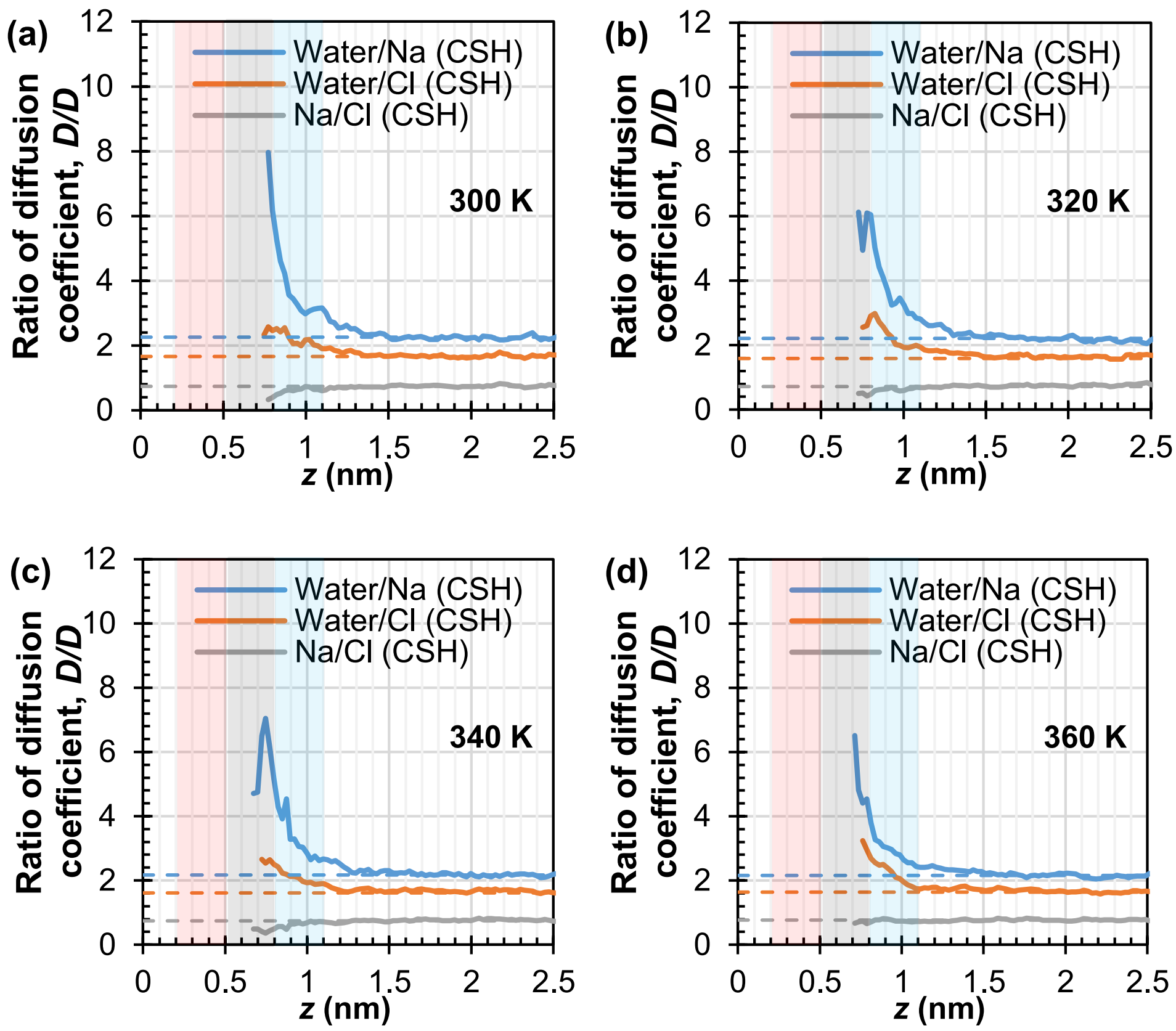

**Figure S6.** Evolution of the diffusion coefficient ratio for water/Na, water/Cl, and Na/Cl across the nanopore at different temperatures: (a) 300 K, (b) 320 K, (c) 340 K, and (d) 360 K. Dashed lines in the figures represent the corresponding bulk values for the 2 M NaCl solution at each temperature.

**Section S8 The number density distribution of different solution species at different temperatures**

The number density distributions of different solution species at various temperatures are presented in **Figure S7**. The overall trends at elevated temperatures are consistent with the 300 K results presented in **Figure 5**a of the main article, showing clear deviations from bulk solution values near the solid-liquid interface. These deviations reflect interfacial structuring or ordering, with number densities gradually recovering to bulk-like values away from the surface. Notably, as temperature increases, a greater number of $Na^+$ ions appear to adsorb onto the gel surface, resulting in reduced $Na^+$ density near the center of the nanochannel. This enhanced interfacial accumulation of $Na^+$ promotes increased secondary adsorption of $Cl^-$ ions due to electrostatic attraction. The resulting interfacial charge buildup generates a stronger electrostatic potential normal to the gel surface, aligning the dipole orientation of nearby water molecules, as shown in **Figure S8**, and decreasing the overall mobility of the interlayer liquid. Additionally, **Figure S7** shows that while the water density profiles recover to their bulk values when $z > \sim 1.0$ nm, the $Na^+$ and $Cl^-$ profiles only fully relax beyond $z \approx \sim 1.5$ nm.

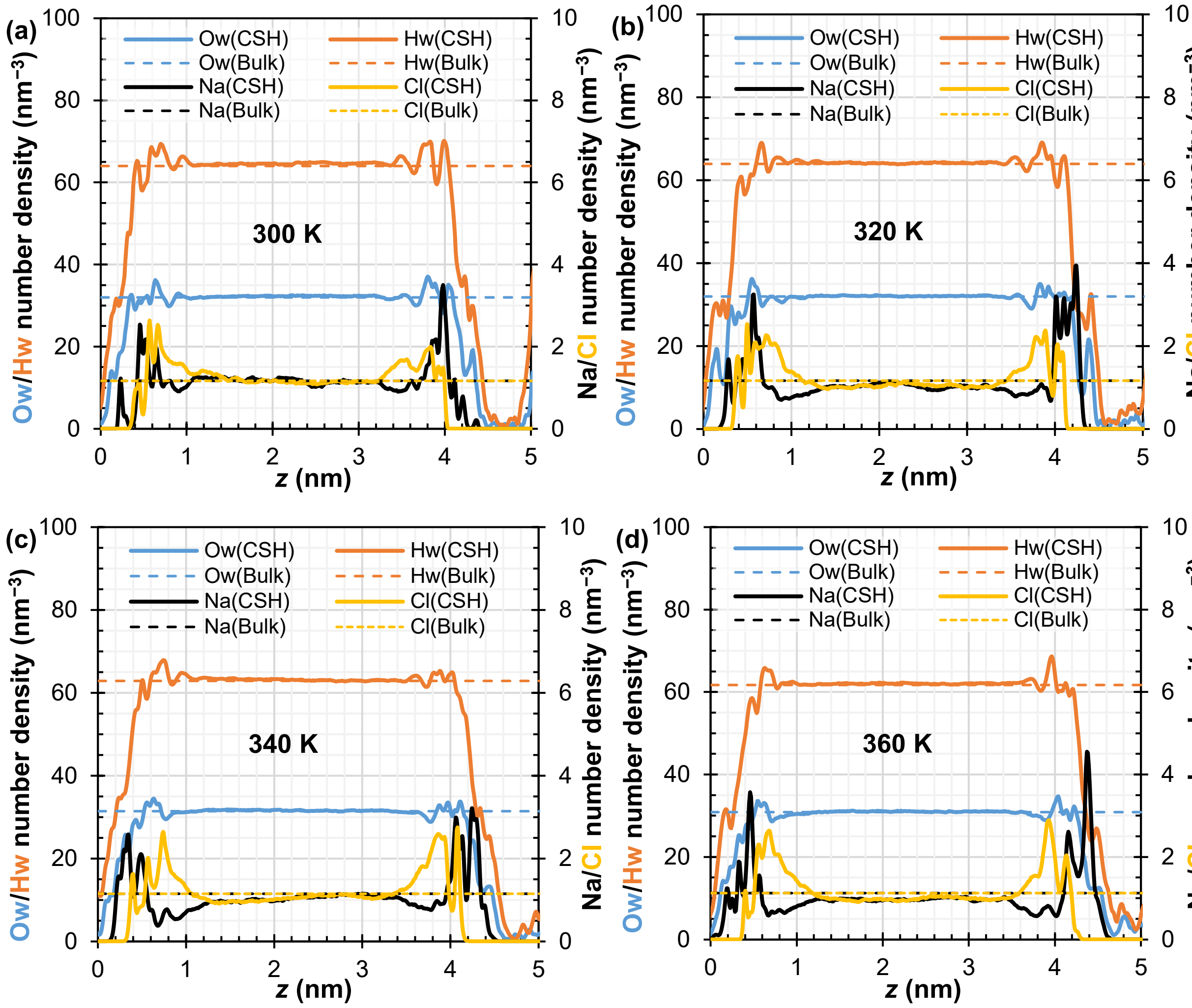

**Figure S7**. The number density distribution of different solution species at various temperatures: (a) 300 K, (b) 320 K, (c) 340 K, and (d) 360 K. Ow = Water oxygen atom; Hw = water hydrogen atom.

## Section S9 The dipole orientation distribution of nanoconfined water molecules at different temperatures

**Figure S8** shows (a) the definition of the angle ($\varphi$) between a water dipole vector and the surface normal and (b) the dipole orientation distribution of nanoconfined water molecules at different temperatures. **Figure S8**b reveals that the dipole orientation of interlayer water significantly deviates from its bulk value ($\varphi = 90°$) near the solid interface and gradually recovers to its bulk value beyond $z \approx 1.5$ nm. The results also show that with increasing temperature, water dipoles tend to align more strongly towards the C–S–H surface (i.e., smaller $\varphi$ values), indicating enhanced attraction of water oxygen atoms to the solid interface.

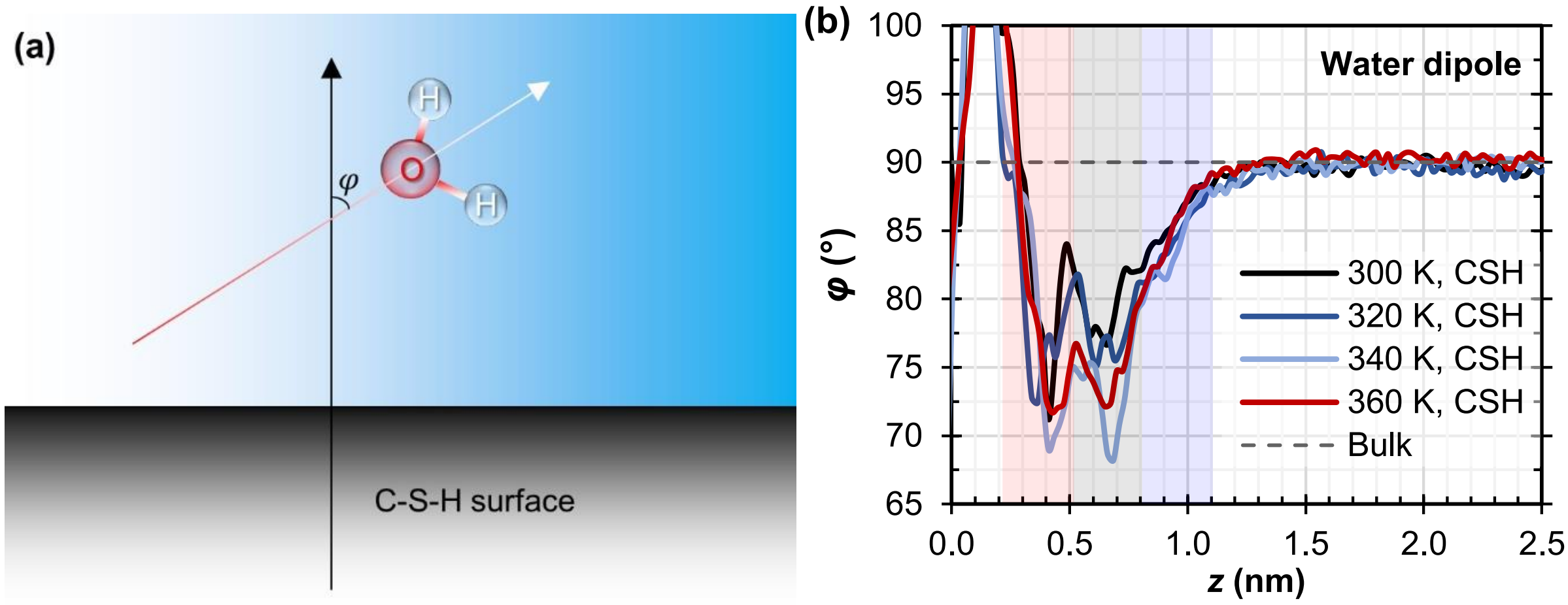

**Figure S8**. (a) Schematic illustrating the dipole orientation of a water molecule and the angle ($\varphi$) relative to the surface normal. (b) Spatial distribution of water dipole orientation angles at different temperatures, showing how alignment varies with distance from the C–S–H pore surface.

**Section S10 The spatial distribution of the activation energy barrier and intrinsic mobility**

The spatial distribution profiles of activation energy ($E_a$) and intrinsic mobility ($D_0$) are derived from diffusivity profile at different temperatures in **Figure S2** and presented in **Figure S9**. These profiles show that ion and water diffusivities approach zero near the solid-liquid interface, suggesting that they are strongly adsorbed/immobilized on the C–S–H gel surface. In this interfacial region, diffusion behavior may deviate from the classical Arrhenius relation due to the potential complexity of surface-mediated transport mechanisms, e.g., hopping or gliding between different adsorption sites, as reported for other material surfaces, including metal and semiconductor.[8, 9] In such cases, molecules or ions are deeply trapped in local energy wells with high adsorption energies. While desorption from these sites might still follow Arrhenius behavior (involving a single energy barrier), lateral surface diffusion may not, especially when the surface exhibits roughness, heterogeneous or disordered features with variable binding strengths.

This interpretation aligns with the assumptions underlying the Arrhenius equation, which describes kinetics governed by a single, well-defined energy barrier.[10] In the center of the nanochannel, where the solution approximates bulk conditions, the ions and water molecules within each spatial bin are close to uniformly and randomly distributed, experiencing similar local coordination environments. As a result, they exhibit a consistent activation energy, $E_a$, and follow Arrhenius behavior. However, near the C–S–H surface, local coordination environments vary significantly from molecule to molecule due to the disordered and chemically heterogeneous nature the C–S–H gel surface, together with its surface roughness. As a result, the effective $E_a$ may differ across different trajectories, reflecting a distribution of energy barriers rather than a single value, thereby leading to deviations from the ideal Arrhenius behavior. Moreover, because diffusion events near the surface occur much less frequently than in the channel center, limited simulation timescales may fail to capture rare hopping or surface diffusion events, further contributing to apparent non-Arrhenius behavior.

In addition, for the interlayer region where ion and water diffusivity approaches zero ($D \approx 0$), the logarithmic value ($\ln D \rightarrow -\infty$) becomes numerically unstable, introducing significant error into the linear fitting. By comparing the left and right interfacial regions and evaluating their goodness-of-fit ($R^2$), it is found that the Arrhenius relation holds reasonably well for $Na^+$ ions at $z > 0.65$ nm,

Cl⁻ ions at $z$ > 0.85 nm, and water at $z$ > 0.25 nm. Therefore, only these ranges are included in the Arrhenius analysis presented in **Figure 7** of the main text.

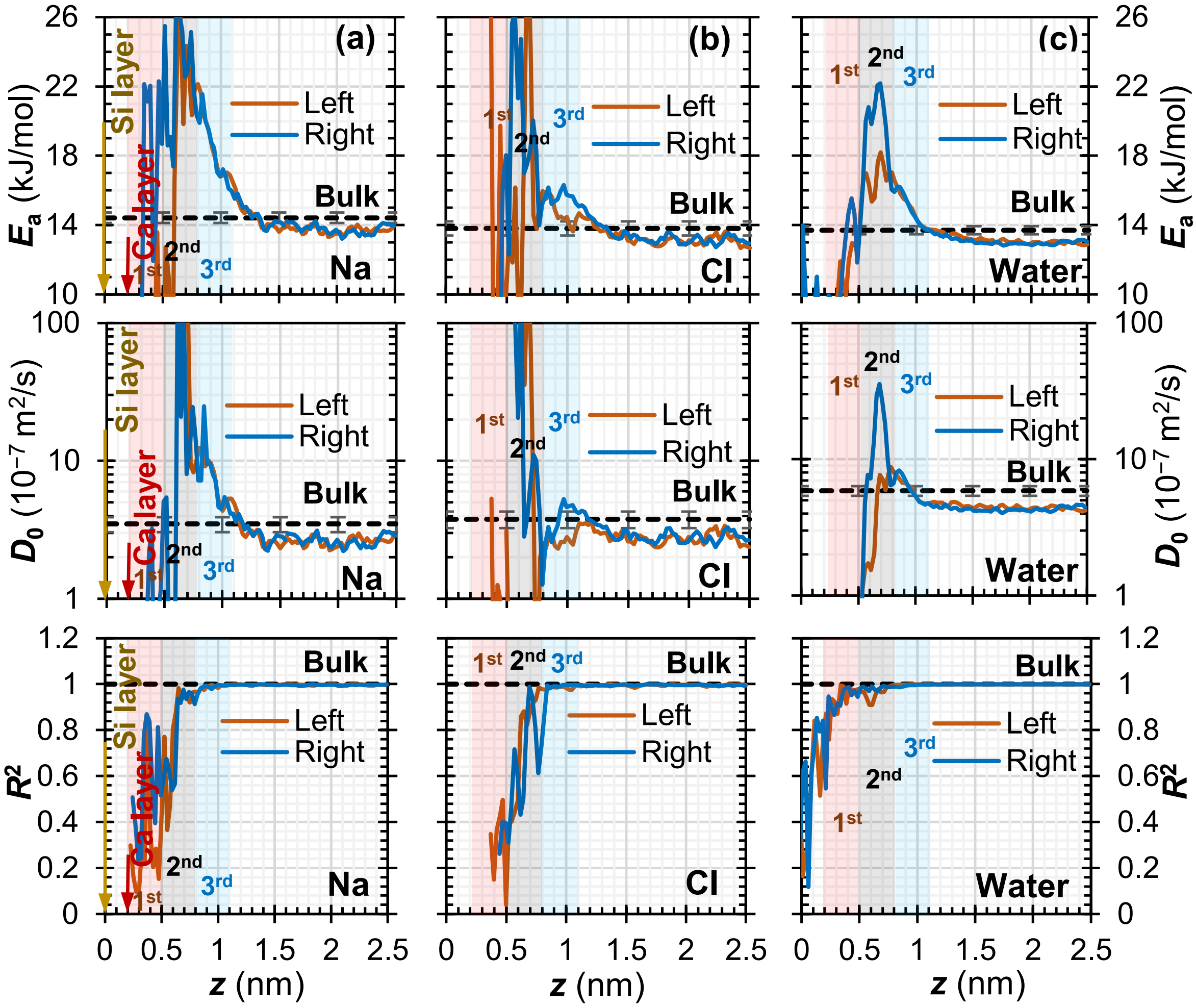

**Figure S9**. Spatial distributions of the activation energy barrier (top), the intrinsic mobility (middle), and corresponding goodness-of-fit (bottom) for (a) $Na^+$, (b) $Cl^-$, and (c) water diffusion. Both the left and right solid-liquid interfaces are shown for comparison. $z$ = 0 nm indicates the average position of surface silicon (Si) atoms. Dashed lines with error bars in the top and middle panels represent the corresponding bulk values from NaCl solutions, while dashed lines in the bottom panel indicate perfect linear fitting against Arrhenius relation ($R^2 = 1$).

**Section S11 The partial radial distribution function (RDF) at different temperatures**

The partial RDFs for various atom-atom pairs in the interlayer solution at 300 K are presented below in **Figure S10**, revealing that: (1) the nearest neighbors of Na-X pairs with attractive interactions include O and Cl atoms; (2) the nearest neighbors of Cl-X pairs with attractive interactions include H and Ca atoms; (3) the nearest neighbors of Ow-X pairs with attractive interactions include H, Ca, and Na atoms; (4) the nearest neighbors of Hw-X pairs with attractive interactions include O and Cl atoms. Therefore, in the main text and below, only the nearest neighbors with attractive interactions are presented. The partial RDFs for all temperatures are shown below from **Figure S11** to **Figure S14**.

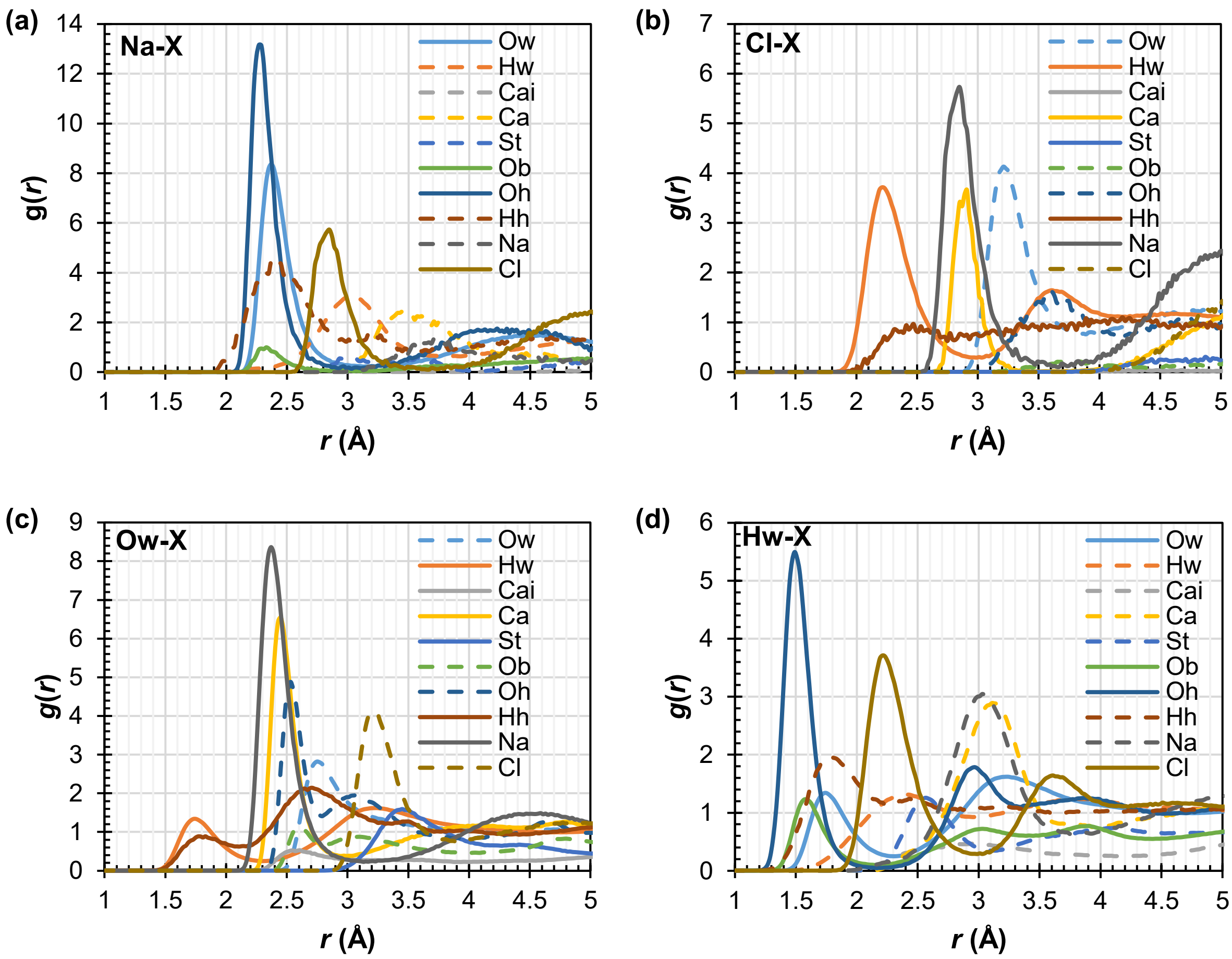

**Figure S10.** Partial radial distribution functions (RDFs) of interlayer solution species at 300 K for (a) Na-X, (b) Cl-X, (c) Ow-X, and (d) Hw-X pairs, where the first element denotes the central atom type while the second one “X” represents the surrounding atom type. Note that the intramolecular interactions, specifically Ow–Hw and Hw-Ow (within water molecules) are excluded from the analysis. Atom labels are defined as follows: water oxygen (Ow), water hydrogen (Hw), intralayer calcium (Cai), interlayer calcium (Ca), silicon (St), bridging oxygen (Ob), hydroxyl oxygen (Oh), hydroxyl hydrogen (Hh), sodium ion (Na), and chloride ion (Cl).

**Figure S11**a displays the total RDFs averaged over all atom-atom pairs for bulk NaCl solution and that confined within the 4-nm C–S–H nanochannel at 300 K, where distinct peaks correspond to contributions from individual partial RDFs shown in **Figure S11**b–h. **Figure S11**f shows the coordination environment of silicate and calcium atoms in the C–S–H gel, where the bond distances for St–Ob, Ca–Ob, and Cai–Ob are ~1.55 Å, ~2.43 Å, and ~2.35 Å, respectively, consistent with the experimental data by the total X-ray scattering.[11] The corresponding results for 320, 340 and 360 K are shown in **Figures S12, S13 and S14**, respectively.

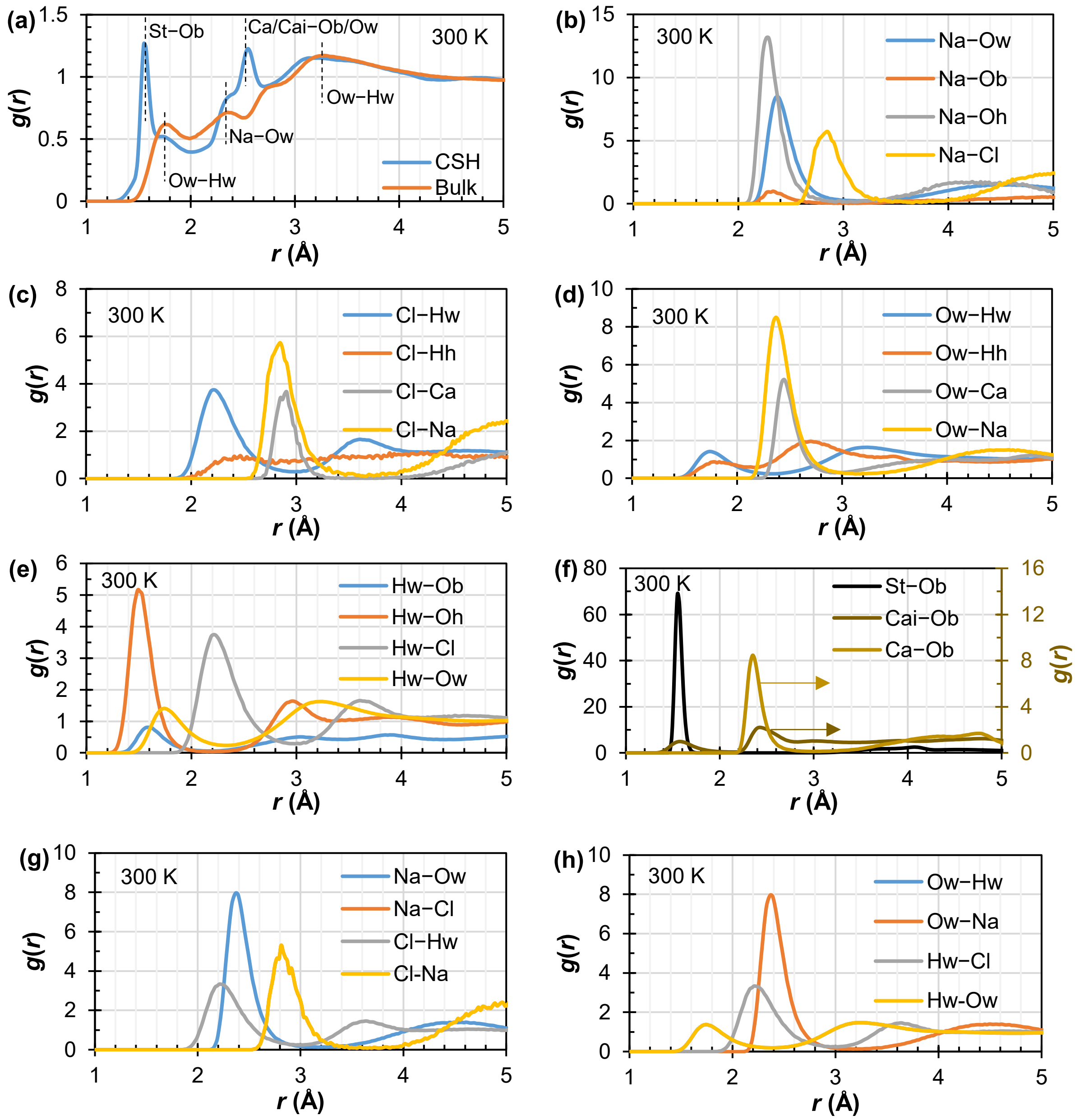

**Figure S11**. The total and partial radial distribution functions (RDFs) for (a–f) nanoconfined and (g–h) bulk solution at 300 K.

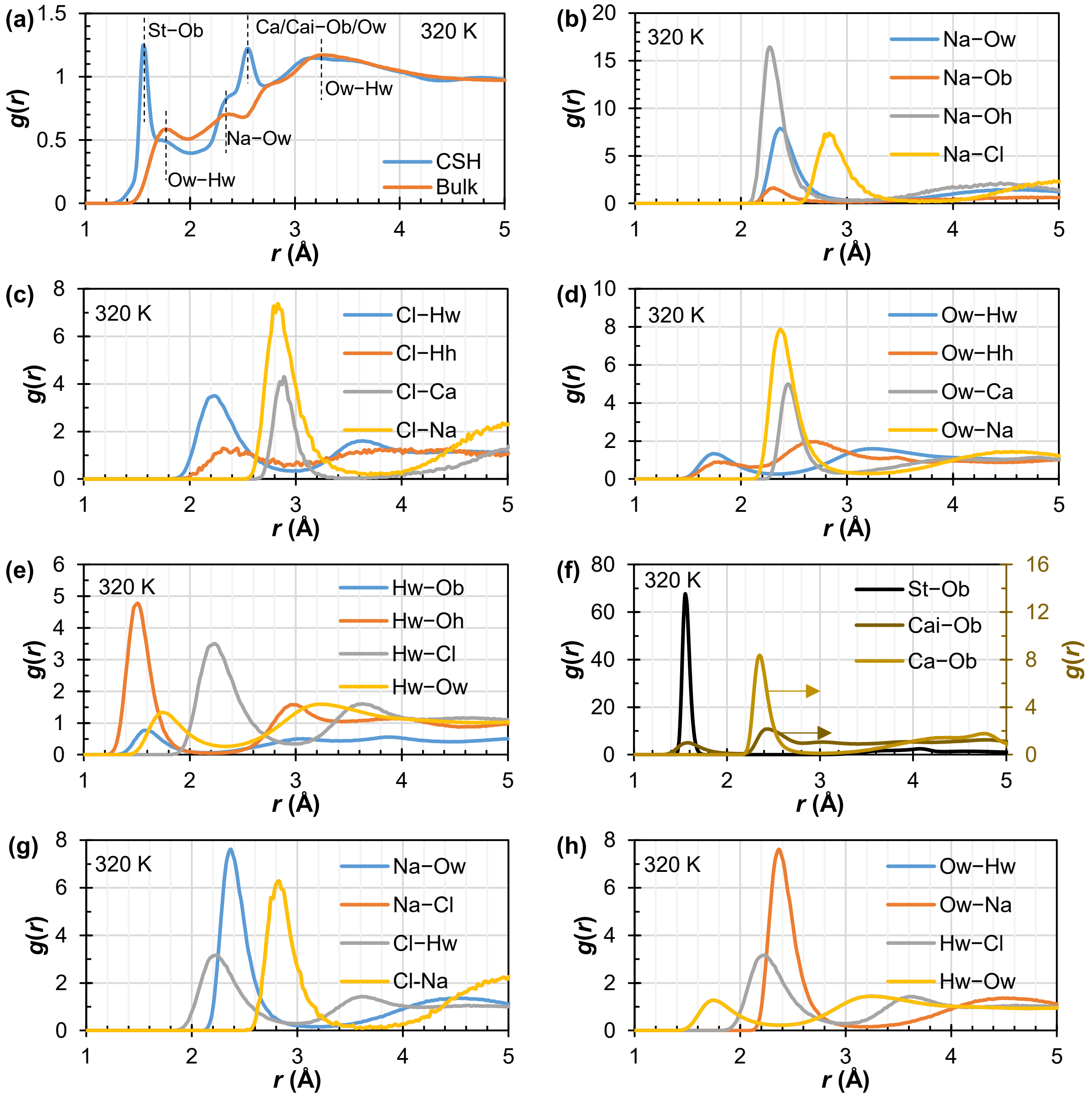

**Figure S12**. The total and partial radial distribution functions (RDFs) for (a–f) nanoconfined and (g–h) bulk solution at 320 K.

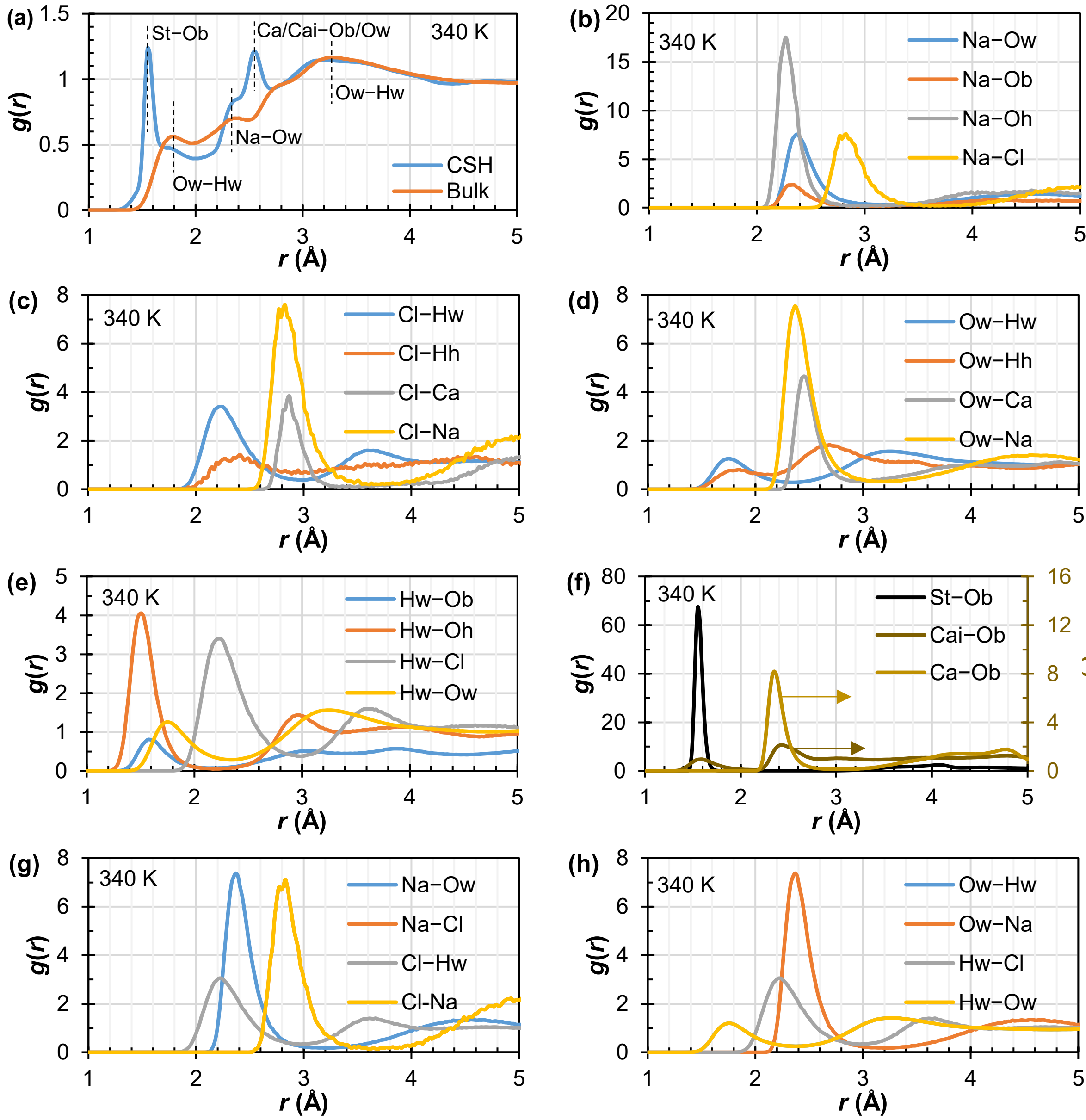

**Figure S13**. The total and partial radial distribution functions (RDFs) for (a–f) nanoconfined and (g–h) bulk solution at 340 K.

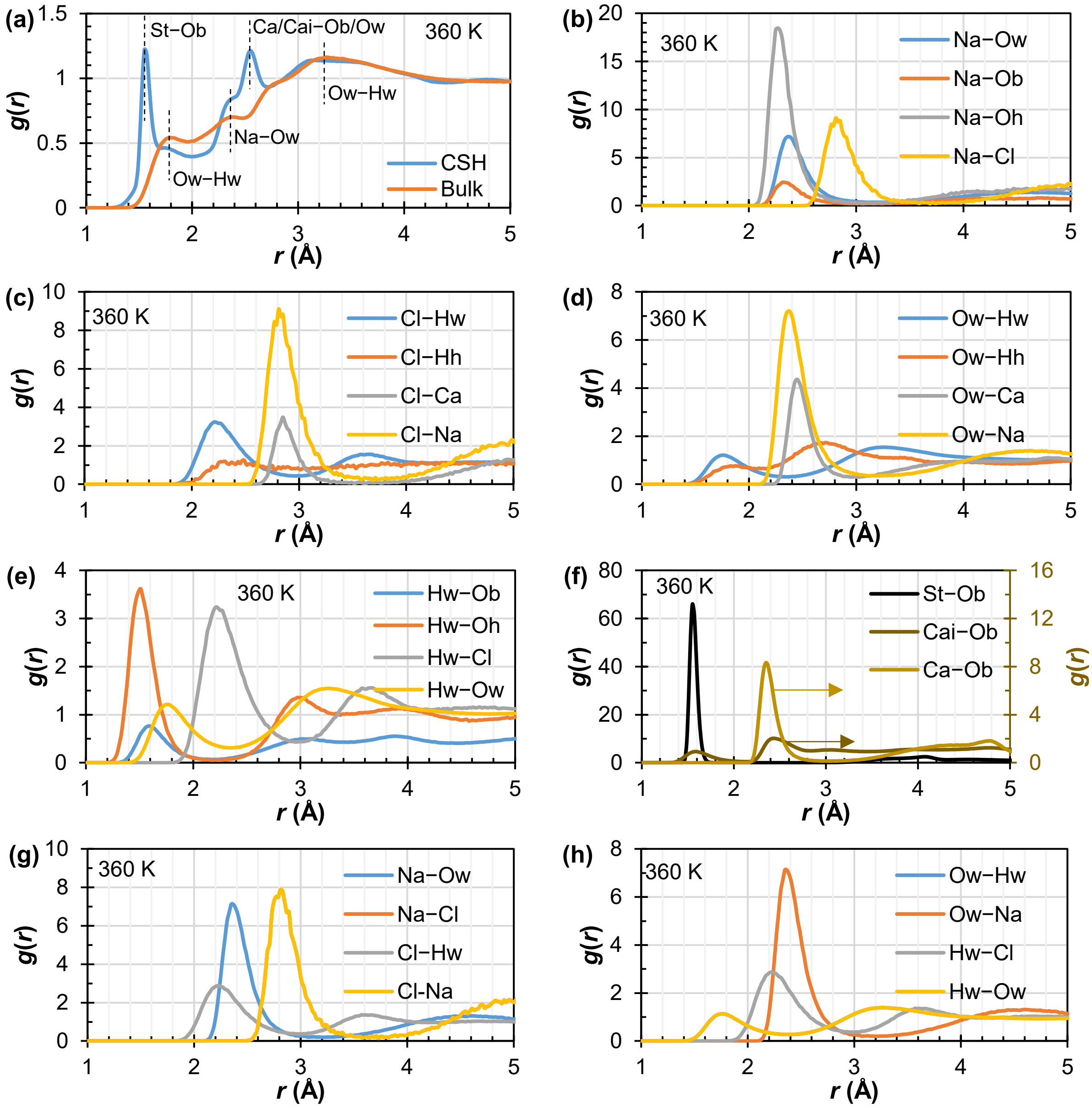

**Figure S14**. The total and partial radial distribution functions (RDFs) for (a–f) nanoconfined and (d–h) bulk solution at 360 K.

**Section S12 Cutoff distances used to calculate coordination numbers for different atom pairs**

**Table S4** summarizes the cutoff radii used to calculate coordination numbers (CNs) for various atom pairs. These cutoff values were initially determined from the exact position of the first minimum (valley) in the corresponding partial RDFs, as shown in **Figures S11–14**. Given the minor variations across different temperatures and between bulk and nanoconfined conditions, averaged cutoff values for the same atom pairs are used for calculating CN in the main text to ensure consistency and enable direct comparison.

**Table S5** shows the average coordination numbers (CNs) for different atom pairs in the interlayer solutions by integrating the RDF profiles up to their respective cutoff radii, determined from the exact first minima of the RDFs. The average CN determined using the average cutoff radii are shown in **Table 4** in the main article. **Figure S15** provides a direct comparison between the values in **Table S5** and **Table 4,** showing only trivial differences between the two approaches and leading to the same observations. For example, (1) In the bulk solution, increasing temperature leads to ion and water molecule dehydration and enhances ion pairings. A similar trend is observed under nanoconfined conditions. (2) Moreover, enhanced $Na^+$ binding to the surface is observed under nanoconfined conditions with increasing temperature, as dehydrated sodium ions preferentially coordinate with surface oxygen atoms. (3) The quantified CNs are consistent with previously reported experimental and computational data, as referenced in the caption of **Table S5**.

**Table S4.** Summary of cutoff distances used to calculate the coordination numbers of Na, Cl, and water molecules for different atom-atom pairs within their first coordination shells. The cutoff distances at various temperatures are first determined from the exact first minima of the corresponding partial RDF profiles. These values are then averaged for each atom pair, which are then used in the main text to determine the corresponding coordination number for each atom pairs (**Table 4** in the main text), irrespective of temperature or nanoconfinement, to ensure consistent and direct comparison.

| Cutoff Distance | | Cutoff distance: $r_{\mathrm{c,Na-j}}$ | | | | Cutoff distance: $r_{\mathrm{c,Cl-j}}$ | | | | Cutoff distance: $r_{\mathrm{c,Ow-j}}$ | | | | Cutoff distance: $r_{\mathrm{c,Hw-j}}$ | | | |
|---|---|---|---|---|---|---|---|---|---|---|---|---|---|---|---|---|---|
| NaCl solutions | *T* (K) | Ow | Ob | Oh | Cl | Hw | Hh | Ca | Na | Hw | Hh | Ca | Na | Ob | Oh | Cl | Ow |
| Bulk | 300 | 3.17 | 0 | 0 | 3.65 | 2.95 | 0 | 0 | 3.65 | 2.39 | 0 | 0 | 3.17 | 0 | 0 | 2.95 | 2.39 |
| | 320 | 3.13 | 0 | 0 | 3.61 | 3.01 | 0 | 0 | 3.61 | 2.37 | 0 | 0 | 3.13 | 0 | 0 | 3.01 | 2.37 |
| | 340 | 3.21 | 0 | 0 | 3.61 | 2.97 | 0 | 0 | 3.61 | 2.37 | 0 | 0 | 3.21 | 0 | 0 | 2.97 | 2.37 |
| | 360 | 3.21 | 0 | 0 | 3.73 | 2.97 | 0 | 0 | 3.73 | 2.39 | 0 | 0 | 3.21 | 0 | 0 | 2.97 | 2.39 |
| Nanoconfined | 300 | 3.13 | 2.99 | 3.03 | 3.65 | 2.99 | 2.75 | 3.43 | 3.65 | 2.33 | 2.17 | 2.93 | 3.13 | 2.15 | 2.15 | 2.99 | 2.33 |
| | 320 | 3.23 | 2.97 | 3.01 | 3.69 | 2.97 | 2.89 | 3.65 | 3.69 | 2.33 | 2.11 | 3.01 | 3.23 | 2.21 | 2.19 | 2.97 | 2.33 |
| | 340 | 3.19 | 2.99 | 3.01 | 3.61 | 2.99 | 2.89 | 3.49 | 3.61 | 2.35 | 2.15 | 2.97 | 3.19 | 2.21 | 2.17 | 2.99 | 2.35 |
| | 360 | 3.11 | 3.07 | 3.17 | 3.69 | 2.95 | 2.77 | 3.61 | 3.69 | 2.35 | 2.07 | 3.01 | 3.11 | 2.23 | 2.23 | 2.95 | 2.35 |
| Average | | 3.18 | 3.01 | 3.06 | 3.65 | 2.98 | 2.83 | 3.55 | 3.65 | 2.38 | 2.13 | 2.98 | 3.18 | 2.20 | 2.19 | 2.98 | 2.38 |

**Table S5.** Calculated coordination numbers (CNs) based on the cutoff distances determined from the exact first RDF minima shown in **Table S4**. The CNs obtained for bulk solutions at 300 K are consistent with literature-reported values for Na–Ow (5.4[12], 5.2–5.6[13], ~5.2[14]), Cl–Hw (~6.24[14], ~6.4[15]), Na–Cl (0.159[16]), and Ow–Hw (1.81 for pure water, calculated by integrating the corresponding RDF in ref.[17] and using the estimated number density for water hydrogen atom at 1 g/cm$^3$, 0.0668/Å$^3$) pairs in aqueous solutions.

| CN | | Na | | | | Cl | | | | Water | | | | | | | |
|---|---|---|---|---|---|---|---|---|---|---|---|---|---|---|---|---|---|
| | *T* (K) | Ow | Ob | Oh | Cl | Hw | Hh | Ca | Na | Hw | Hh | Ca | Na | Ob | Oh | Cl | Ow |
| Bulk | 300 | 5.55 | 0 | 0 | 0.20 | 6.54 | 0 | 0 | 0.20 | 1.64 | 0 | 0 | 0.20 | 0 | 0 | 0.12 | 0.82 |
| | 320 | 5.45 | 0 | 0 | 0.25 | 6.56 | 0 | 0 | 0.25 | 1.60 | 0 | 0 | 0.20 | 0 | 0 | 0.12 | 0.80 |
| | 340 | 5.43 | 0 | 0 | 0.28 | 6.38 | 0 | 0 | 0.28 | 1.57 | 0 | 0 | 0.20 | 0 | 0 | 0.12 | 0.78 |
| | 360 | 5.33 | 0 | 0 | 0.32 | 6.25 | 0 | 0 | 0.32 | 1.56 | 0 | 0 | 0.19 | 0 | 0 | 0.11 | 0.78 |
| Nanoconfined | 300 | 5.23 | 0.13 | 0.33 | 0.21 | 6.39 | 0.06 | 0.12 | 0.21 | 1.40 | 0.03 | 0.14 | 0.19 | 0.05 | 0.06 | 0.12 | 0.70 |
| | 320 | 5.05 | 0.24 | 0.43 | 0.27 | 6.15 | 0.09 | 0.15 | 0.27 | 1.39 | 0.03 | 0.14 | 0.18 | 0.05 | 0.06 | 0.11 | 0.70 |
| | 340 | 4.81 | 0.38 | 0.46 | 0.27 | 6.18 | 0.09 | 0.13 | 0.27 | 1.39 | 0.03 | 0.13 | 0.17 | 0.05 | 0.05 | 0.11 | 0.70 |
| | 360 | 4.54 | 0.46 | 0.56 | 0.29 | 5.90 | 0.08 | 0.13 | 0.29 | 1.38 | 0.02 | 0.12 | 0.17 | 0.05 | 0.05 | 0.11 | 0.69 |

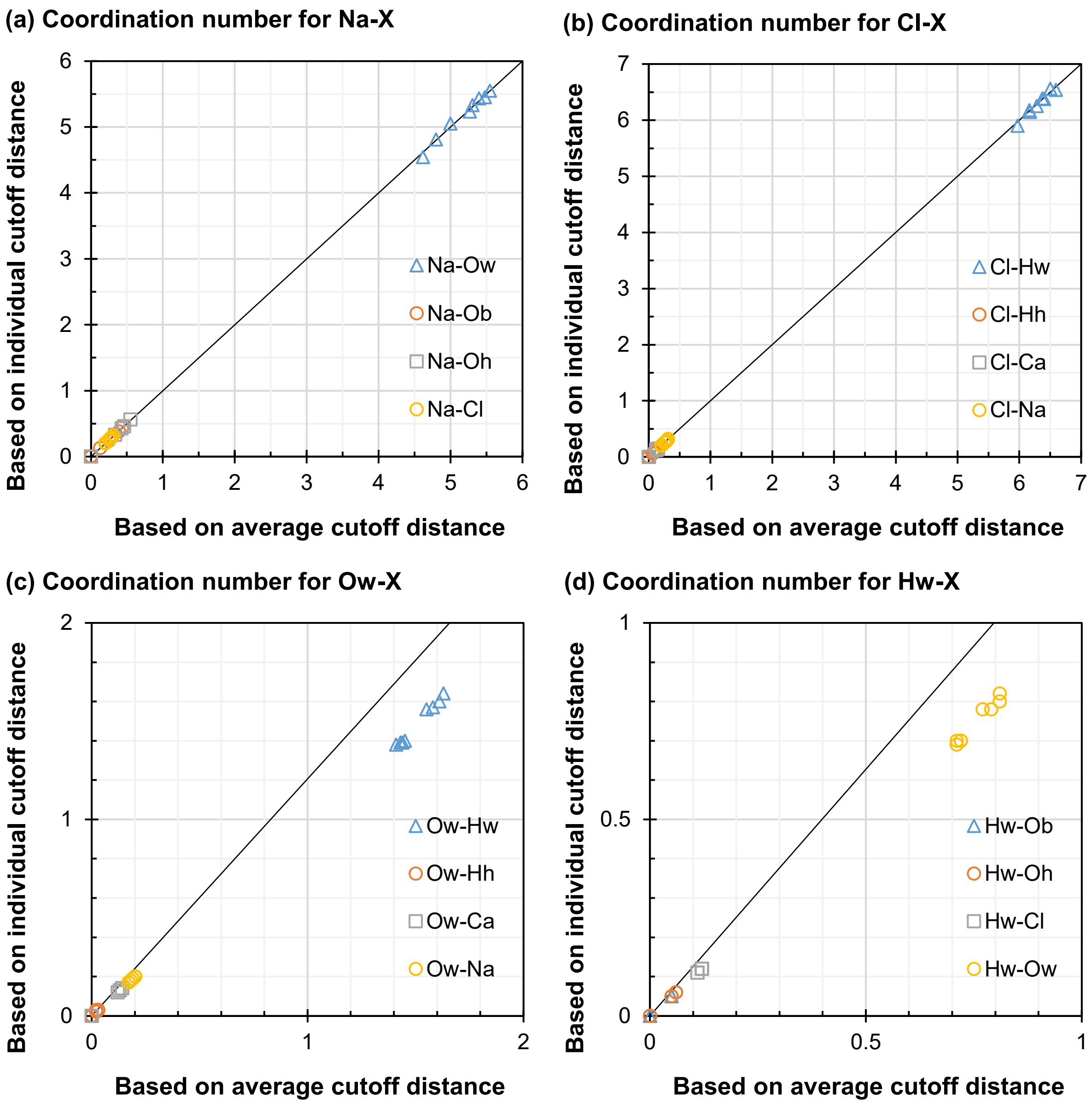

**Figure S15.** Comparison of the calculated average coordination numbers for (a) Na-X, (b) Cl-X, (c) Ow-X, and (d) Hw-X pairs, based on (i) the average cutoff distance in **Table S4** (data given in **Table 4** of the main manuscript) and (ii) individual cutoff distance in **Table S4** (data given in **Table S5** of **Supporting Information**).

## Section S13 The coordination number (CN) distribution at different temperatures

The spatial evolution of coordination numbers (CNs) across the entire channel at different temperatures was calculated from the corresponding spatially resolved partial RDFs and is shown from **Figure S16** to **Figure S19**. These profiles reveal significant variations in coordination behavior near the surface, while convergence towards bulk values is observed beyond approximately three water layers ($z > \sim 1.1$ nm).

The main observations and conclusions at elevated temperatures remain consistent with those at 300 K, as discussed in **Section 3.3.2** of the main text. For example, $Na^+$ undergoes dehydration near the C–S–H surface, and its immobilization is primarily driven by strong adsorption onto bridging oxygen (Ob) and hydroxyl (Oh) sites on the gel surface. Similarly, $Cl^-$ also dehydrates near the surface and is mainly stabilized through strong interactions with surface $Ca^{2+}$ sites, along with secondary contributions from enhanced coordination with $Na^+$ and hydrogen atoms from surface hydroxyl groups (Hh).

Near the surface, the hydrogen-bonding network among water molecules is significantly disrupted. Water molecules in the first hydration layer ($z < \sim 0.5$ nm) exhibit strong interactions with surface $Ca^{2+}$ and Oh sites, leading to their immobilization. These surface interactions weaken in the second hydration layer ($\sim 0.5$ nm $< z < \sim 0.8$ nm), where coordination involving Ow–$Na^+$, Hw–$Cl^-$, and inter-water hydrogen bonding becomes more prominent. In the third hydration layer ($\sim 0.8$ nm $< z < \sim 1.1$ nm), the primary deviation from bulk water behavior is due to a persistent increase in Hw–$Cl^-$ coordination, suggesting residual surface influence beyond the immediate interface.

Furthermore, while minor discrepancies exist between the profiles on the left and right sides—attributable to the inherent amorphousness and structural asymmetry of the C–S–H gel surfaces—the primary trends remain consistent.

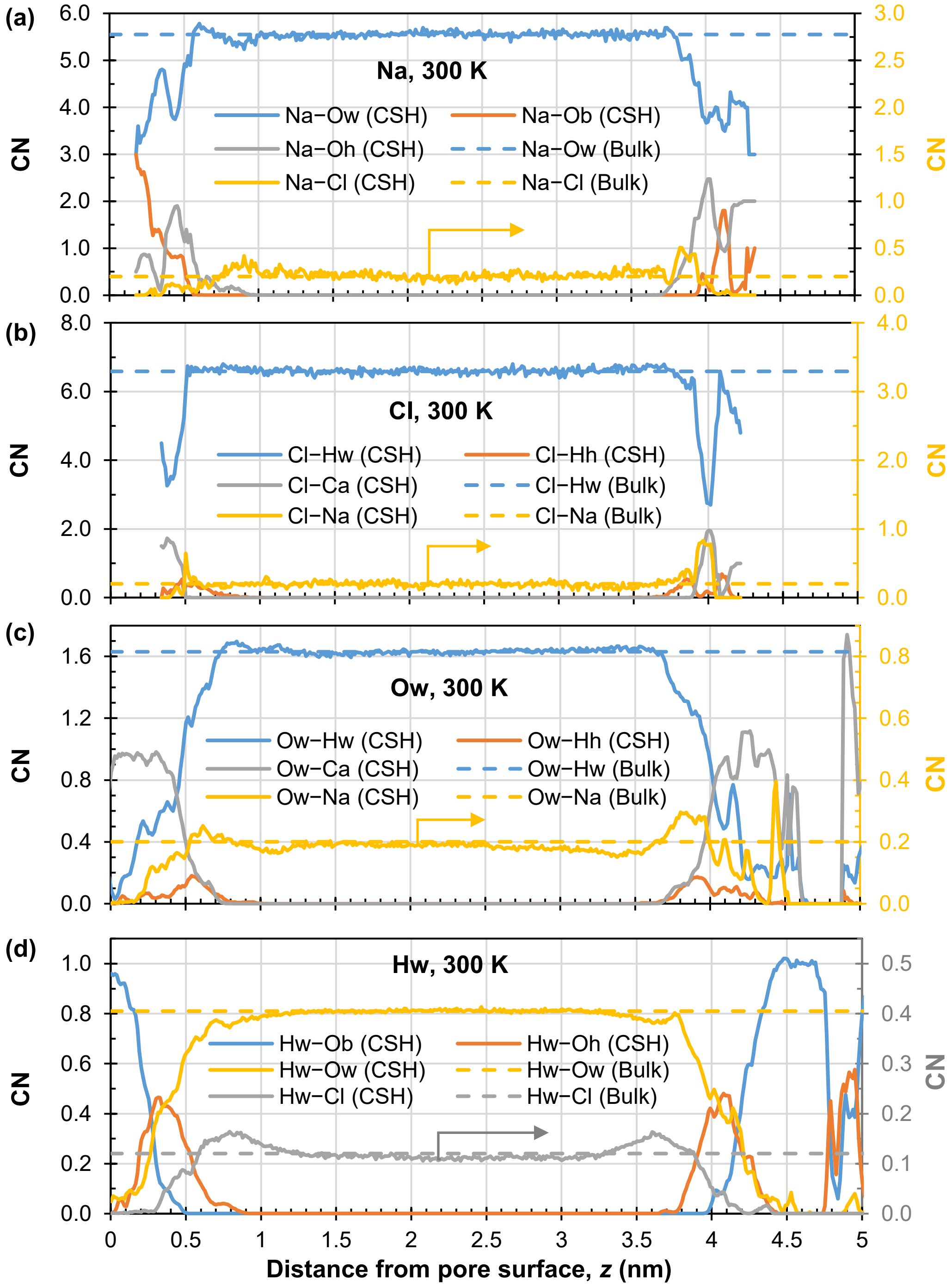

**Figure S16**. Spatial evolution of coordination numbers (CN) across the entire C–S–H nanopore, for (a) $Na^+$, (b) $Cl^-$, (c) Ow, and (d) Hw, with respect to their nearest neighbors at 300 K. The horizontal dashed lines represent the corresponding CNs values in the bulk NaCl solutions.

In addition, the coordination numbers for Ow-Na and Hw-Cl at the channel center decrease relative to their bulk values as temperature increases, indicating a progressively weakened coordination environment in this region at increasing temperature. This effect may have contributed to the reduced activation energy barrier in the middle of the nanochannel relative to its bulk as observed in **Figure 7** and discussed in **Section 3.2.2** in the main text.

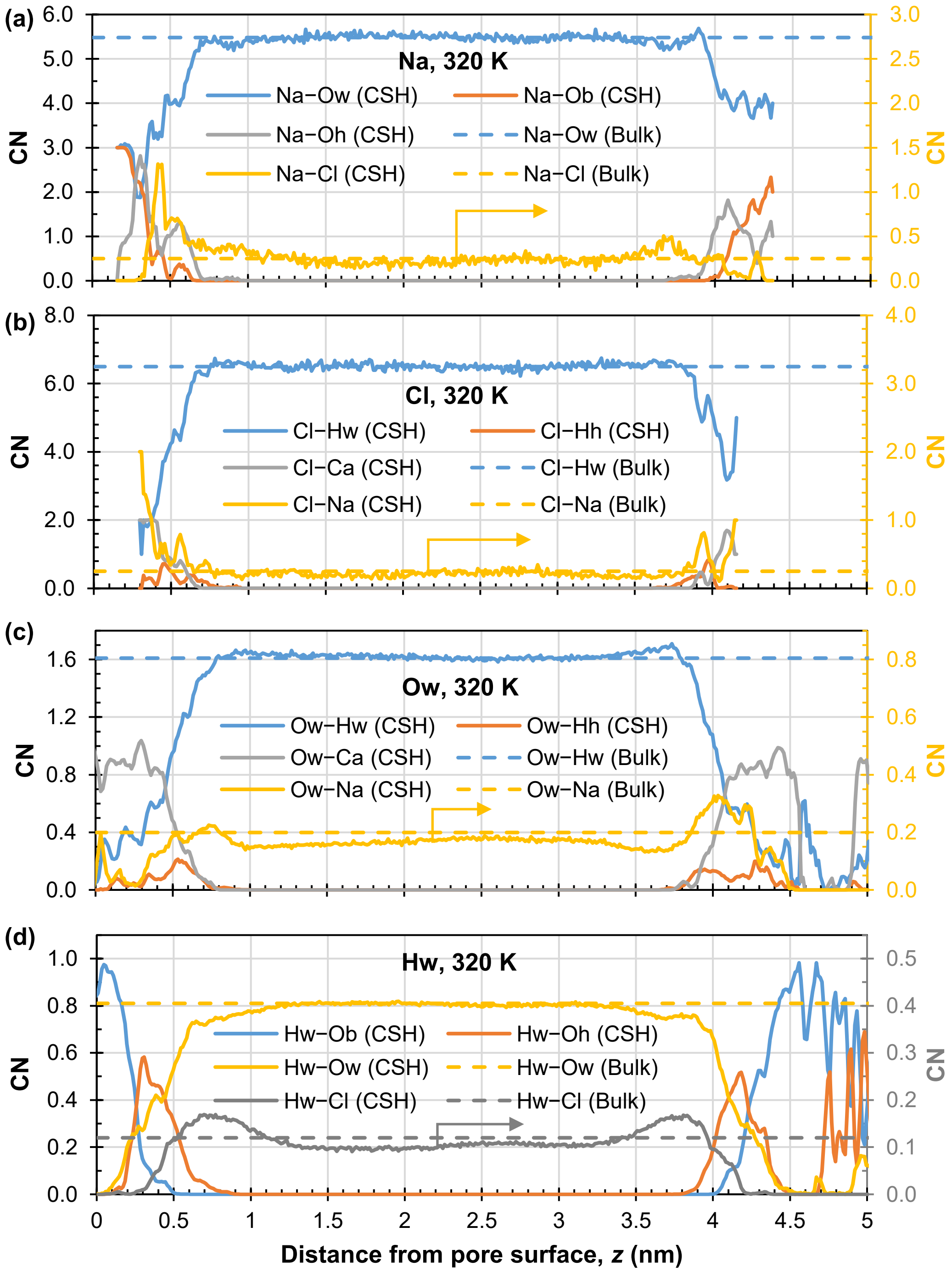

**Figure S17**. Spatial evolution of coordination numbers (CN) across the C–S–H nanopore, for (a) $Na^+$, (b) $Cl^-$, (c) Ow, and (d) Hw, with respect to their nearest neighbors at 320 K. The horizontal dashed lines represent the corresponding CNs values in the bulk NaCl solutions.

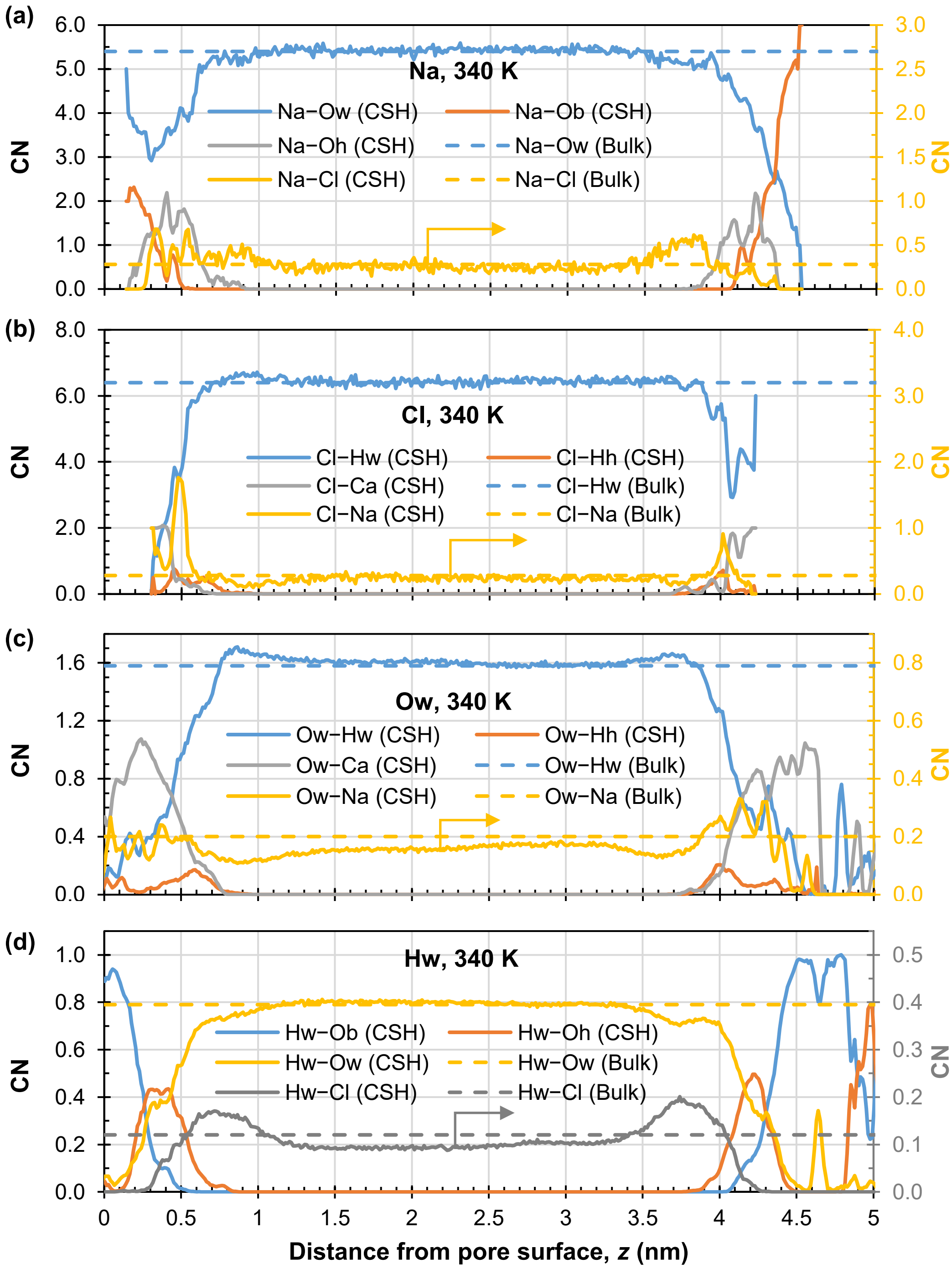

**Figure S18**. Spatial evolution of coordination numbers (CN) across the C–S–H nanopore, for (a) $Na^+$, (b) $Cl^-$, (c) Ow, and (d) Hw, with respect to their nearest neighbors at 340 K. The horizontal dashed lines represent the corresponding CNs values in the bulk NaCl solutions.

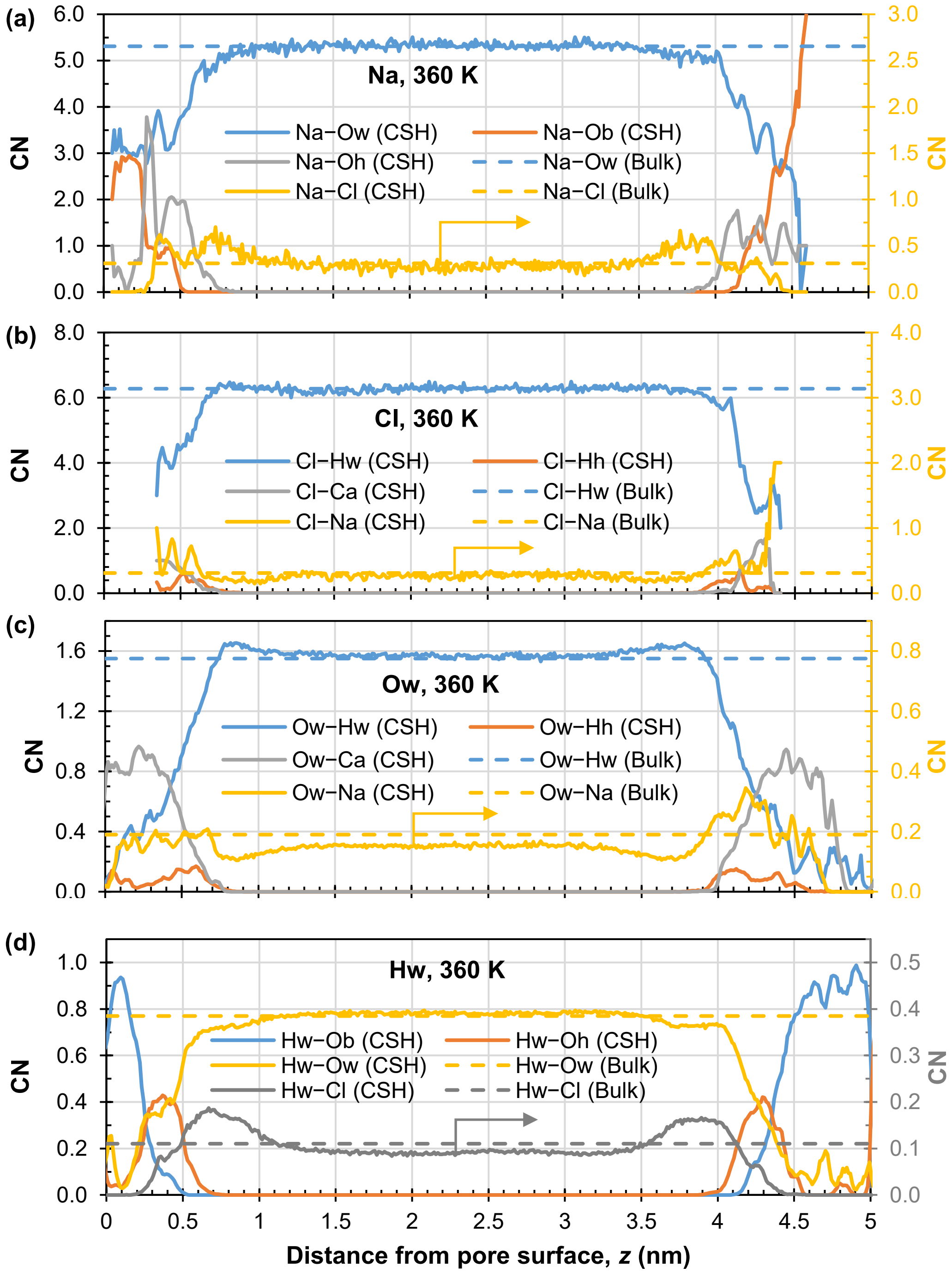

**Figure S19**. Spatial evolution of coordination numbers (CN) across the C–S–H nanopore, for (a) $Na^+$, (b) $Cl^-$, (c) Ow, and (d) Hw, with respect to their nearest neighbors at 360 K. The horizontal dashed lines represent the corresponding CNs values in the bulk NaCl solutions.

**Section S14 The total coordination strength (TCS) distribution at different temperatures**

As introduced in **Section 3.4** of the main text, the total coordination strength (TCS) distributions for each nanoconfined solution species at different temperatures are calculated and shown from **Figure S20** to **Figure S23**. It is found that near the C–S–H surface, the TCS values for both ions and water are significantly elevated, indicating stronger local structural constraints. The TCS values decrease progressively with increasing distance from the surface, approaching bulk-like values beyond $z$ = ~1.0 nm.

A detailed breakdown of the individual contributions to the TCS of each species at 300 K is also presented in **Figure S20**, where panels (a)–(c) show results based on the force method while panels (d)–(f) use the energy method (see **Section S2** on the methods). The elevated TCS near the surface is mainly due to (i) Na-Oh and Na-Ob interactions for $Na^+$, (ii) Cl-Ca interactions for $Cl^-$, and (iii) Ow-Ca, Hw-Ob and Hw-Oh interactions for water molecules.

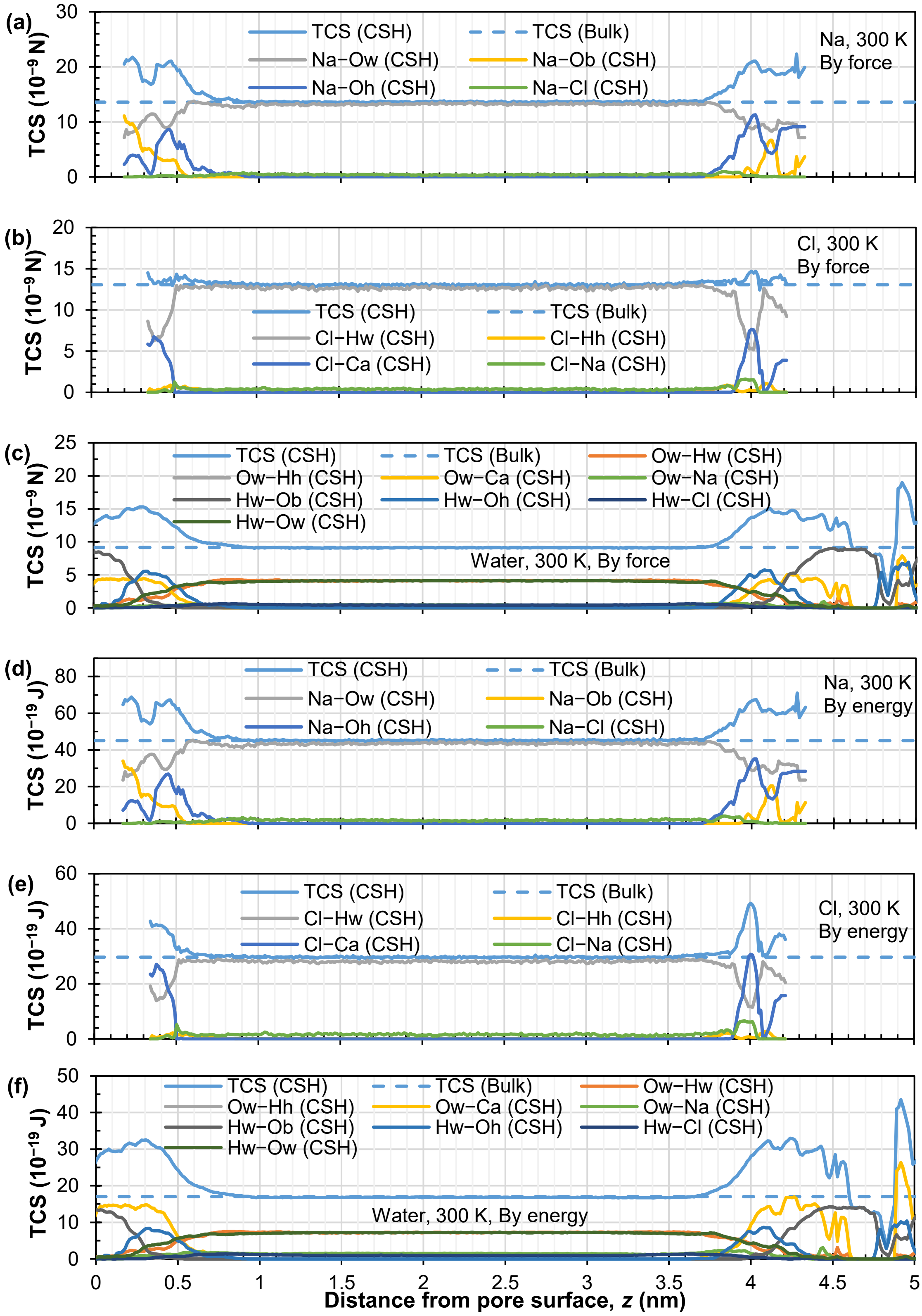

**Figure S20**. Total coordination strength (TCS) distributions of nanoconfined solution species at 300 K, calculated using (a–c) the bond force and (d–f) the bond energy method. Individual contributions to the overall TCS for each species are also presented in each figure.

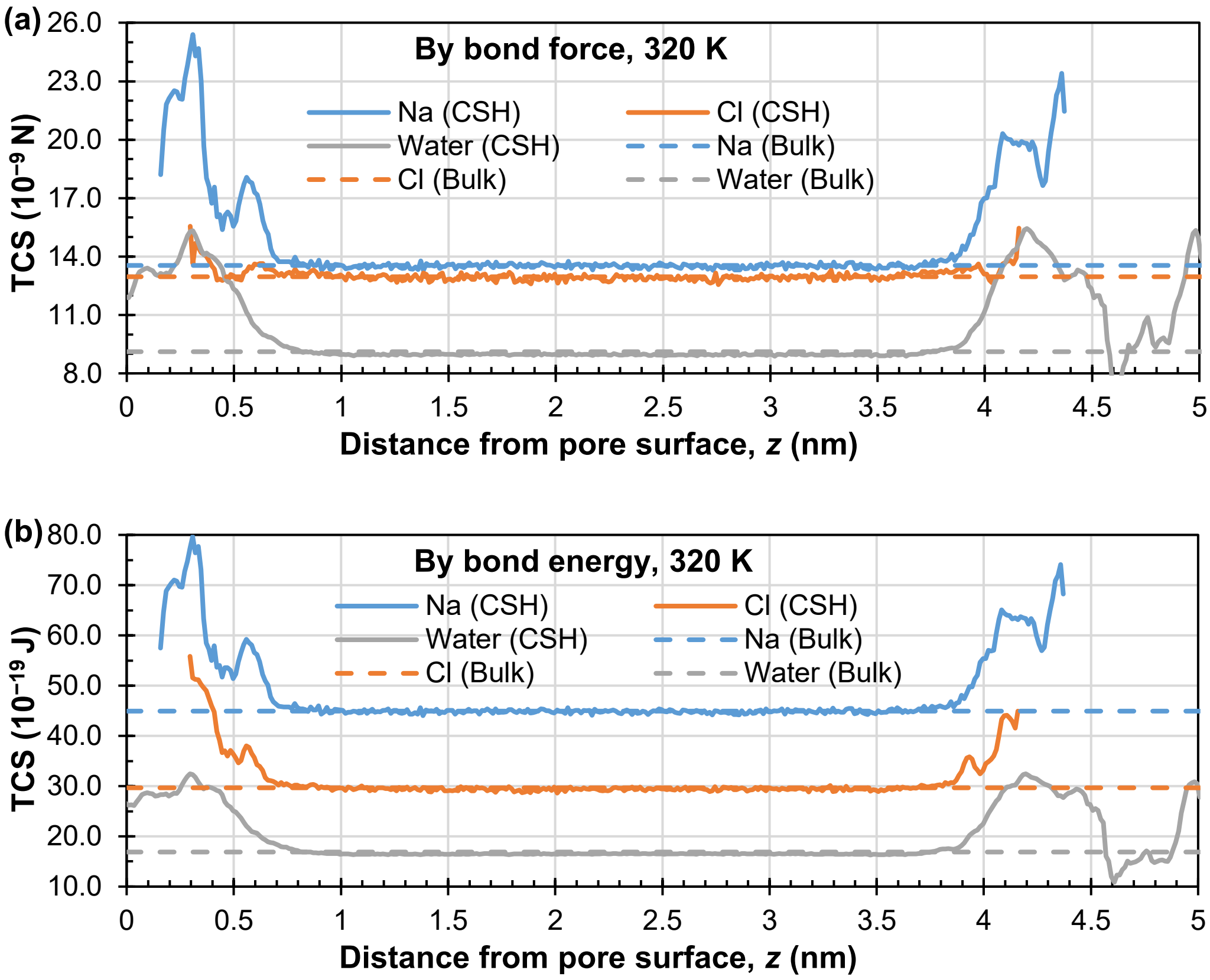

**Figure S21**. The total coordination strength (TCS) distribution of nanoconfined solution species at 320 K, calculated by (a) the bond force and (b) the bond energy method.

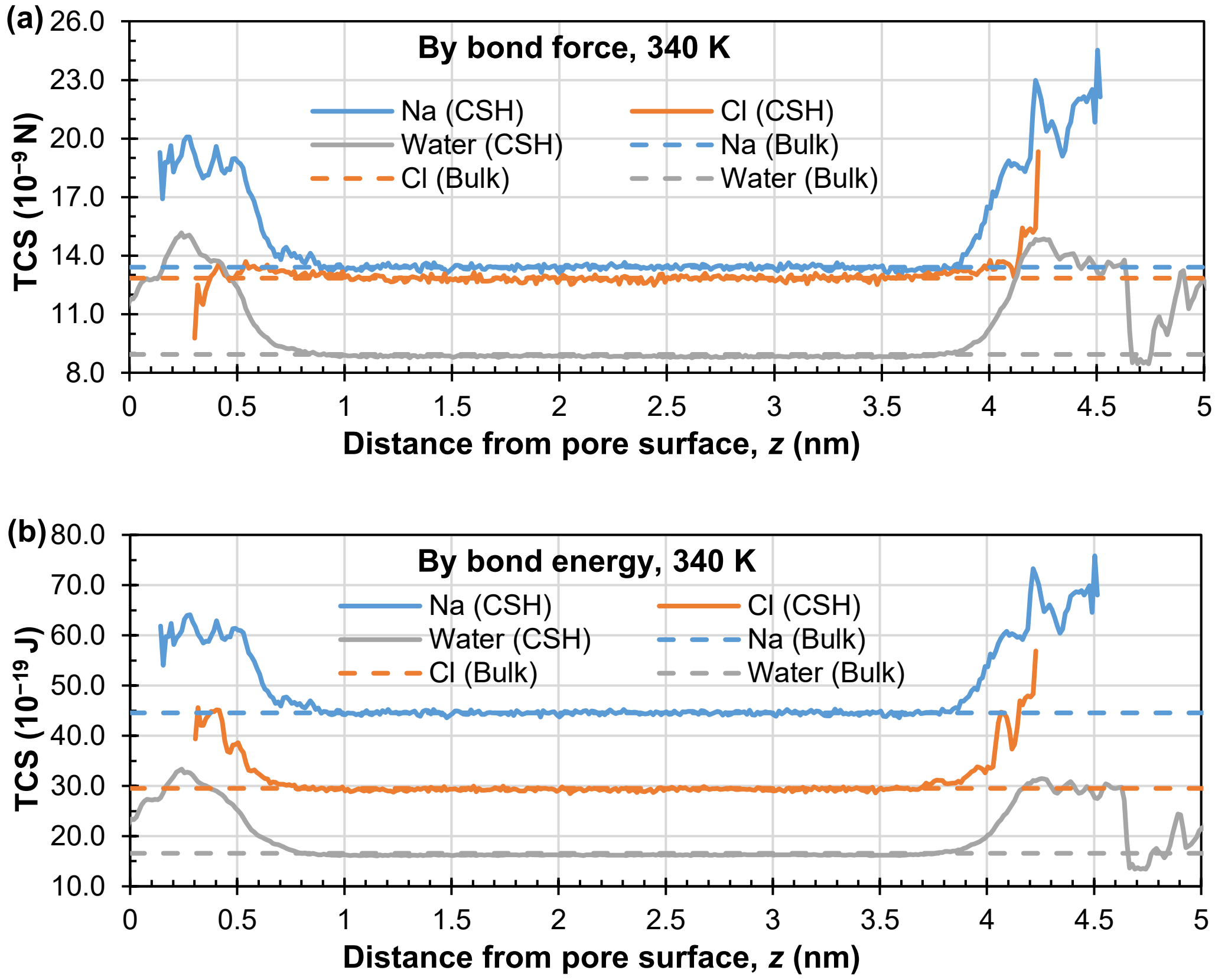

**Figure S22**. The total coordination strength (TCS) distribution of nanoconfined solution species at 340 K, calculated by (a) the bond force and (b) the bond energy method.

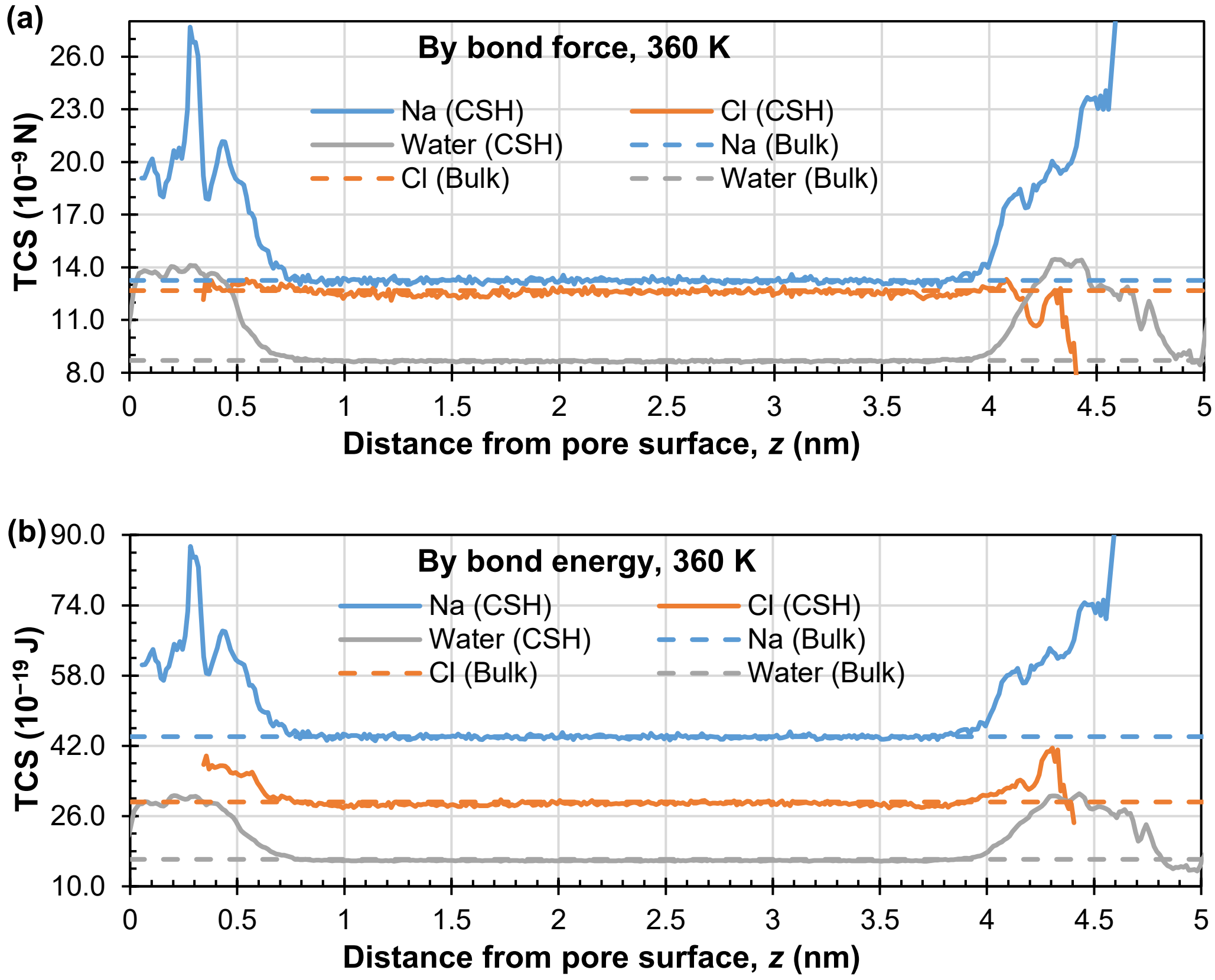

**Figure S23**. The total coordination strength (TCS) distribution of nanoconfined solution species at 360 K, calculated by (a) the bond force and (b) the bond energy method.

**Section S15 The correlation between the diffusivity and the total coordination strength (TCS) calculated by the bond force**

The total coordination strength (TCS) results calculated using bond energy are presented in **Figure 10** in the main text. The corresponding results calculated using the bond force method are shown below in **Figure S24**. The overall trends and observations are consistent with those discussed in the main text.

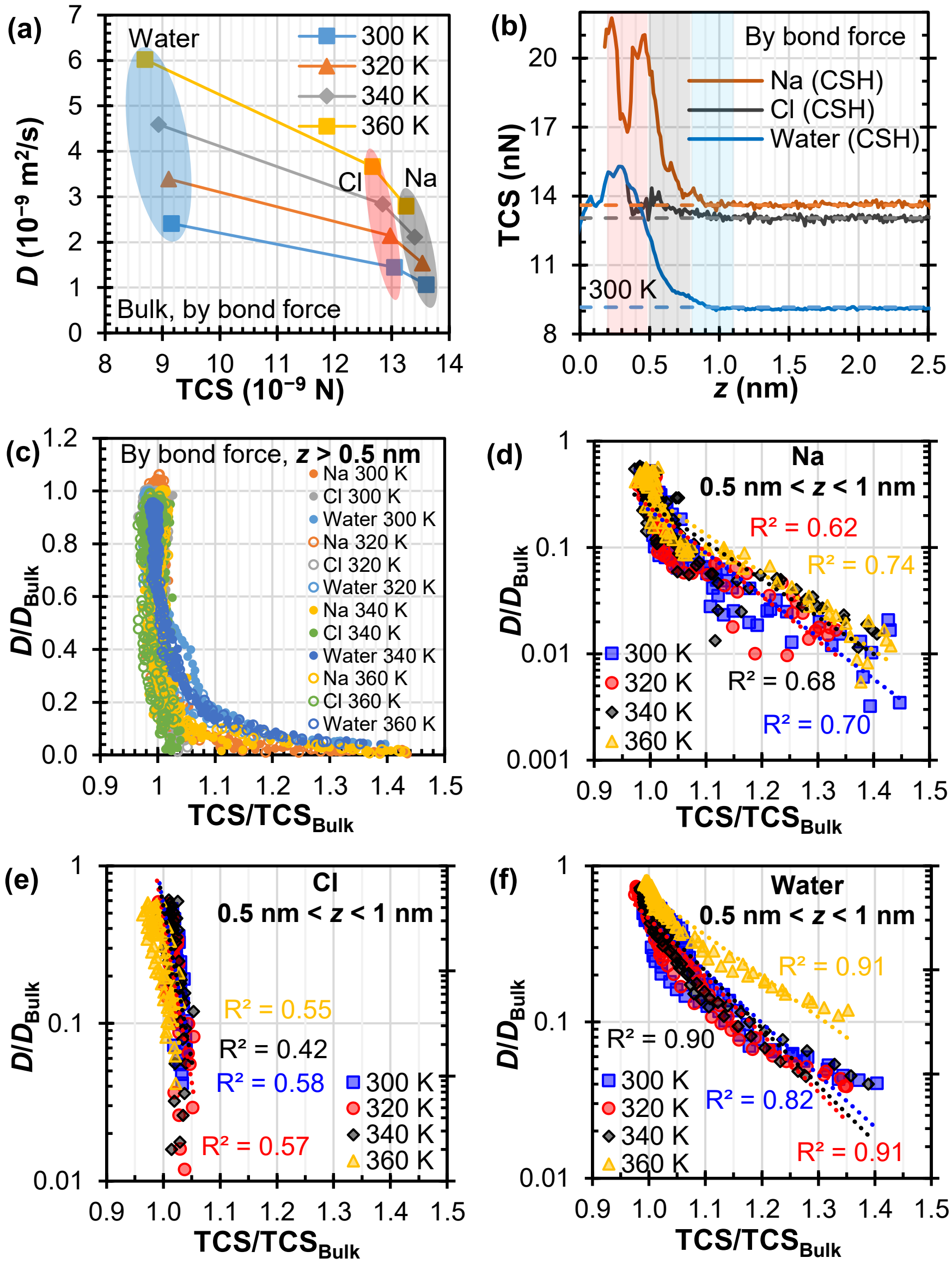

**Figure S24.** (a) Correlation between the total coordination strength (TCS) and the diffusion coefficient for $Na^+$, $Cl^-$, and water molecules under bulk conditions. (b) Spatial profiles of the TCS for each solution species across the C–S–H nanochannel at 300 K. (c) Dependence of normalized diffusivity ($D/D_{bulk}$) on normalized TCS ($TCS/TCS_{bulk}$) for nanoconfined $Na^+$, $Cl^-$, and water molecules across all temperatures studied. (d–f) Exponential correlation between normalized diffusivity and normalized TCS for (d) $Na^+$, (e) $Cl^-$, and (f) water molecules within the near-surface region (0.5 nm < $z$ < 1.0 nm). $R^2$ values in (d–f) were obtained from exponential regression fits.

**Section S16 Analytical solution of the one-dimensional steady-state Darcy–Brinkman equation**

Under an applied constant pressure gradient in the *x* direction ($-\frac{\mathrm{d}p}{\mathrm{d}x}$), the spatial distribution of fluid velocity profile along the *z*-direction, $u(z) \equiv u_x(z)$, near the solid-fluid interface can be described by the one-dimensional steady-state Darcy-Brinkman equation for an isotropic medium in a semi-infinite domain (liquid region, $z \geq 0$)[18]:

$$\eta \frac{\mathrm{d}^2 u(z)}{\mathrm{d}z^2} - \eta\alpha^2 u(z) = -\frac{\mathrm{d}p}{\mathrm{d}x},\ z \geq 0 \tag{S8}$$

where, $\eta$ is the dynamic viscosity of the fluid, $\alpha$ is the Brinkman damping parameter related to the permeability $K$ via $K = \frac{1}{\alpha^2}$, $-\frac{\mathrm{d}p}{\mathrm{d}x}$ is the applied pressure gradient driving the fluid flow, and $u(z)$ the axial velocity profile along the $z$-direction. When a partial slippage boundary condition is applied at the solid-fluid interface, i.e., $z = 0$,

$$u(0) = L_s \left.\frac{du(z)}{dz}\right|_{z=0}, \tag{S9}$$

where, $L_s$ is the slippage length. The analytic solution for the velocity profile is:

$$u(z) = u_{z\to+\infty}\left(1 - \frac{e^{-\alpha z}}{1+\alpha L_s}\right), \tag{S10}$$

where, $u_{z\to+\infty}$ denotes the asymptotic fluid velocity far from the surface, corresponding to the bulk flow regime.

**Section S17 The empirical equation fitting of water and ion diffusivity profiles**

**Figure S25** shows the fitted diffusivity profiles at $z > \sim 1.0$ nm for all cases using **Equation** (7) in the main text. It is found that the empirical model accurately captures the spatial distribution of water and ion diffusivity for $z > \sim 1.0$ nm, while deviating in the near-surface region ($z < \sim 1.0$ nm), consistent with observations in **Figure 11** of the main text.

To further assess the robustness of the empirical fitting model inspired by the one-dimensional steady-state Darcy–Brinkman equation, we extended the fitting range beyond the originally chosen boundary of $z > 1.0$ nm. In particular, fittings were performed for $z > 0.75$ nm and $z > 0.5$ nm. The resulting curves are presented in **Figure S26** and **Figure S27**. The extended fittings show that the functional form is able to capture the overall decay of the diffusion coefficient ($D$) from the pore center toward the surface and provide seemingly good agreement across the entire region above 0.5 nm. However, systematic deviations are observed within 1 nm, especially for $Cl^-$ and water at 360 K. These discrepancies originate from the strong interfacial effects within the regime of $z < 1.0$ nm, including enhanced adsorption and local structuring of solution species at the C–S–H surface. Thus, the adoption of a fixed 1.0 nm interfacial layer in the main text is not arbitrary but is supported by structural analyses, such as the total coordination strength (TCS), which demonstrate that solution species within $z < 1.0$ nm exhibit markedly stronger local coordination compared to those in the pore interior. This provides a consistent physical basis for treating the near-surface region separately from the pore interior in the analysis of transport behavior.

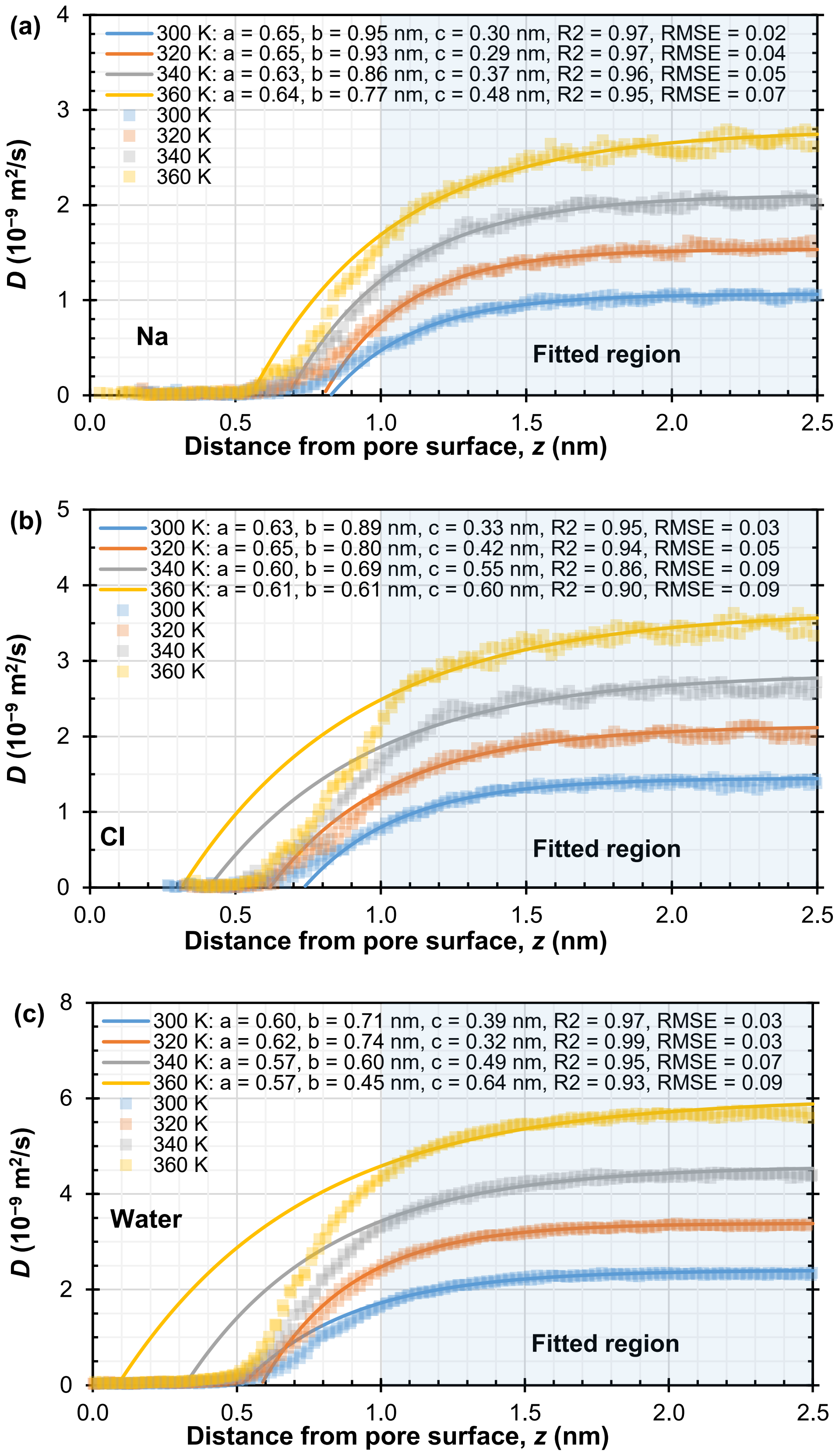

**Figure S25**. Empirical equation (**Equation** (7)) fitting of spatial diffusivity profiles within $z > 1$ nm for (a) $Na^+$, (b) $Cl^-$, and (c) water at different temperatures. The goodness-of-fit ($R^2$) along with the fitting parameters are given in each figure.

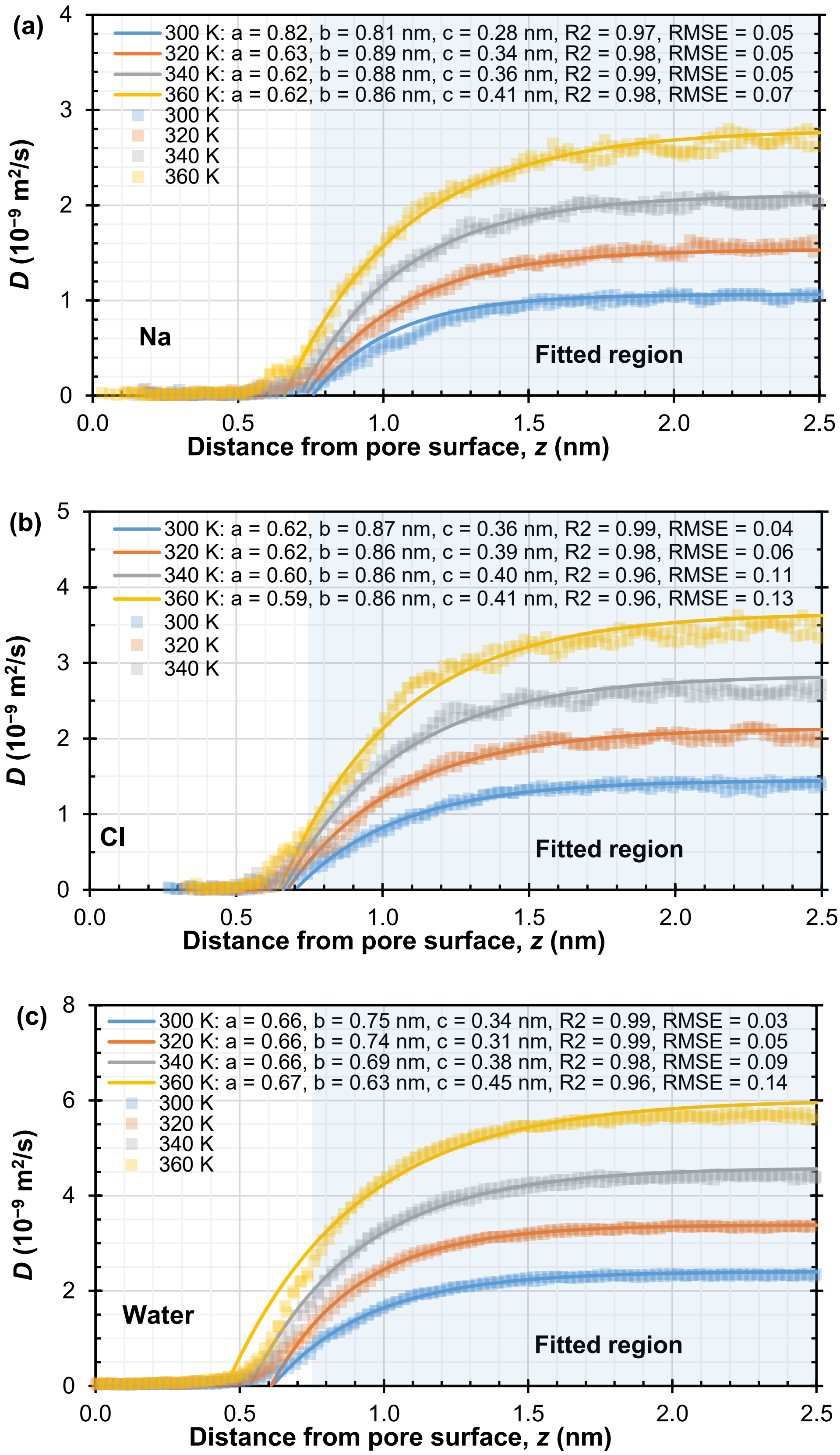

**Figure S26.** Empirical equation (**Equation** (7)) fitting of spatial diffusivity profiles within $z > 0.75$ nm for (a) $Na^+$, (b) $Cl^-$, and (c) water at different temperatures. The goodness-of-fit ($R^2$) along with the fitting parameters are given in each figure.

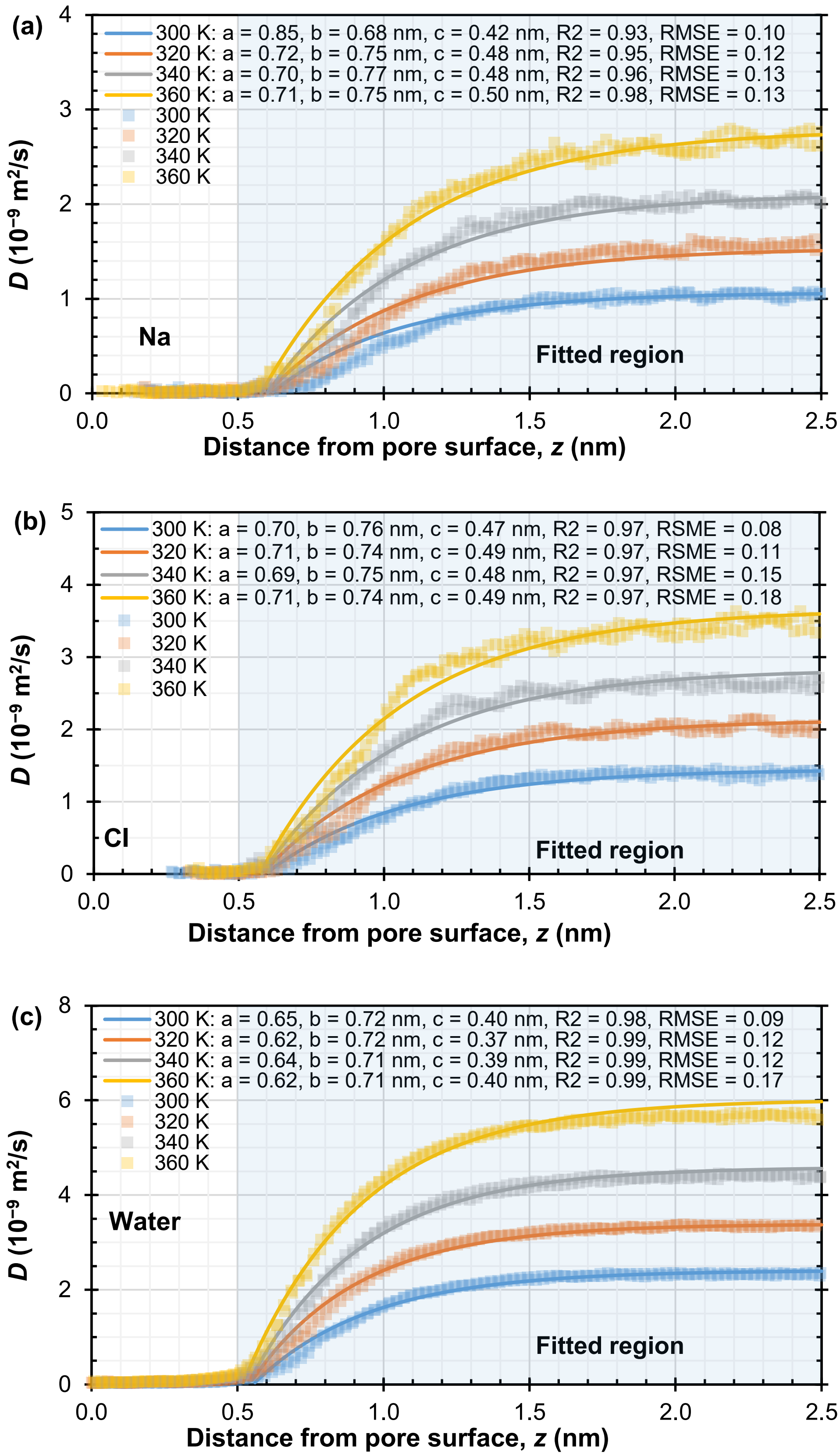

**Figure S27.** Empirical equation (**Equation** (7)) fitting of spatial diffusivity profiles within $z > 0.5$ nm for (a) $Na^+$, (b) $Cl^-$, and (c) water at different temperatures. The goodness-of-fit ($R^2$) along with the fitting parameters are given in each figure.

## Section S18 References